%% file: main.tex
\documentclass[acmsmall]{acmart}

\AtBeginDocument{%
  \providecommand\BibTeX{{%
    \normalfont B\kern-0.5em{\scshape i\kern-0.25em b}\kern-0.8em\TeX}}}

\usepackage{tikz,pgfplots}
\usepackage{standalone}
\usepackage[listings,skins,breakable]{tcolorbox}
\usepackage{multirow}
\usepackage{subcaption}
\usepackage{makecell}
\usepackage{hyperref}
\usepackage{pifont}
\usepackage{balance}
\usepackage{booktabs}
\usepackage{pgfplots}
\usepackage{amsmath}
\usepackage[ruled, linesnumbered, vlined, commentsnumbered]{algorithm2e}
\usepackage{xcolor,colortbl}
\usepackage{adjustbox}

\newtcolorbox{quotebox}{colback=teal!10,boxrule=0.4pt,colframe=black,fonttitle=\bfseries,top=2pt,bottom=2pt}
\newtcolorbox{expbox}{colback=red!10,boxrule=0.4pt,colframe=black,fonttitle=\bfseries,top=2pt,bottom=2pt}
\definecolor{steel}{rgb}{0, 0.2, 0.9} 
\newcommand{\approach}{\texttt{MNAL}}

\DeclareMathAlphabet\mathbfcal{OMS}{cmsy}{b}{n}

\pgfplotsset{compat=1.11,
   /pgfplots/xbar legend/.style={
   /pgfplots/legend image code/.code={%
      \draw[##1,/tikz/.cd,yshift=-0.3em]
       (0cm,0cm) rectangle (3pt,0.8em);},
  },
}

\acmJournal{TOSEM}
\acmVolume{37}
\acmNumber{4}
\acmArticle{111}
\acmMonth{8}

\newtcbox{\mytag}{nobeforeafter,colframe=blue,colback=blue!30!white,boxrule=0.7pt,arc=0pt,
 boxsep=-3pt,left=6pt,right=6pt,top=4pt,bottom=5pt,tcbox raise base}
\renewcommand{\mytag}[1]{}

\newtcbox{\mytagTwo}{nobeforeafter,colframe=blue,colback=blue!30!white,boxrule=0.7pt,arc=0pt,
 boxsep=-3pt,left=6pt,right=6pt,top=4pt,bottom=5pt,tcbox raise base}

\begin{document}

\title{Human-Machine Co-boosted Bug Report Identification with Mutualistic Neural Active Learning}

\author{Guoming Long}
\authornote{Guoming Long is also supervised in the IDEAS Lab.}
\affiliation{%
  \institution{University of Electronic Science and Technology of China}
  \city{Chengdu}
  \country{China}}
  
\author{Shihai Wang}
\authornote{Shihai Wang is also supervised in the IDEAS Lab.}
\affiliation{%
  \institution{University of Electronic Science and Technology of China}
  \city{Chengdu}
  \country{China}}

\author{Hui Fang}
\affiliation{%
  \institution{Loughborough University}
  \city{Loughborough}
  \country{United Kingdom}}
\email{h.fang@lboro.ac.uk}

\author{Tao Chen}
\authornote{Corresponding author: Tao Chen, t.chen@bham.ac.uk.}
\affiliation{%
  \institution{IDEAS Lab, University of Birmingham}
  \city{Birmingham}
  \country{United Kingdom}}
\email{t.chen@bham.ac.uk}

\begin{abstract}
Bug reports, encompassing a wide range of bug types, are crucial for maintaining software quality. However, the increasing complexity and volume of bug reports pose a significant challenge in sole manual identification and assignment to the appropriate teams for resolution, as dealing with all the reports is time-consuming and resource-intensive. In this paper, we introduce a cross-project framework, dubbed Mutualistic Neural Active Learning (\approach), designed for automated and more effective identification of bug reports from GitHub repositories boosted by human-machine collaboration. \approach~utilizes a neural language model that learns and generalizes reports across different projects, coupled with active learning to form neural active learning. A distinctive feature of \approach~is the purposely crafted mutualistic relation between the machine learners (neural language model) and human labelers (developers) when enriching the knowledge learned. That is, the most informative human-labeled reports and their corresponding pseudo-labeled ones are used to update the model while those reports that need to be labeled by developers are more readable and identifiable, thereby enhancing the human-machine teaming therein. We evaluate \approach~using a dataset of $1,275,881$ reports from over $127,000$ software projects against the state-of-the-art approaches, baselines, and different variants. The results indicate that, remarkably, \approach~achieves up to 95.8\% and 196.0\% effort reduction in terms of readability and identifiability during human labeling, respectively, while resulting in a better performance (e.g., F1-score) in bug report identification. Additionally, our \approach~is model-agnostic since it is capable of improving the model performance with various underlying neural language models. To further verify the efficacy of our approach, we conducted a qualitative case study involving 10 human participants, who rate \approach~as being more effective while saving more time and monetary resources. The dataset and code are made publicly available at \textcolor{blue}{\texttt{\url{https://github.com/ideas-labo/MNAL}}}. 




\end{abstract}



\received{20 February 2007}
\received[revised]{12 March 2009}
\received[accepted]{5 June 2009}

\maketitle

\input{sec/intro}

\input{sec/bg}
\input{sec/method}

\input{sec/exp}

\input{sec/result}

\input{sec/why}
\input{sec/threat}

\input{sec/related}

\input{sec/conclusion}

\section*{Data Availability Statement}

The dataset and replication package supporting this study are publicly
available at Zenodo: \textcolor{blue}{\url{https://zenodo.org/records/19193244}}.

\section*{Acknowledgment}
This work was supported by a NSFC Grant (62372084) and a UKRI Grant (10054084).

\bibliographystyle{ACM-Reference-Format}
\bibliography{references}

\end{document}

%% file: sec/intro.tex
\section{Introduction}


Software bugs can cause devastating consequences for developers and end-users, including crashes, data loss, and performance degradation \cite{DBLP:journals/sncs/TranLNH20, DBLP:journals/hcis/LeeS20,DBLP:conf/icse/WangChen26,irap2026}, especially for the modern AI systems~\cite{DBLP:conf/ease/LongC22}. To identify and fix bugs in software systems, modern version control systems, such as GitHub, allow practitioners to formally submit posts, serving as a critical channel for users to report issues, glitches, or malfunctions they have encountered. As such, bug reporting has been playing a pivotal role in the software development lifecycle~\cite{DBLP:journals/tse/ZimmermannPBJSW10}. However, those submitted reports might not be bug related; they might well be feature requests, documentation updates, or even reports of a false bug alarm simply incurred by an inappropriate usage. In practice, services like GitHub allow the submitter to label the issues. However, this feature is unreliable because:

\begin{itemize}
    \item It has been reported that a considerable amount of submitted GitHub issues are unlabeled due to, e.g., the submitter being a newcomer to the project or lack of experiences~\cite{DBLP:conf/icsm/KallisSCP19,DBLP:journals/infsof/WangZCX22}. In a recent large-scale study over GitHub~\cite{DBLP:journals/infsof/WangZCX22}, it has been found that 54.4\% of the issues studied have no label at all. 
    \item Even if an issue has a label, the quality varies~\cite{DBLP:journals/infsof/PanichellaCS21,DBLP:conf/kbse/HooimeijerW07,DBLP:journals/tse/ZimmermannPBJSW10}. For example, a feature request could turn out to be a bug or a bug issue might turn out to be a false alarm.
\end{itemize}

All the above urge the need to automatically predict/verify whether a submitted issue report is bug-related or not, even if it comes with a submitter-provided label. Therefore, the engineering task, namely bug report identification that distinguishes bug related reports from the others, is crucial for ensuring the effectiveness of bug triage~\cite{10.1145/3750040}.

While it can be the most ideal situation to predict multiple types of issues, it is known that those multiple class predictions can be highly inaccurate and hence lack practical significance~\cite{DBLP:conf/apsec/LongCC22}, since the possible number of classes is large. Therefore, here in this work, we focus on a binary classification problem: identifying whether an issue report is indeed bug-related, which is more likely to achieve high accuracy and hence being useful. The outcome would still be significant because often these bug-related reports are the most concerned ones that should have been dealt with before the other types. In practice identifying whether an issue report is indeed bug-related could be time-consuming, and doing so for every issue report submitted is costly~\cite{DBLP:journals/chinaf/ZhangWHXZM15}. 

Even with such a simplified scenario, identifying bug related reports is nevertheless not a trivial task. This is because bug reports are often written in a way that is complex, ambiguous, and incomprehensive, making the identification complicated, e.g., Mani \textit{et al.}~\cite{DBLP:conf/sigsoft/ManiCSD12} show that there could be up to 332 sentences per report on average in which most contents are highly vague. Such a complication significantly slows down the speed and efficiency of a developer to understand and classify the reports. Indeed, Herzig \textit{et al.}~\cite{DBLP:conf/icse/HerzigJZ13} report that two experienced developers can merely identify around 78 reports in terms of whether they are bug related within one full working day. Further, the volume of submitted reports for software projects increases drastically~\cite{DBLP:journals/infsof/WangWH23,DBLP:journals/soco/FangWLYAM21}. For example, the escalating trend of bug reports is exemplified by the Firefox bugzilla project, where the annual number of bug reports surged from 9962 in 2007 to 15,652 in 2014, marking an increase of $1.5\times$ over the seven-year period~\cite{DBLP:journals/smr/DuRLJY23}. All the above render solely manual report identification expensive and inefficient.


Existing work has leveraged machine learning models~\cite{DBLP:journals/tse/FanXLH20,DBLP:journals/smr/ZhouTGG16,kukkar2018supervised,DBLP:conf/kbse/SunLKJ11,DBLP:conf/sigsoft/ZhouS17}, and more recently neural language models~\cite{DBLP:conf/qrs/XiaLJW19,DBLP:conf/issre/HeXF0YL20,DBLP:conf/issre/ZhengZTCCWS21,DBLP:journals/tr/DuZXZT22,DBLP:journals/infsof/ChoLK22}, to comprehensively parse and analyze the information from the reports to predict whether they are bug related, hence automating bug report identification. Since those models are data-driven, it is crucial that there are a good amount of labeled samples to train and update them, i.e., previously known cases of bug related and non-bug related reports. It is a common practice to assume a static setting: train a model once and rely on it to identify the newly submitted reports. However, simply assuming that the developers can label a large number of submitted reports beforehand for training the model is unrealistic due to the expensive human analysis process, therefore a considerable amount of the available reports are often unlabeled. Even if there exists a large number of labeled reports, the training over all of them can be resource- and time-consuming, especially when a complex language model is used \cite{DBLP:journals/infsof/NeysianiBA20,DBLP:journals/access/ZaidiALWL20}. Of course, a naive approach is to randomly select some labeled reports to update the model (or ask the developers to label them). Yet, this does not guarantee the quality of those samples, hence whether they are positive to the model quality is still questionable. 

To overcome the lack of timely labeled reports and excessive training overhead, recent efforts have been relying on active learning~\cite{DBLP:journals/tse/TuYM22,DBLP:journals/infsof/WuZCZYM21,DBLP:journals/infsof/GeFQGQ22,DBLP:conf/kbse/WangWCW16,DBLP:conf/iwpc/ThungLL15}. In a nutshell, active learning is a learning paradigm that allows the model to continually query humans to label a small number of representative new samples for improving model performance, achieving human-machine teaming~\cite{DBLP:conf/pkdd/CiravegnaPBMG23}. Instead of relying solely on pre-labeled datasets, the active learning chooses which reports should be labeled next, typically prioritizing those samples that are more challenging, uncertain, or likely to have a significant impact on improving its performance~\cite{DBLP:journals/infsof/GeFQGQ22}.

The intuition of the above is simple: since a complete manual identification is too labor intensive while the fully automatic bug report identification using learning models relies too heavily on high-quality labeled data and their quantity, why not take an approach that lies between the two extremes and retains the best of both worlds? This fits exactly with the concept behind active learning.  However, existing work adopts vanilla active learning procedure to bug report identification, hence limiting its true potential, because 1) they are restricted to a single software project, requiring complete retraining for application on alternative projects which is not resource efficient; 2) when querying the new reports for labeling, they ignore the efforts that the developers need to complete labels, which often results in highly difficult reports and hence negatively impacts the result and time-to-update the model---a typical consequence of \textit{cognitive fatigue} ~\cite{tharwat2023survey}; 3) since in bug report identification, there are often a much larger number of unlabeled reports than their labeled counterparts \cite{DBLP:journals/corr/abs-2202-06149}, existing work has not been able to fully exploit the valuable information from the unlabeled pool of reports given the human-labeled ones. To sum up, using active learning can benefit from mitigating the cost of labor through optimizing the annotation process, making it more efficient and less burdensome by not only improving model accuracy but also reducing the annotators' labeling effort. While public history data is available for many open-source projects to train general models, training highly accurate models for specific projects can greatly benefit from our method. Using active learning can efficiently train tailored and precise models for a variety of individual projects.

To address the challenges and limitations in identifying bug reports from GitHub repositories above, we propose \textbf{\underline{M}}utualistic \textbf{\underline{N}}eural \textbf{\underline{A}}ctive \textbf{L}earning, or \approach, a cross-project framework that enhances model consolidation through human-machine collaboration. Our approach leverages a neural language model to generalize reports across different projects, paired with active learning that involves human-labeled samples in the training process. This creates a mutualistic relationship between humans (developers who label the data) and machines (the neural language model that generalizes the knowledge from the reports), enriching the learning process. This mutualistic relation mirrors a symbiotic relationship where both parties benefit, as the model selects the most informative reports for labeling, reducing the effort required from developers while improving performance.

Specifically, the main contributions of this work are as follows:

\begin{itemize}
    \item \textbf{Neural Active Learning Framework}:
    \begin{itemize}
        \item We propose a neural active learning framework that facilitates human-machine co-boosted bug report identification and cross-project handling.
    \end{itemize}

    \item \textbf{Mutualistic Relation in Active Learning}:
    \begin{itemize}
        \item \textbf{Model Improvement}: The model is enhanced through human-labeled reports and a new method for generating pseudo-labeled reports.
        \item \textbf{Developer Assistance}: With effort-aware uncertainty sampling, developers are only asked to label more readable and identifiable reports, reducing their workload.
    \end{itemize}

    \item \textbf{Comprehensive Evaluation}:
    \begin{itemize}
        \item We evaluate \approach~on datasets with 1,275,881 reports from over 127,000 GitHub projects, comparing it against five baselines and several variants using six metrics for performance and labeling efforts.
        \item A case study involving 10 human participants provides qualitative evaluation of the human involvement in \approach.
    \end{itemize}
\end{itemize}

The experimental results are encouraging, demonstrating that \approach:

\begin{itemize}
    \item \textbf{Performance and Efficiency}:
    \begin{itemize}
        \item Achieves significant improvements in readability (98.1\%) and identifiability (194.7\%) compared to random sampling, and 97.2\% and 283.6\% improvements compared to uncertainty sampling, all statistically significant.
        \item Considerably enhances model performance using pseudo-labeled reports.
    \end{itemize}

    \item \textbf{Model-Agnostic Benefits}:
    \begin{itemize}
        \item Improves performance and efficiency across various neural language models, with readability and identifiability improvements of 78.6\% and 171.5\%, respectively, supported by statistically significant results.
    \end{itemize}

    \item \textbf{Reduction in Labeling Efforts}:
    \begin{itemize}
        \item Significantly reduces the efforts required for labeling compared to state-of-the-art active learning approaches, with improvements of up to 95.8\% in readability and 196.0\% in identifiability.
        \item Human participants report high satisfaction, with qualitative readability improvements of 74.7\% and identifiability improvements of 64.8\%, and a reduction in time and monetary resources needed for labeling by threefold.
    \end{itemize}
\end{itemize}

The remainder of this paper is organized as follows: Section~\ref{sec:bg} provides an overview of the background and challenges. Section~\ref{sec:method} presents our detailed designs of \approach~and their rationals. Section~\ref{sec:exp} outlines the specifics of the experiment settings that evaluate \approach. Sections~\ref{sec:result},~\ref{sec:why}, and~\ref{sec:threat} discuss the experimental results, any additional aspects related to \approach, and threats to validity, respectively. Section~\ref{sec:related} positions our approach within the related work. Finally, Section~\ref{sec:conclusion} presents the conclusion.

%% file: sec/bg.tex
\section{Preliminaries}
\label{sec:bg}

This section specifies the necessary backgrounds and challenges that motivate this work. 


\begin{figure}[t!]
  \centering
  \includegraphics[width=0.8\columnwidth]{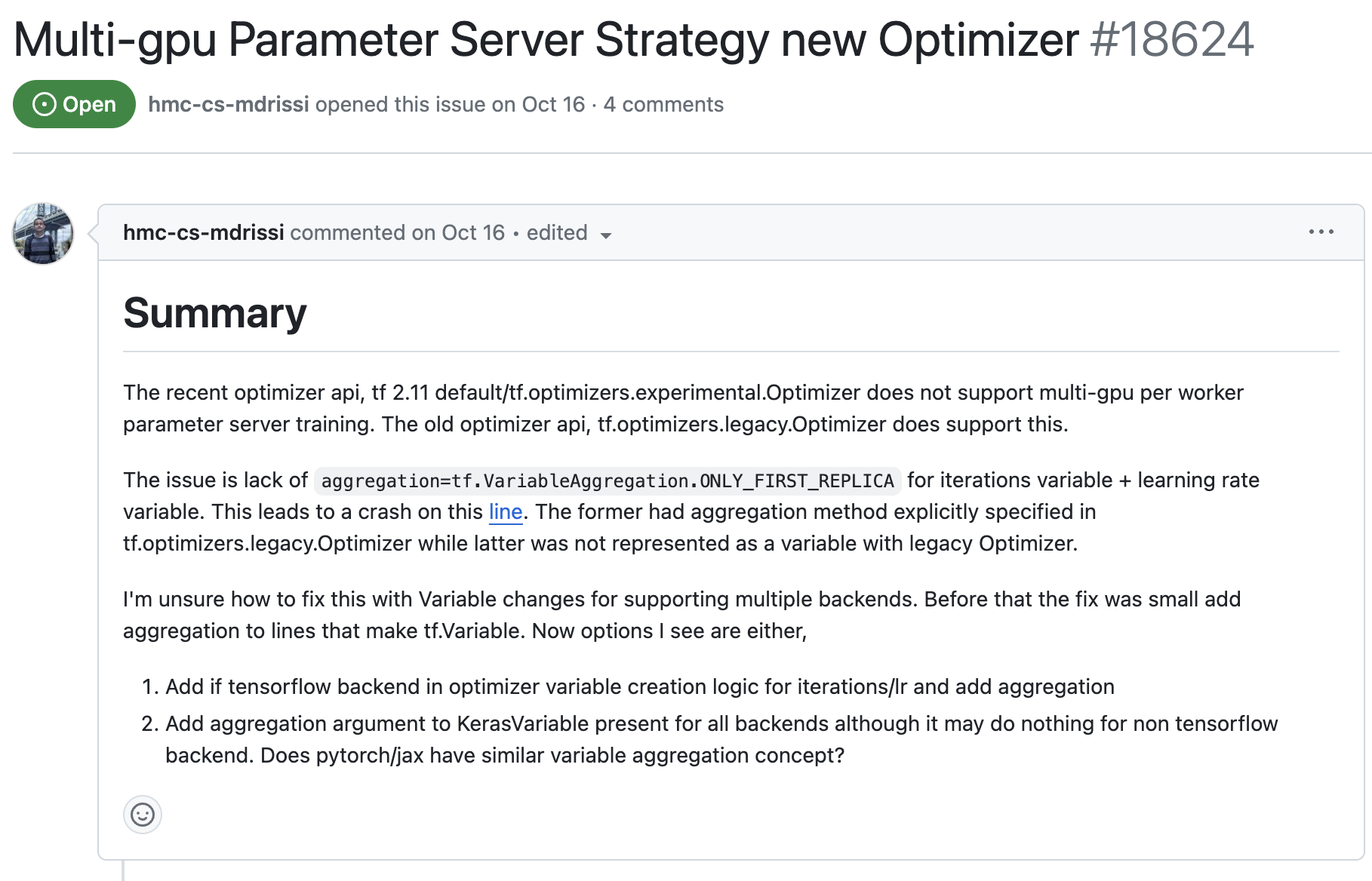}
  \caption{An example of the submitted reports from GitHub.}
  \label{fig:exp}
\end{figure}

\subsection{Bug Report Identification}

Modern software repositories like GitHub are prevalent, providing the perfect venue for documenting and managing information about software. Typically, when users or developers encounter issues with the software, they generate bug reports to document these anomalies~\cite{DBLP:journals/ase/LiWSBCT22}. As shown in Figure~\ref{fig:exp}, a typical report contains a summarized title and a description with details about the encountered problem, steps to reproduce the issue, and other relevant information that can assist developers in diagnosing and resolving the problem. However, not all the submitted reports reflect real bugs; some of them might falsely report a nonexistent bug; some may merely be feature requests, documentation updates, or question-asking, which are non-bug related~\cite{neysiani2019new}. 

A crucial step in bug triage is to identify the bug related reports amongst the others, hence developers can focus on investigating those with a high priority, or otherwise any unnecessary delay in analyzing the potential bug in a report might lead to devastating consequences. Yet, identifying bug related reports is not easy, since the report itself can be vague, imprecise, and with very lengthy content, all of which can require a significant amount of effort to understand and also increase the chance of human misjudgment~\cite{DBLP:conf/icse/HerzigJZ13}. Further, as software projects grow and attract a larger user base, the number of submitted reports can grow exceptionally, rendering manual identification impractical~\cite{DBLP:conf/tase/MengWWW22}. Therefore, automatic bug report identification is in high demand, which is considered a necessity for real bugs to be timely and promptly fixed in current software engineering practice~\cite{DBLP:journals/kais/FuZL13}.





\subsection{Neural Language Model}


Because the submitted reports from the software repository are written by humans in various ways, they often exhibit characteristics similar to natural language. Therefore, neural language models, such as BERT~\cite{DBLP:conf/naacl/DevlinCLT19}, CodeBERT~\cite{DBLP:conf/emnlp/FengGTDFGS0LJZ20}, RoBERTa~\cite{DBLP:journals/corr/abs-1907-11692}, and RTA~\cite{DBLP:conf/icse/FangZTJXS23}, can be promising solutions for automatically identifying bug reports~\cite{article2021}. In a nutshell, neural language models are based on neural networks for processing text data. In contrast to classic language models, neural language models typically embed text inputs into a vectorized latent space. This space is then processed using an attention mechanism, allowing the model to focus on specific parts of the input when making predictions. In this way, neural language models are equipped with the ability to better understand the semantics of the text data.

Taking BERT---a most conventional pre-trained natural language processing model---as an example, its strength is the ability to capture contextualized representations of words, taking into account the surrounding words bidirectionally. This bidirectional context helps grasp the nuances and semantics of natural language. In the context of automatic bug report identification, BERT can be leveraged to understand the language pattern associated with software bugs. Mathematically, the representation of a report in BERT can be expressed as $R = \{w_1, w_2, ..., w_n\}$ where $w_n$ are the words in a bug report $R$. BERT is pre-trained on general natural language corpora (e.g., Wikipedia information) over several phases, including tokenization, embedding, multi-head self-attention with Transformer encoders, and masking words, after which the parameters are updated with data sourced from the downstream task (i.e., bug report identification). To identify whether $R$ is bug related, BERT outputs a binary prediction.

\subsection{Active Learning}



Since the labeling process of reports itself is expensive and labor-intensive, the available number of labeled data samples used to update/train the model in a timely manner is often dramatically smaller compared with the number of unlabeled ones. In addition, even if there exists a good amount of labeled reports, training/fine-tuning all of them requires a complex enough model and lengthy training, leading to a drastic increase in the requirements of resources and time. For example, \citeauthor{DBLP:journals/tosem/ZhaoSLZWKHY22}~\cite{DBLP:journals/tosem/ZhaoSLZWKHY22} spent four hours for updating/training a deep neural network on 4,000 bug reports. If this model were to be trained on the million-level datasets (similar to those used in this paper), it would take approximately 53 days. Simply updating the neural language model on some randomly selected data samples works only under the assumption that all available data is equally informative. However, in real-world scenarios, especially in the context of software report analytics, this is seldom the case. In fact, a nontrivial amount of reports are much less informative than others, and hence blindly training on any data can lead to poorly-performed models~\cite{DBLP:journals/jcst/YangLXHS17}. Therefore, it is important to determine which reports are the most representative for labeling, hence making better use of the developers' efforts. To that end, active learning, which queries the most representative samples for labeling and updating the underlying model, presents a promising solution~\cite{DBLP:journals/sncs/TranLNH20,DBLP:journals/infsof/GeFQGQ22}. 

Given an initial set of labeled data $\mathbfcal{D}_{l}$ which trains a model $f_0$ together with a pool of unlabeled data $\mathbfcal{D}_{u}$, active learning improves the performance of a model by querying a set of $k$ representative, unlabeled data samples $\mathbfcal{D}_{q}$ for the human to label ($|\mathbfcal{D}_{q}| \leq |\mathbfcal{D}_{u}|$), which are then used to update a new model $f_t$, at timestep $t \in \{0,...,T\}$. $\mathbfcal{D}_{q}$ will be removed from $\mathbfcal{D}_{u}$ and added to $\mathbfcal{D}_{l}$. The goal is to improve the accuracy of those updated models $f_t$ throughout the timesteps using a separate test set.

Active learning can be broadly divided into two categories: stream-based and pool-based~\cite{cacciarelli2023active}. In the former, each data sample is drawn from some distribution in a streaming manner and the active learning needs to decide immediately whether to query the human to label this sample or not ($k=1$ and $|\mathbfcal{D}_{u}|=1$). The latter, which is a more realistic setting and is the focus of this work, allows flexible access to a pool of unlabeled data ($k>1$ and $|\mathbfcal{D}_{u}|>1$). Pool-based active learning fits our problem better since it is often the case that new reports accumulate rapidly~\cite{DBLP:journals/infsof/WangWH23,DBLP:journals/jcst/KanwalM12,rathor2023technological}.

Among various pool-based strategies, three notable ones include uncertainty sampling, margin sampling, and confidence-based sampling. Uncertainty sampling focuses on selecting samples for which the model is least confident in its predictions. For instance, it might prioritize samples where the predicted probability is close to 0.5, indicating high uncertainty. Margin sampling chooses samples based on the margin between the model's top two predictions, targeting those with small margins as they are less confidently predicted. It is worth noting that in binary classification tasks, margin sampling effectively becomes equivalent to uncertainty sampling. This equivalence arises because, in binary tasks, the margin between the two top predictions is directly related to the model’s uncertainty. Confidence-based sampling, on the other hand, selects samples where the model's confidence in its predictions is high, often in a reverse manner where the model queries examples it is confident about but uncertain if they are correctly labeled. By applying these strategies, active learning ensures that the most informative and useful samples are prioritized for labeling, thus enhancing the efficiency and effectiveness of the training process.

\subsection{Challenges}

However, several challenges still exist when adopting active learning to train a reliable and generalizable bug report identification model: 

\begin{itemize}
    \item \textbf{Challenge 1: Working on Cross-Project Reports.} There could be a limited number of reports for a single project to learn and generalize, therefore, utilizing reports from different projects to update the models and make predictions therein, is of high demand since building or updating a model for each individual project is inefficient. However, it is challenging to handle cross-project reports while effectively exploiting information from the newly queried and labeled reports made by the developers.

    \item \textbf{Challenge 2: Considering Developers' Labeling Effort} Classic active learning primarily aims to improve accuracy, but little attention has been paid to easing the efforts required to label the data samples. This, if ignored, can cause severe consequences of \textit{cognitive fatigue} in bug report identification with active learning~\cite{tharwat2023survey}: there can be a significant delay in updating the model or the reliability of labeling might be compromised when the reports to be labeled by the software developers are always highly complex.

    \item \textbf{Challenge 3: Extracting Meaningful Information from Unlabeled Reports.} Indeed, the labeled reports can be useful for improving the model, but they might still be insufficient considering the largely imbalanced ratio between the number of unlabeled reports and the amount of reports that can be labeled. It would be beneficial, but challenging to further extract meaningful information from the unlabeled ones to enrich the data for updating/training.
\end{itemize}

In the following, we delineate \approach~that is specifically designed to overcome the challenges above. 

%% file: sec/method.tex
\section{Synergizing Mutualistic Relation with Neural Active Learning for Identifying Bug Reports}
\label{sec:method}

In this section, we elaborate on the key properties and detailed designs of the \approach~framework.

\subsection{Key Properties}

\begin{figure}[!t]
\centering

\begin{subfigure}{.8\columnwidth}
  \centering
  \includegraphics[width=\linewidth]{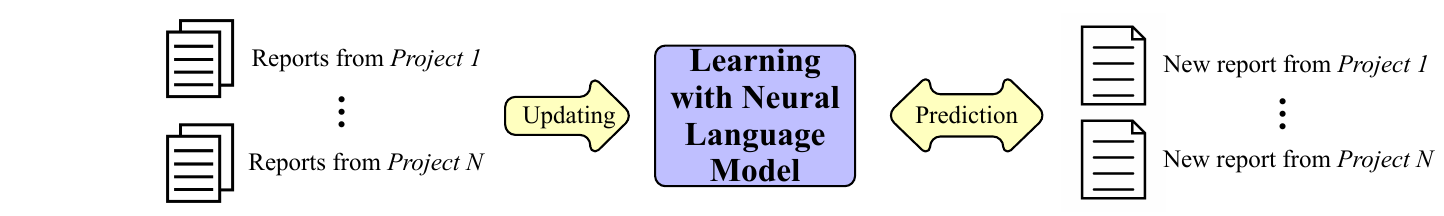} 
  \caption{Learning with a neural language model}
\end{subfigure}

\begin{subfigure}{.8\columnwidth}
  \centering
  \includegraphics[width=\linewidth]{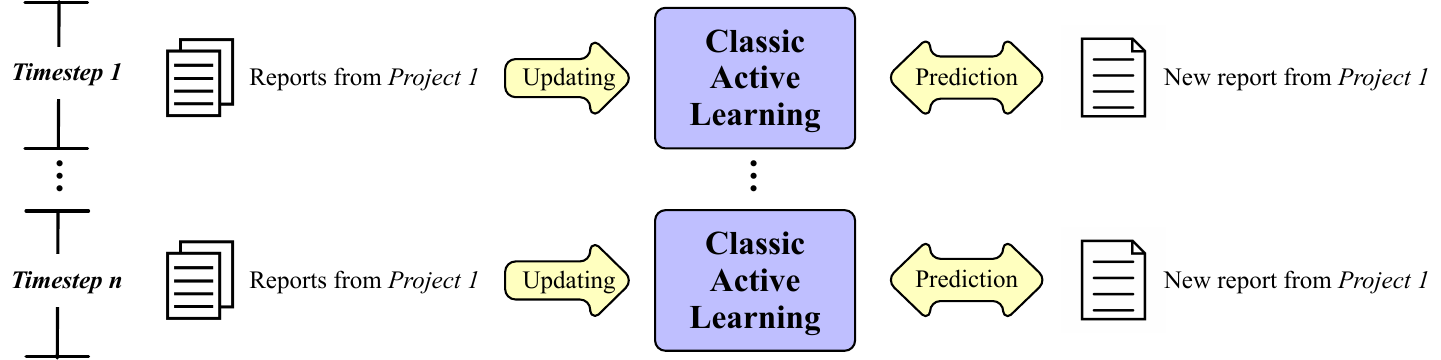} 
  \caption{Classic active learning}
\end{subfigure}

\begin{subfigure}{.8\columnwidth}
  \centering
  \includegraphics[width=\linewidth]{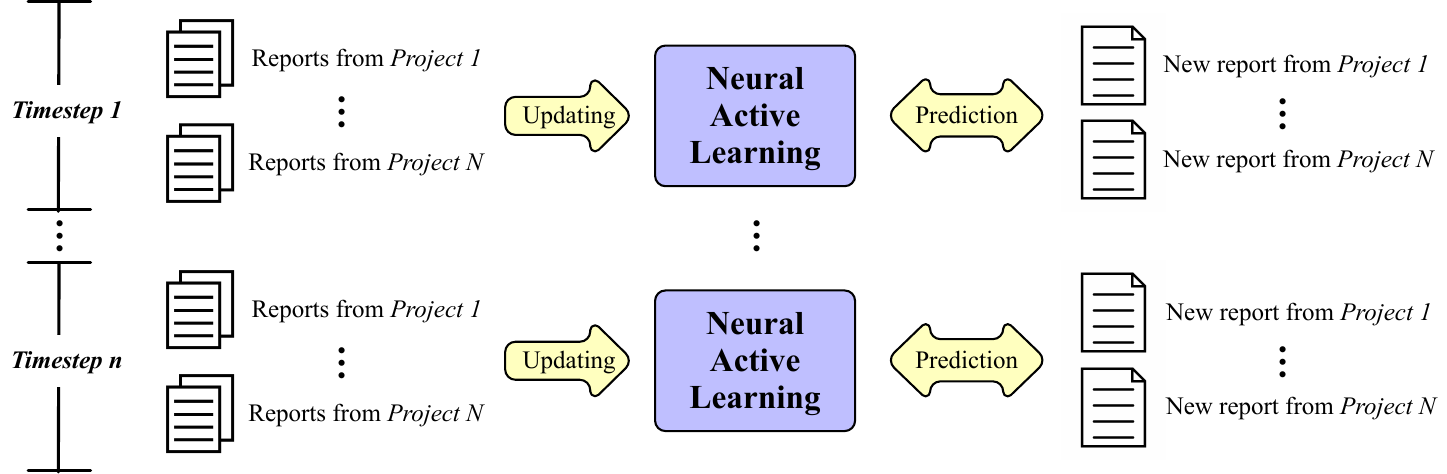} 
  \caption{Neural active learning}
\end{subfigure}

  \caption{Illustrating the difference among learning with a neural language model, classic active learning, and neural active learning for bug report identification.}
     \label{fig:nal}
\end{figure}



\subsubsection{Neural Active Learning} 

From the literature, neural language models~\cite{DBLP:conf/qrs/XiaLJW19,DBLP:conf/issre/HeXF0YL20,DBLP:conf/issre/ZhengZTCCWS21,DBLP:journals/tr/DuZXZT22,DBLP:journals/infsof/ChoLK22} and active learning~\cite{DBLP:journals/tse/TuYM22,DBLP:journals/infsof/WuZCZYM21,DBLP:journals/infsof/GeFQGQ22,DBLP:conf/kbse/WangWCW16,DBLP:conf/iwpc/ThungLL15} have been independently studied for identifying bug reports, but they are rarely considered together.
\approach~combines a neural language model (e.g., BERT) with active learning, achieving the strength of both semantic understanding and human-machine teaming. In particular, this unique combination poses both challenges and opportunities: the limited set of newly human-labeled reports might not be sufficient for the neural language model to improve. Yet, the neural language model provides the embedding of the report in a latent space, serving as the foundation that allows us to design more sophisticated methods to enrich the data throughout active learning (in the \textit{Pseudo-labeling} component). Since the neural language model operates on the texts of the reports, \approach~seamlessly works cross-projects, including both the continuous update with humans and prediction (addressing \textbf{Challenge 1}), e.g., it might query the reports to developers specialized in different projects and identifies the newly submitted reports for those projects. It is also worth noting that \approach~is model-agnostic, i.e., by design, it can be paired with any neural language model as long as the model supports latent embedding of the reports.

Figure~\ref{fig:nal} illustrates the difference between neural active learning, learning with a neural language model, and classic active learning alone. As can be seen, the neural language model supports cross-project training/prediction but it neither handles the model change nor actively makes queries to the developers. Active learning deals with the continuous update and the queries to humans well but it does not usually work on cross-project reports, as it is often underpinned by a statistical classifier. As such, the process needs to be repeated from scratch for each project. Neural active learning, in contrast, combines the best of both worlds. The ability of neural active learning ability to make active queries while leveraging advanced neural network models instead of traditional statistical classifiers, which are most commonly used with active learning.

\begin{figure}[t!]
  \centering
  \includegraphics[width=\columnwidth]{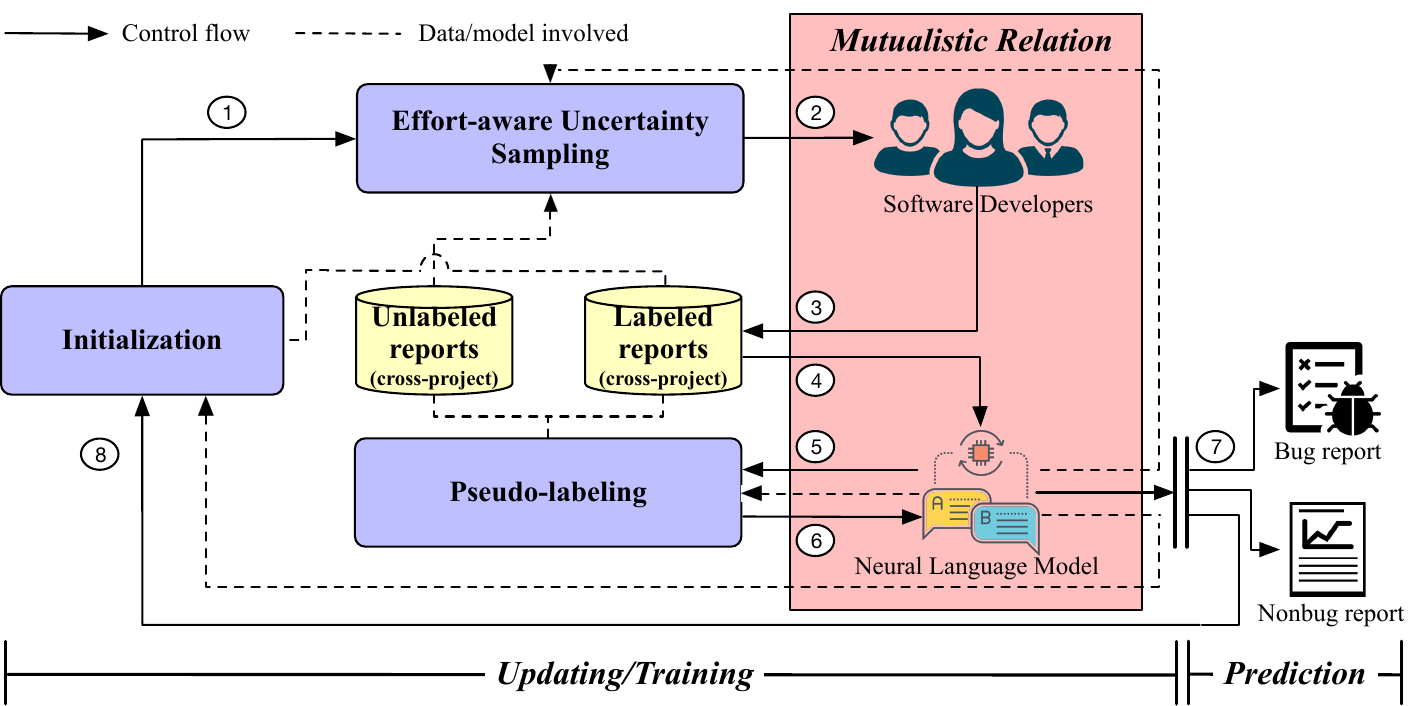}
  \caption{The workflow of \approach~that supports enhanced human-machine teaming. The serial numbers refer to the order of workflow.}
  \label{fig:arch}
\end{figure}

\subsubsection{Mutualistic Relation} 

While active learning can be considered a common iterative learning approach for humans and machines to work as a team during training, \approach~goes beyond the classic active learning by introducing the mutualistic relation between humans (software developers) and machines (neural language model) when enriching the knowledge learned, enabling a better human-machine teaming (addressing \textbf{Challenge 2} and \textbf{Challenge 3}). As can be seen from Figure~\ref{fig:arch}, this enables a bidirectional benefit:

\begin{itemize}
    \item On one hand, the effort-aware uncertainty sampling would select unlabeled reports that require an acceptable level of effort to label (the \textit{Effort-aware Uncertainty Sampling} component). This is for the benefit of the software developers, making their jobs easier. 
    \item On the other hand, the \textit{Pseudo-labeling} component enriches the data to train the neural language model, thanks to its ability to represent the report in a latent space. These, together with the newly human-labeled reports, are expected to improve the learning accuracy and efficiency, thereby being beneficial for the model. 
\end{itemize}

Such a correlation between software developers and the neural language model is a typical mutualistic relation---they ``work together in a team'' to enrich the knowledge learned about identifying bug reports, each benefiting from the relationship while enhancing the efficiency of human-machine co-boosted bug report identification.


\subsection{Overview}

The iterative workflow of our proposed framework, \approach, is illustrated in Figure~\ref{fig:arch} and Algorithm~\ref{alg:mnal}. This process is designed to create an effective human-machine team by integrating active learning with a pseudo-labeling strategy. The workflow, as detailed by the serial numbers in the figure, can be understood in three main phases:

\paragraph{Phase 1: Active Sampling of Informative Reports (Steps \textcircled{\small{1}}-\textcircled{\small{2}})}

The cycle begins with an initialized model trained on previously labeled data.
\begin{itemize}
    \item[\textbf{ \textcircled{\small{1}}:}] The workflow is triggered. A neural language model is initialized using the pool of all available labeled reports which corresponds to the Algorithm lines 2-3. The system directs the pool of unlabeled reports to the \texttt{Effort-aware Uncertainty Sampling} module.
    \item[\textbf{ \textcircled{\small{2}}:}] This step is to use \texttt{Effort-aware Uncertainty Sampling} to intelligently sample a query size of $k$ representative unlabeled reports (from within or cross-project).  for (one or more) software developers of the relevant projects to decide whether they are bug related (Algorithm line 4).
\end{itemize}

\paragraph{Phase 2: Human-Machine Collaborative Labeling (Steps \textcircled{\small{3}}-\textcircled{\small{6}})}
This phase represents the core ``Mutualistic Relation'' where human expertise and machine intelligence collaborate to expand the labeled dataset efficiently.
\begin{itemize}
    \item[\textbf{\textcircled{\small{3}}:}] Software developers invest their effort to manually inspect and label the challenging reports sampled in Step~\textcircled{\small{2}}. These new, high-quality human-provided labels are added to the \texttt{Labeled reports} dataset (Algorithm line 5).
    \item[\textbf{ \textcircled{\small{4}}:}] The updated \texttt{Labeled reports} dataset are immediately used to fine-tune the \texttt{Neural Language Model}. This step ensures the model learns directly from expert knowledge (Algorithm line 6)
    \item[\textbf{ \textcircled{\small{5}}:}] The newly human-labeled reports are then employed in the \texttt{Pseudo-labeling} module to assign labels to the most similar unlabeled reports (Algorithm line 7).
    \item[\textbf{ \textcircled{\small{6}}:}] The pseudo-labeled reports are incorporated with the labeled reports in labeled pool to train the neural language model together (Algorithm lines 8–10).
\end{itemize}

\paragraph{Phase 3: Model Consolidation and Iteration (Steps \textcircled{\small{7}}-\textcircled{\small{8}})}
In the final phase, the model is consolidated and prepared for both prediction and the next cycle of learning.
\begin{itemize}
    \item[\textbf{ \textcircled{\small{7}}:}] The resulting robust and more capable model is then deployed for the \texttt{Prediction} task, automatically classifying new, incoming reports as ``Bug report'' or ``Nonbug report''.
    \item[\textbf{ \textcircled{\small{8}}:}] The process is designed to be iterative. The expanded set of labeled reports (both human- and pseudo-labeled) forms the basis for the \texttt{Initialization} of the next learning cycle, allowing the model to continuously improve over time as more reports are processed.
\end{itemize}

Particularly, in the pool of labeled reports, any incorrect labels due to human mistakes in the labeling or inaccurate pseudo-labeling can be corrected when discovered.

\input{algo/mnal-learning}

\subsection{Initialization} 
Initializing \approach~can use two distinct strategies: cold start and warm start. These strategies determine how the initial language model is created and subsequently updated to fulfill the task of bug report identification. The choice between cold and warm starts hinges on the availability and adequacy of a pre-trained neural language model, which can be case-dependent. 


\subsubsection{Cold Start}

When \approach~is used completely from scratch, an initial language model needs to be trained by using the available labeled reports from any project, thanks to the semantic attention to the naturalness of language/code in the reports. Noteworthily, the number of reports used in the initialization needs not be consistent with the query size and they can come from cross-project since it is important to provide a good general foundation. Most commonly, the initial number of labeled reports is set according to the balance between the training overhead and anticipated model performance.

The cold start strategy benefits from creating a robust initial model capable of handling diverse bug reports from the outset. However, this approach involves a significant initial computational cost and time investment. The model's performance heavily relies on the quality and variety of the labeled reports used for training. A well-balanced and diverse training set can lead to a more generalized and effective model, but acquiring such a dataset can be resource-intensive.

\subsubsection{Warm Start}
If the neural language model can be pre-trained, we do not need any initial training since the interactive model updating process within \approach~can naturally serve as the fine-tuning process. In this way, we gradually specialize \approach~to identify bug reports for the specific problem of bug report identification. For the subsequent timesteps following the first one, however, any newly labeled reports or reports with corrected labels in the pool of labeled reports after the last update would be used to update the model. Noteworthily, such a fine-tuning process is much more lightweight compared with training the neural language model from scratch.

The Warm Start strategy offers significant advantages in terms of efficiency, as it builds on an already pre-trained model, reducing initial computational requirements. This approach allows for quicker adaptation to specific datasets and problems, enabling faster convergence and deployment. However, the pre-trained model's initial quality and relevance to the specific task are crucial. If the pre-trained model is not well-aligned with the target domain, the performance gains might be limited, necessitating more extensive fine-tuning.



\subsection{Effort-aware Uncertainty Sampling}
\label{sec:sampling}


\subsubsection{Uncertainty}

In classic (pool-based) active learning, at every timestep, the key is to decide what data samples (unlabeled reports), from the pool of unlabeled data, should be sent to the human for labeling. This is important, as to what extent those samples are representative determines the accuracy improvement in active learning. From the literature on exploiting active learning in bug report identification and other software engineering tasks, the most common criterion for choosing the data sample labeling is uncertainty \cite{DBLP:conf/iwpc/ThungLL15,DBLP:journals/infsof/WuZCZYM21,DBLP:journals/infsof/GeFQGQ22}:
\begin{equation}
 U(x) = -\sum_{y} p(y|\mathbf{\overline{x}}) \log_2(p(y|\mathbf{\overline{x}}))
 \label{eq:unc}
\end{equation}

The above is basically a measurement of information entropy~\cite{jaynes1957information} whereby $\mathbf{\overline{x}}$ represents the embedded and vectorized input text of an unlabeled report $x$; $y$ stands for the label of whether the report is bug related or not being predicted by the neural language model; $p(y|\mathbf{\overline{x}})$ represents the conditional probability of a possible label $y$ provided the input $\mathbf{\overline{x}}$ extracted from the neural language model (e.g., at the \texttt{softmax} layer of BERT). A larger entropy indicates higher uncertainty that the current language model knows about $\mathbf{\overline{x}}$. 

The motivation behind the use of uncertainty in active learning is intuitive: since the underlying language model often performs poorly on the types of data samples that it has never seen before, it would be best if we could train it with the data that is currently known the least. Since the uncertainty measurement of a report simply reflects the information the current language model knows about a report, we should choose the top $k$ ones with the highest uncertainty and send them to the software developers for labeling in the active learning.

\subsubsection{What is the limitation?}

While selecting only the most uncertain data samples has been proven to be helpful for model accuracy~\cite{DBLP:conf/iwpc/ThungLL15}, it still relies on humans to provide reliable labels. This, however, can be problematic in the other aspects of bug report identification because: 

\begin{itemize}
    \item We found that the most uncertain reports are likely to be rather lengthy, especially when the report consists of both natural language and code snippets~~\cite{DBLP:conf/sigsoft/ManiCSD12}.
    \item When the reports are complex, the reliability of the labeled report becomes an issue \cite{DBLP:journals/tse/WuZXL22,DBLP:conf/qrs/SassoML16,DBLP:journals/tse/ZimmermannPBJSW10}.
\end{itemize}

The above would inevitably require large efforts from the software developers to label the reports for the benefits of active learning due to cognitive fatigue. To tackle this unwanted issue, when recommending the data samples (reports) for human-labeling, it is essential to consider the effort required, which is a challenge that the \approach~aims to address via introducing a mutualistic relation. That is, in addition to providing labeled data for the benefits of the neural language model, the process should also be effort-aware, thereby taking into account the benefits of software developers who actually label the reports. In particular, we exploit and propose two effort-aware metrics in the sampling process, namely readability~\cite{farr1951simplification} and identifiability, of which the details are explained in the following sections.


\begin{figure}[!t]
\centering

\begin{subfigure}{.38\columnwidth}
  \centering
  \includegraphics[width=\linewidth]{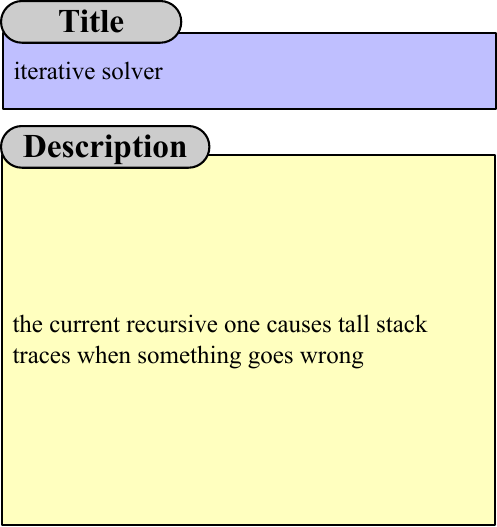} 
  \caption{High readability but low identifiability}
\end{subfigure}
~\hspace{1cm}
\begin{subfigure}{.38\columnwidth}
  \centering
  \includegraphics[width=\linewidth]{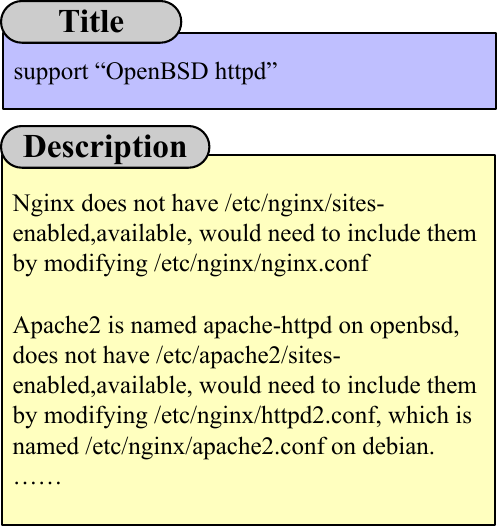} 
  \caption{High identifiability but low readability}
\end{subfigure}

  \caption{Excerpt of the exampled reports with different scores of readability and identifiability.}
     \label{fig:obj-exp}
\end{figure}

\subsubsection{Readability}

In addition to the uncertainty, in \approach, we measure the readability of a report as part of the human effort for labeling when choosing which should be labeled by software developers. In essence, our goal is to assess the ease with which a piece of text can be comprehended by its readers. To that end, we use the Flesch reading-ease score~\cite{farr1951simplification}---a well-known metric from the psychology domain---to calculate the readability score of the reports, which can be formally defined as:    
\begin{equation}
    R(x) = 206.83 - 1.015 {w \over s_e} - 84.6 {s_y \over w}
\end{equation}
whereby there are two important terms:

\begin{itemize}
    \item \textbf{Sentence Structure:} The first term, $w \over s_e$, evaluates the sentence structure through a ratio between the number of words ($w$) and the number of sentences ($s_e$) in the unlabeled report $x$. A lower ratio implies a report with shorter content and/or more concise sentences, rendering it more readable for the software developers. In contrast, a higher ratio suggests a longer report or one with much longer sentences, which are often more difficult to read. 

    \item \textbf{Word Complexity:} $s_y \over w$ is the second term that assesses the ratio between the number of syllables ($s_y$) and the number of words. Intuitively, a higher proportion of syllables against the total number of words indicates that the words are generally long and complex, hence the readability of the report tends to be low. On the other hand, a higher amount of shorter words often renders the report easier to read.
\end{itemize}

Clearly, a higher value of $R(x)$ is preferred. Figure~\ref{fig:obj-exp} shows two contrasted examples of reports with high and low readability, respectively.

\subsubsection{Identifiability}
While the readability is a good indicator of the required human effort for labeling in \approach, a more readable report does not necessarily mean that it is also easier to be labeled by the software developers. This is because readability does not reflect the amount of information a report contains for performing the labeling. Taking the example from Figure~\ref{fig:obj-exp} again, clearly, the readability of Figure~\ref{fig:obj-exp}a is excellent since it is short, but it contains very little information that can barely help one to identify whether it is bug related.

To mitigate this issue, we further introduce a metric namely identifiability, aiming to ensure that the chosen reports for labeling are not only understandable but also more distinguishable in terms of their bug related nature. This can be formally defined as:
    \begin{equation}
        I(x) = {{T_r + T_i} \over w}
    \end{equation}
where $w$ is again the number of words in an unlabeled report $x$; $T_r$ and $T_i$ denote the common ``\textit{Bug Report Relevant Terms}'' and ``\textit{Bug Report Irrelevant Terms}'', respectively\footnote{Note that not all the terms in a report can fit into these two categories.}, as identified in the literature~\cite{DBLP:conf/re/MaalejN15, DBLP:conf/msr/HindleEGM11}:

\begin{itemize}
    \item \textbf{Bug Report Relevant Terms:}
    Bug reports are expected to be concise and informative, containing essential terms that describe the bug and issue. Here, the most common terms are compiled from the literature including \texttt{\textcolor{blue}{``error''}}, \texttt{\textcolor{blue}{``bug''}}, \texttt{\textcolor{blue}{``reproduce''}}, \texttt{\textcolor{blue}{``issue''}}, \texttt{\textcolor{blue}{``behavior''}}, \texttt{\textcolor{blue}{``debug''}}, \texttt{\textcolor{blue}{``failed''}}, \texttt{\textcolor{blue}{``expected''}}, and \texttt{\textcolor{blue}{``crash''}}.

    \item \textbf{Bug Report Irrelevant Terms:}
    In contrast, reports may also include terms that are less relevant to the bugs. These terms normally appear in non-bug reports such as those describing feature enhancements, questions, or documentation, and hence serve as strong indications that the report is not bug related. In this work, we also extracted the terms from existing studies, including \texttt{\textcolor{blue}{``add''}}, \texttt{\textcolor{blue}{``would''}}, \texttt{\textcolor{blue}{``like''}}, \texttt{\textcolor{blue}{``use''}}, \texttt{\textcolor{blue}{``feature''}}, \texttt{\textcolor{blue}{``request''}}, \texttt{\textcolor{blue}{``support''}}, \texttt{\textcolor{blue}{``improvement''}}, \texttt{\textcolor{blue}{``want''}}, and \texttt{\textcolor{blue}{``documentation''}}.
\end{itemize}
    
Among all the terms in a report, if there is a high proportion for either (or both) of the above categories of terms, then we say the report would contain more helpful information for the software developers to label whether it is bug related. Figure~\ref{fig:obj-exp}b is a typical example of high identifiability (although with low readability) since it contains a larger proportion of the terms that are irrelevant to bugs. Otherwise, the identifiability is said to be low, e.g., Figure~\ref{fig:obj-exp}a.

\subsubsection{Quality-Effort Score}

The final score of an unlabeled report $x$ (and its embedding $\mathbf{\overline{x}}$) being used in our effort-aware uncertainty sampling is measured by an acquisition function that aggregates the (normalized) uncertainty, readability, and identifiability, defined as:
\begin{equation}
score(x) = U(x) + R(x) + I(x)
\label{eq:score}
\end{equation}
in which the uncertainty indicates the potential quality improvement of the underlying language model using the newly-labeled reports while readability and identifiability reflect the efforts required by the software developers to label those reports. The higher the value of the score, the better. Since all three metrics can come with drastically different scales, we use \textit{max-min standardization} to make them commensurable, in which the bounds are their extreme values found so far throughout the neural active learning in \approach\footnote{The bounds might be updated across different timesteps.}.Given the time constraints, we opted to use an initial setting where all weights of uncertainty, readability, and identifiability are set to 1, which yielded good experimental results. This choice was pragmatic and allowed us to demonstrate the feasibility and benefits of our approach without extensive parameter tuning. There is no specific preference towards any of the three objectives (uncertainty, readability, and identifiability), i.e., all of them are equally important. This is the reason that in the quality score calculation within \approach, they are given the same weight, and hence none of them is prioritized against the others.

Ultimately, the aim of the effort-aware uncertainty sampling is to recommend top $k$ (where $k$ in this experiment is 300, 500, or 700) unlabeled reports measured by the quality-effort score for human labeling. Since such a recommendation considers the labeling effort, it contributes to the benefit of the human side within the mutualistic relation in \approach. An algorithmic illustration can be found in Algorithm~\ref{alg:sampling}.

\input{algo/sampling}

\begin{figure}[!t]
\centering

\begin{subfigure}{.49\columnwidth}
  \centering
  \includegraphics[width=\linewidth]{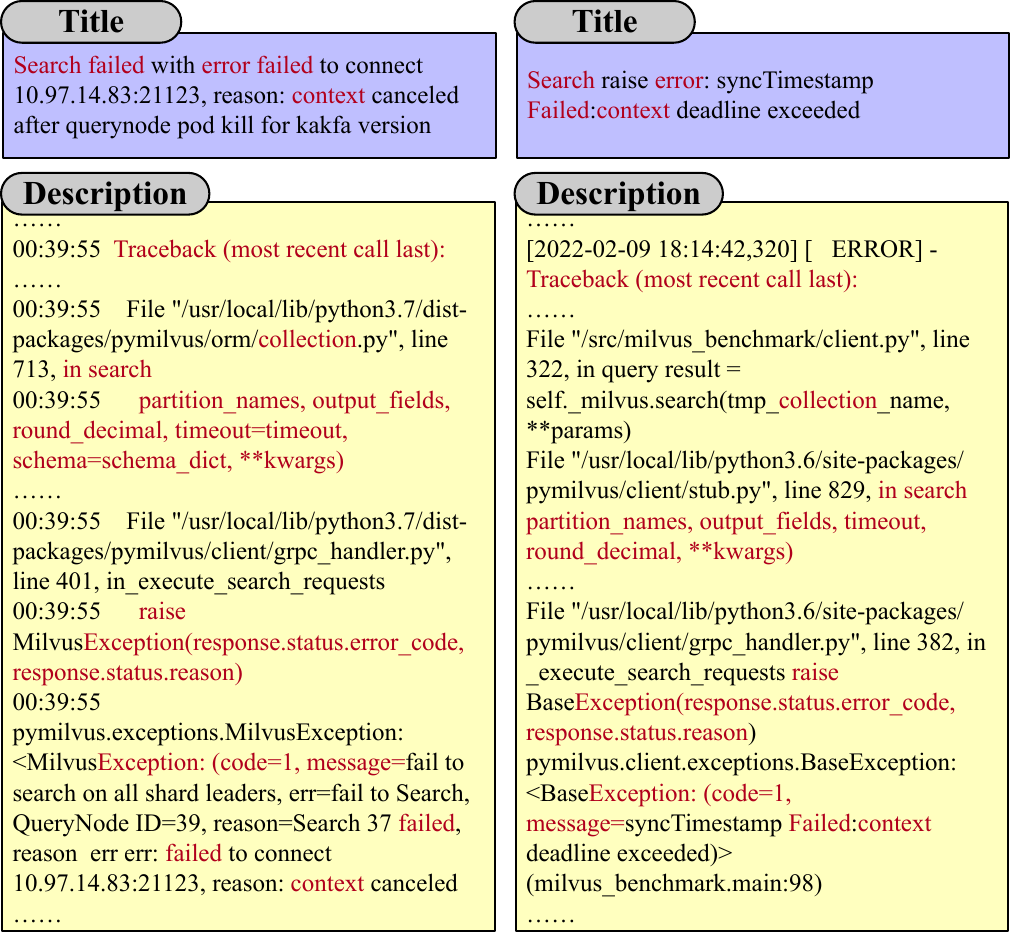} 
  \caption{Similar bug related reports}
\end{subfigure}
~\hfill
\begin{subfigure}{.49\columnwidth}
  \centering
  \includegraphics[width=\linewidth]{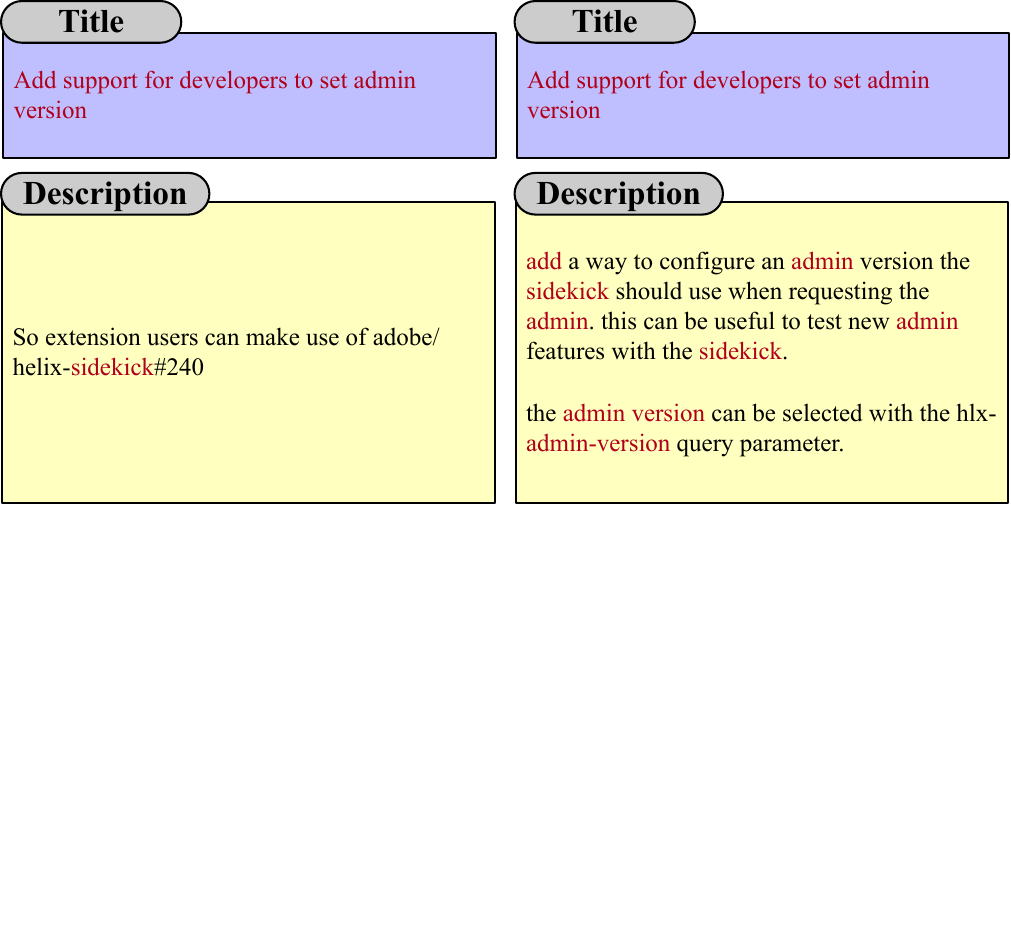} 
  \caption{Similar non-bug related reports}
\end{subfigure}

  \caption{Excerpt of the exampled reports that are the most similar and with the same label. The red texts highlight the parts that cause both reports to have high semantic and syntactical similarity.}
     \label{fig:pseudo-exp}
\end{figure}

\subsection{Pseudo-labeling}

Indeed, given the nature of active learning, the reports labeled by the software developers can serve as the beneficial effect to a machine learner contributed by humans---the other end of the mutualistic relation. However, using the human-labeled reports alone to consolidate the neural language model can easily suffer a difficult trade-off: if we use too few newly human-labeled reports, then it is hard for the updated model to be effective as they might only add a little extra information; this is especially true for updating/fine-tuning a neural language model within neural active learning. In contrast, if we assume that there are many reports that can be queried, the software developers might be overwhelmed by the labeling task, even though the efforts might have already been reduced via our effort-aware uncertainty sampling.

\input{algo/pseudo-label}

To mitigate the above, we hypothesize that those human-labeled reports, which bear strong domain understanding as being processed by the software developers, contain rich information that can guide us to extract useful information from the pool of unlabeled reports. In particular, those unlabeled reports that are more similar to the human-labeled reports are more likely to belong to the same label. For example, Figure~\ref{fig:pseudo-exp} respectively shows two pairs of bug related and non-bug related reports with the closest distance searched during the pseudo-labeling process, which tends to be semantically or syntactically similar. Therefore, our goal is to pseudo-label those most similar unlabeled reports and incorporate them to train the neural language model together with the human-labeled ones. In this way, we can increase the number of labeled reports in learning without asking software developers to label more during the active learning process, hence boosting the mutualistic relation therein. Since the accurate ``measurement of similar reports'' relies on a comprehensive representation that quantifies the report texts, this can only be made possible within the unique context of neural active learning. 

Specifically, as shown in Algorithm~\ref{alg:pseudo}, our pseudo-labeling process in \approach~follows the steps below:

\begin{figure}[!t]
\centering
\begin{subfigure}{0.6\columnwidth}
  \centering
  \includegraphics[width=\linewidth]{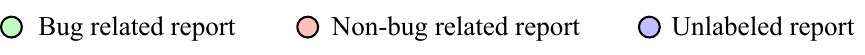} 
\end{subfigure}
\begin{subfigure}{.32\columnwidth}
  \centering
  \includegraphics[width=\linewidth]{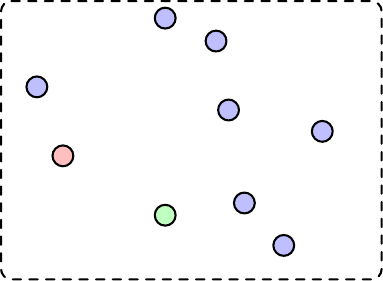} 
  \caption{Embed both human-labeled and unlabeled reports}
\end{subfigure}
~\hfill
\begin{subfigure}{.32\columnwidth}
  \centering
  \includegraphics[width=\linewidth]{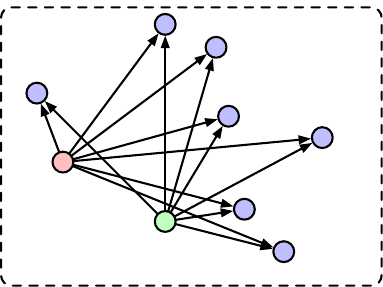} 
  \caption{Examine the distance of all human-labeled/unlabeled pairs}
\end{subfigure}
~\hfill
\begin{subfigure}{.32\columnwidth}
  \centering
  \includegraphics[width=\linewidth]{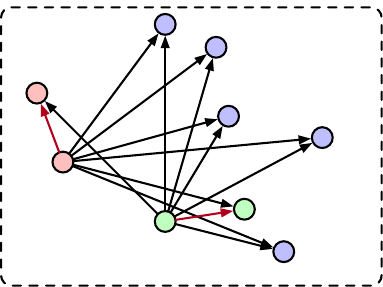} 
  \caption{Label the nearest unlabeled report to each human-labeled one}
\end{subfigure}
  \caption{Illustrating the key steps of the pseudo labeling process in \approach~under a projected 2D space. The points represent the vectorized embedding of the reports produced by the neural language model, e.g., a fine-tuned BERT.}
     \label{fig:pseudo-label}
\end{figure}

\begin{enumerate}
    \item To fully exploit the valuable data samples obtained so far for the next step, we firstly update the neural language model $\mathbfcal{M}$ by training/fine-tuning it with all the available sets of labeled reports (including the newly human-labeled ones) at line 1. This will create an intermediate version of the model solely used for pseudo-labeling.

    \item We feed the newly human-labeled reports, $\mathbfcal{D}_q$, and all the remaining pool of unlabeled reports ($\mathbfcal{D}_u$) into $\mathbfcal{M}$. This is to obtain the vectorized embedding of those reports for measuring their semantic similarity, for which we use the last encoding layer of the neural language model (e.g., a fine-tuned BERT), as in Figure~\ref{fig:pseudo-label}a and lines 3-5. This is only plausible with the power of a neural language model that can meaningfully represent the report texts in a latent space.

    \item As shown in Figures~\ref{fig:pseudo-label}b and~\ref{fig:pseudo-label}c, for each newly human-labeled report $r_i$, we seek to find its most similar unlabeled counterpart $r_s$ from all possible unlabeled report $r_j$ (lines 6-14):
    \begin{equation}
        r_s = {\arg}\min_{r_j\in \mathbfcal{D}_u} {\lVert \mathbf{\overline{v}_i} - \mathbf{\overline{v}_j} \rVert _2}
	\label{eq:distance}
     \end{equation} 
    whereby {$\lVert \cdot \rVert _2$ denotes the Euclidean distance} between the {768-dimensional} vectors $\mathbf{\overline{v}_i}$ and $\mathbf{\overline{v}_j}$, {which are the BERT-based embeddings} for $r_i$ and $r_j$, respectively. Finally, we assign the label of $r_i$ to the identified $r_s$ as its pseudo label.
\end{enumerate}

The above process will be repeated in the next timestep during the mutualistic neural active learning.

%% file: algo/mnal-learning.tex
\begin{algorithm}[t!]
	\DontPrintSemicolon
	
	\caption{Pseudo code for the learning in \approach}
	\label{alg:mnal}
	\KwIn{A pool of unlabeled reports $\mathbfcal{D}_u$; a pool of all labelled reports $\mathbfcal{D}_l$ and a pre-trained neural language model $\mathbfcal{M}$}

   \For{$t \in \{0,...,T\}$}
   {
        \If{$\mathbfcal{M}=\emptyset$ or $\mathbfcal{D}_l$ has been updated since last timestep} 
        {
          $\mathbfcal{M}\leftarrow$ update a neural language model using $\mathbfcal{D}_l$\\
        }
        $\mathbfcal{D}_q\leftarrow$\textsc{EUSampling}($\mathbfcal{M}$,$\mathbfcal{D}_u$) \tcc*[h]{\textcolor{blue}{effort-aware uncertainty sampling the top $k$ unlabeled reports.}}\\
        $\mathbfcal{D}_q\leftarrow$ the developers label all the queried reports\\

        $\mathbfcal{D}_u=$ remove all reports in $\mathbfcal{D}_q$ from $\mathbfcal{D}_u$\\
        $\mathbfcal{D}_p\leftarrow$ \textsc{pseudoLabeling}($\mathbfcal{D}_l$,$\mathbfcal{D}_q$,$\mathbfcal{D}_u$) \tcc*[h]{\textcolor{blue}{pseudo-labeling additional $k$ unlabeled reports.}}\\
        $\mathbfcal{D}_u=$ remove all reports in $\mathbfcal{D}_p$ from $\mathbfcal{D}_u$\\
        $\mathbfcal{D}_l$ = $\mathbfcal{D}_l \cup \mathbfcal{D}_q \cup \mathbfcal{D}_p$\\
        update $\mathbfcal{M}$ with $\mathbfcal{D}_l$
   }

	
\end{algorithm}

%% file: algo/sampling.tex
\begin{algorithm}[t!]
	\DontPrintSemicolon
	
	\caption{\textsc{EUSampling} function}
	\label{alg:sampling}
	\KwIn{All remaining unlabeled reports $\mathbfcal{D}_u$; the current neural language model $\mathbfcal{M}$}
     \KwOut{The $k$ unlabeled reports to be queried $\mathbfcal{D}_{q}$}

   \For{$\forall r_{i} \in \mathbfcal{D}_u$}
   {
        $s_i$ = get the effort-accuracy score of $r_{i}$ using $\mathbfcal{M}$ and normalization via Equation~\ref{eq:unc}-\ref{eq:score}\\
        \If{$|\mathbfcal{D}_q| < k$ or $s_i >$ the score of the at least one entry of $\mathbfcal{D}_q$}    
        {
             \If{$|\mathbfcal{D}_q| = k$} {
               remove the last entry of $\mathbfcal{D}_q$\\
              } 
              $\mathbfcal{D}_q\leftarrow r_i$\\
            sort $\mathbfcal{D}_q$ descendingly\\
        }
       
   }

    \Return $\mathbfcal{D}_{q}$\\
	
\end{algorithm}

%% file: algo/pseudo-label.tex
\begin{algorithm}[t!]
	\DontPrintSemicolon
	
	\caption{\textsc{pseudoLabeling} function}
	\label{alg:pseudo}
	\KwIn{Historically labeled reports $\mathbfcal{D}_{l}$; the set of newly human-labeled reports $\mathbfcal{D}_q$; all remaining unlabeled reports $\mathbfcal{D}_u$}
     \KwOut{The set of pseudo-labeled reports $\mathbfcal{D}_{p}$ for further consolidating the neural language model}

   $\mathbfcal{M}\leftarrow$ update an intermediate version of the neural language model $\mathbfcal{M}$ using all labeled reports so far $\mathbfcal{D}_{l}\cup \mathbfcal{D}_q$\\
   \For{$\forall r_{i} \in \mathbfcal{D}_q$}
   {
        $\mathbfcal{D'}_q\leftarrow \langle r_{i},\mathbf{\overline{v}_i} \rangle =$ get embedding vector $\mathbf{\overline{v}_i}$ when predicting $r_{i}$ using $\mathbfcal{M}$\\
   }

      \For{$\forall r_{j} \in \mathbfcal{D}_u$}
   {
        $\mathbfcal{D'}_u\leftarrow \langle r_{j},\mathbf{\overline{v}_j} \rangle =$ get embedding vector $\mathbf{\overline{v}_j}$ when predicting $r_{j}$ using $\mathbfcal{M}$\\
   }
    \For{$\forall \langle r_{i},\mathbf{\overline{v}_i} \rangle  \in \mathbfcal{D'}_q$}   
    {
         $d_s=-1$\\
         $r_s=\emptyset$\\
         \For{$\forall \langle r_{j},\mathbf{\overline{v}_j} \rangle \in \mathbfcal{D'}_u$}    
         {
             $d_{ij}=$ calculate the 2-norm distance between $\mathbf{\overline{v}_i}$ and $\mathbf{\overline{v}_j}$\\
             \If{$d_s=-1$ or $d_{ij}<d_s$} 
             {
                $r_s = r_j$\\
                $d_s = d_{ij}$\\
             }
             
         }
          $\mathbfcal{D}_{p}\leftarrow r_s$ with the label of $r_i$\\
    }

    \Return $\mathbfcal{D}_{p}$\\
	
\end{algorithm}

%% file: sec/exp.tex
\section{Experimental Setup}
\label{sec:exp}

In this section, we elaborate on the details of our experiment settings.

\subsection{Research Questions}
The research questions we aim to answer in this study are:
\begin{itemize}
     \item \textbf{RQ1:} How effective is the effort-aware uncertainty sampling?
     \item \textbf{RQ2:} How useful is the pseudo-labeling?
     \item \textbf{RQ3:} To what extent can \approach~improve an arbitrary neural language model?
     \item \textbf{RQ4:} How effective is \approach~against the state-of-the-art approaches for identifying bug reports?
     \item \textbf{RQ5:} What do the human developers think about \approach~in terms of the effort reduction for labeling reports?
\end{itemize}

We explore \textbf{RQ1} to assess the effectiveness and implication of querying unlabeled reports that are both good at improving accuracy and require a reasonable amount of effort to label. In \textbf{RQ2}, we ask how pseudo-labeled reports, which leverage the unique embedding of a neural language model, can complement the prediction performance in an active learning setting. Both \textbf{RQ1} and \textbf{RQ2} help to confirm the usefulness of the mutualistic relation in \approach. Since the design of \approach~is agnostic to the underlying neural language model, we examine \textbf{RQ3} to understand how well it can improve a neural language model compared with the case where the model is trained with data without any sophisticated sampling and pseudo-labeling. To verify the effectiveness of \approach, in \textbf{RQ4}, we compare \approach~against four state-of-the-art active learning approaches for bug report identification, hence confirming the benefit of neural active learning. Finally, in \textbf{RQ5}, we conduct a qualitative study with human involvement to understand the true usefulness of \approach.


\subsection{Datasets}

We use the dataset from NLBSE'23\footnote{https://nlbse2023.github.io/tools/}~\cite{nlbse2023}, which is one of the most comprehensive collections of datasets about repository reports to date. We entails the statistics of this dataset in Table~\ref{tab:dataset} In particular, NLBSE'23 contains many smaller datasets and it documents the data in a more structured manner. It is chosen for our evaluation because of its following unique benefits:

\input{tab/data}

\begin{itemize}
    \item \textbf{Large Volume:} The collection of datasets is of a large volume, containing a total of 1,275,881 labeled reports. A detailed breakdown reveals that 670,951 reports, accounting for 52.6\% of the datasets, are bug reports. The remaining 604,930 reports are non-bug reports. Such a near-balanced distribution provides a robust foundation for training and evaluating \approach, ensuring that neither class is underrepresented.
    \item \textbf{Cross-Project:} The datasets contain reports collected from different software projects (over 127,000)\footnote{https://tinyurl.com/mr3h4my7}, which might be of different domains from GitHub. This renders rigorous evaluation of the ability for using \approach~on different projects plausible, which is an important property for neural language model-based approaches.
    \item \textbf{Diverse Report Formats:} The collected reports are of diverse formats and types, ranging from simple interface glitches to complex backend anomalies. The textual descriptions in these reports vary in length and complexity, with some providing detailed steps to reproduce the issue, while others offer a brief overview. This diversity in content and context makes the datasets particularly challenging and, at the same time, an ideal ground for testing the robustness of \approach.
\end{itemize}



For all experiments, we evaluate the approaches over all data from the datasets, i.e., all of the reports are part of the set of unlabeled reports, thereby we intentionally hide their true labels from the learning approaches. At every timestep, the query size is $k$, meaning that there will be $k$ new reports to be labeled by the human for training. As will be discussed in Section~\ref{sec:diff-k}, we examine different values of $k$. For all experiments, we follow the warm start strategy by using pre-trained neural language models. The process begins by fine-tuning the model on an initial set of $k$ randomly selected labeled reports, which serves as the baseline model for the first timestep of active learning.


\subsection{Experiment Preparation}
Here, we provide details on the experiments and implementation. We outline the necessary processes and setups required to prepare the experiments for \approach~and all other methods being compared:

\begin{itemize}

\item \textbf{Data Preprocessing:} All reports are preprocessed following the steps below:
\begin{enumerate}
    \item  \textit{Symbols Removal:} HTML tags and punctuation marks that appear in the reports are also removed.
    \item  \textit{Stop Words Removal:} Stop words, such as \texttt{\textcolor{blue}{``the''}} and \texttt{\textcolor{blue}{``is''}}, are removed. They appear frequently in natural language but with little contribution to semantic meaning.
    \item \textit{Case Conversion:} All upper case texts are converted into lower case ones.
\end{enumerate}

\item \textbf{Data Splitting:} For training/updating the neural language model or machine learning model (as in the state-of-the-art approaches), we adopt an 80/20 training/validation split when learning each set of $k$ reports. This ensures that the model is trained on a substantial amount of data while also having a separate set for hyperparameter tuning and validation. For testing, we uniformly sample 5,000 reports (with the true labels) from the test datasets based on their project, type, length, \textit{etc}; these are excluded from the unlabeled report from the beginning.


\item \textbf{Query Size:} To examine different cases of labeling the reports in the process of active learning, we run experiments on three different query sizes, i.e., $k=300$, $k=500$, and $k=700$. It is the number of samples to query at each round. indeed, it is still an open question as to how many samples are sufficient (how large the $k$ should be) for learning in a given problem. In this work, we pragmatically examined these three $k$ values, which serves as a good trade-off between the effectiveness they provide and the cost they incur. 



\item \textbf{Human Involvement:} From \textbf{RQ1} to \textbf{RQ4}, we emulate human involvement by assigning the selected reports the true labels from the datasets. For \textbf{RQ5}, we use the judgments given by the human practitioners as the true labels.

\item \textbf{Number of Timesteps and Repetitions:} To emulate a realistic setting, we conduct 10 timesteps for active learning considering the balance between data sufficiency and the time it takes to run the experiment. The initial set of labeled reports for training is selected at random. To account for variability and ensure the robustness of our results, we repeat the entire experiment 10 runs with different seeds wherein each repeat would have different training/updating and testing splits.

\item \textbf{Model and the Parameter Settings:} For the neural language model in \approach, we adopt BERT~\cite{DBLP:conf/naacl/DevlinCLT19} by default, which is pre-trained on Wikipedia data, since it is the most standard and simple one for natural language processing. Such a model will be fine-tuned incrementally with $k$ samples from our downstream task of bug report identification. However, it can be easily replaced with more sophisticated natural language models such as RoBERTa~\cite{DBLP:journals/corr/abs-1907-11692}. We have also evaluated \approach~with different neural language models to confirm its model-agnostic nature (see Section~\ref{sec:rq3}). The hyperparameter settings of those models are either default or tuned to fit the scale of our experiment infrastructure, e.g., for BERT, we use $18$ epochs, $32$ batch size, learning rate of $3 \times 10^{-5}$ and Epsilon factor of $10^{-8}$. They were determined through preliminary fine-tuning using 10\% of the training data as a validation set. For the hyperparameters of other approaches we compared, we set the same values as what was used in their corresponding work.


\end{itemize}

All experiments are implemented using Python 3.10. We leverage the APIs from PyTorch and Transformers to build and train the neural language model in \approach. All experiments are run on a high-performance machine equipped with AMD EPYC 7742 (Rome) 2.25GHz multi-core CPU. For GPU acceleration, we utilize an NVIDIA A100 with 40GB RAM.

\subsection{Metrics}

Since our experiments are both quantitative and qualitative, we design/select different sets of metrics for them.

\subsubsection{Performance (Quantitative)}

We use the following widely used metrics in software repository mining to quantitatively measure the model performance on the testing data at the end of \textit{Timestep} $n$:

\begin{itemize}
 \item \textbf{Precision:} Precision measures the proportion of correctly identified reports among all the reports retrieved, which is defined by:
     \begin{equation}
        Precision = \frac{tp}{tp + fp}
    \end{equation}
 whereby $tp$ and $fp$ are the number of true positive and false positive predictions, respectively. A higher precision signifies that the classification minimizes false positives and predicts information that is highly relevant.

    \item  \textbf{Recall:} Recall measures the ability of a model to identify and retrieve all relevant reports:
     \begin{equation}
        Recall = \frac{tp}{tp + fn}
    \end{equation}
     where $fn$ is the number of false negative predictions. A higher recall score indicates that the model can better reduce false negatives and ensure that a significant portion of relevant data is not overlooked.

    \item \textbf{Accuracy:} Accuracy is a straightforward metric that calculates the ratio of correctly identified reports (both bug and non-bug) to the total number of reports. Formally, it is defined as: 
     \begin{equation}
        Accuracy = \frac{tp+fp}{tp + fp + tn + fn}
    \end{equation}
    whereby $fp$ denotes the number of false positive predictions.

    \item \textbf{F1-score:} The F1-score is a harmonic mean of precision and recall, offering a balanced measure between the model's ability to correctly identify bug reports (precision) and its capability to capture all genuine bug reports within the datasets (recall), which is computed as:
     \begin{equation}
        F1\text{ }score = 2 \times \frac{Precision \times Recall}{Precision + Recall}
    \end{equation}   
    In scenarios where the cost of false positives and false negatives is significant, the F1-score becomes an important metric. Given the critical nature of bug report identification, where overlooking genuine bugs or misclassifying non-bug reports can have substantial implications, the F1-score serves as a cornerstone of our evaluation.

\end{itemize}

\subsubsection{Effort (Quantitative)}

It is also important to measure the efforts required to label the reports as part of \approach. Quantitatively, we do so by directly using the metrics for readability and identifiability of the queried report during \textit{Timestep} $n$, as discussed in Section~\ref{sec:sampling}. 

\subsubsection{Effort (Qualitative)}

Although the metrics for readability and identifiability used within the sampling of \approach~can indicate the difficulty of the reports that required labeling, they do not reflect how humans think about the effort in the real scenario. To reflect that, we design Likert scale questions that ask one to individually rate the readability and identifiability that are queried to the developers. For each attribute, the question covers five qualitative scores where a higher score indicates a more perceived readability or identifiability. We will further discuss this in Section~\ref{sec:rq5}.

In addition to the above, we also measure the time taken and monetary cost for the developers to label the queried reports recommended by \approach.

\subsection{Statistical Validation}
\label{sec:sta}

To ensure the comparative results are meaningful, we leverage the following statistical methods.

\subsubsection{Scott-Knott Test}
For comparing multiple approaches, we apply the widely used Scott-Knott test~\cite{DBLP:journals/tse/MittasA13,scott1974cluster,DBLP:journals/tosem/ChenL23,DBLP:journals/tosem/ChenL23a}---a hierarchical clustering algorithm---to determine if there are statistically meaningful differences among them. In the context of our study, we employ the Scott-Knott test to ascertain if there are significant differences in the values of metrics and hence generate a ranking across different approaches. Formally, Scott-Knott test aims to find the best split by maximizing the difference $\Delta$ in the expected mean before and after each split:
\begin{equation}
    \Delta = \frac{|l_1|}{|l|}(\overline{l_1} - \overline{l})^2 + \frac{|l_2|}{|l|}(\overline{l_2} - \overline{l})^2
\end{equation}
whereby $|l_1|$ and $|l_2|$ are the sizes of two sub-lists ($l_1$ and $l_2$) from list $l$ with a size $|l|$. $\overline{l_1}$, $\overline{l_2}$, and $\overline{l}$ denote their mean.

Recursively, in each iteration, two approaches are placed into different ranks if the results of their 10 repeated runs do not differ much according to the $\hat{A}_{12}$ effect size~\cite{Vargha2000ACA}; otherwise, they are said statistically similar. By leveraging the Scott-Knott test, we ensure that any differences observed in our results are statistically meaningful with indications of the better and worse while not requiring post-hoc correction. 


\subsubsection{Wilcoxon Sign-rank Test} We apply the Wilcoxon Sign-rank test~\cite{Wilcoxon1945IndividualCB} with $a=0.05$~\cite{DBLP:conf/icse/ArcuriB11} to investigate the statistical significance of the pairwise comparisons over all 10 runs where applicable, as it is a non-parametric statistical test that makes little assumption about the data distribution and has been recommended in software engineering research~\cite{DBLP:conf/icse/ArcuriB11}. It also assumes paired comparison between data points which fit our needs, since individual run is completed with independent seed.

%% file: tab/data.tex
\begin{table}[t!]
	\centering
	\caption{Statistics of dataset.}
	\label{tab:dataset}%
    \begin{adjustbox}{width=0.5\textwidth}
	\begin{tabular}{lll}\toprule
	& \textbf{Training Set} & \textbf{Test Set} \\\midrule
    \textbf{Bug report samples} & 670,951 (52.6\%) & 74,781 (52.5\%)\\
    \textbf{Non-bug report samples} & 604,930 (47.4\%) & 67,539 (47.5\%)\\\hline
    \textbf{Total samples} & 1,275,881 (100\%) & 142,320 (100\%)\\
    \bottomrule
	\end{tabular}
    \end{adjustbox}
\end{table}

%% file: sec/result.tex
\section{Results}
\label{sec:result}

Here, we discuss and analyze the results of the experiments in great detail.
\subsection{Effectiveness of Effort-aware Uncertainty Sampling in \approach}

\begin{figure}[!t]
\centering
\small 

\begin{subfigure}{0.6\columnwidth}
  \centering
  \includegraphics[width=0.8\linewidth]{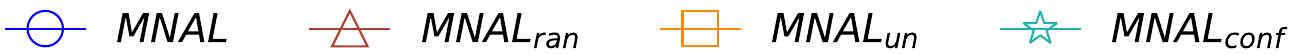} 
\end{subfigure}

\begin{subfigure}[t]{0.3\columnwidth}
  \includegraphics[width=\columnwidth]{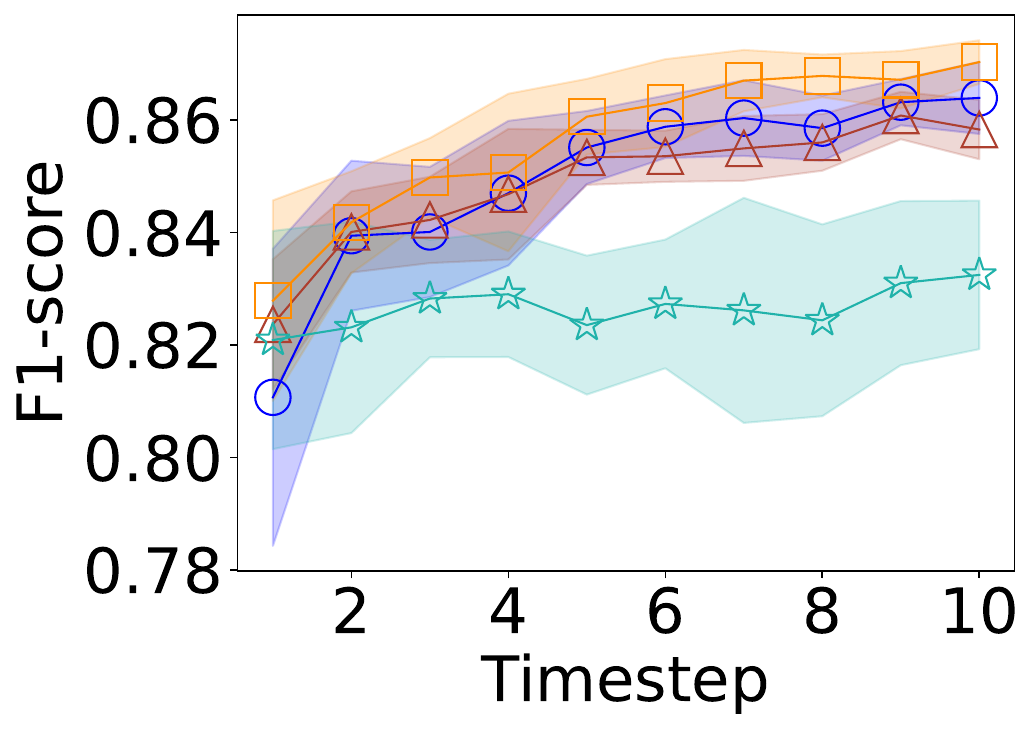}
  \subcaption{Query size 300}
\end{subfigure}%
\hspace{0.02\columnwidth} 
\begin{subfigure}[t]{0.3\columnwidth}
  \includegraphics[width=\columnwidth]{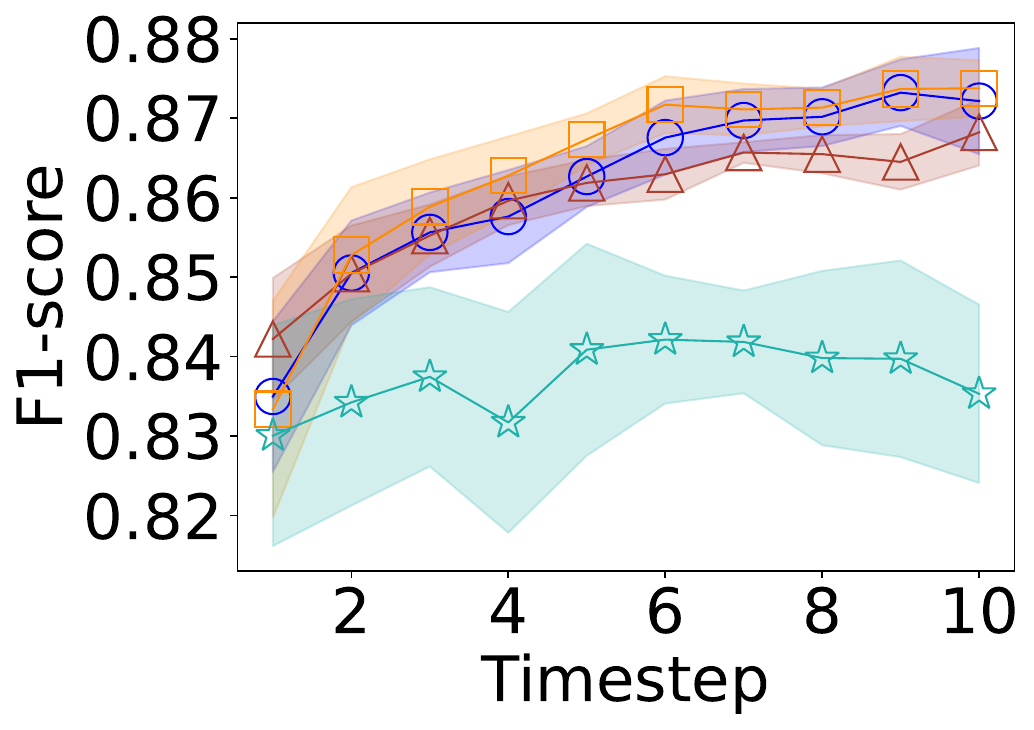}
  \subcaption{Query size 500}
\end{subfigure}%
\hspace{0.02\columnwidth}
\begin{subfigure}[t]{0.3\columnwidth}
  \includegraphics[width=\columnwidth]{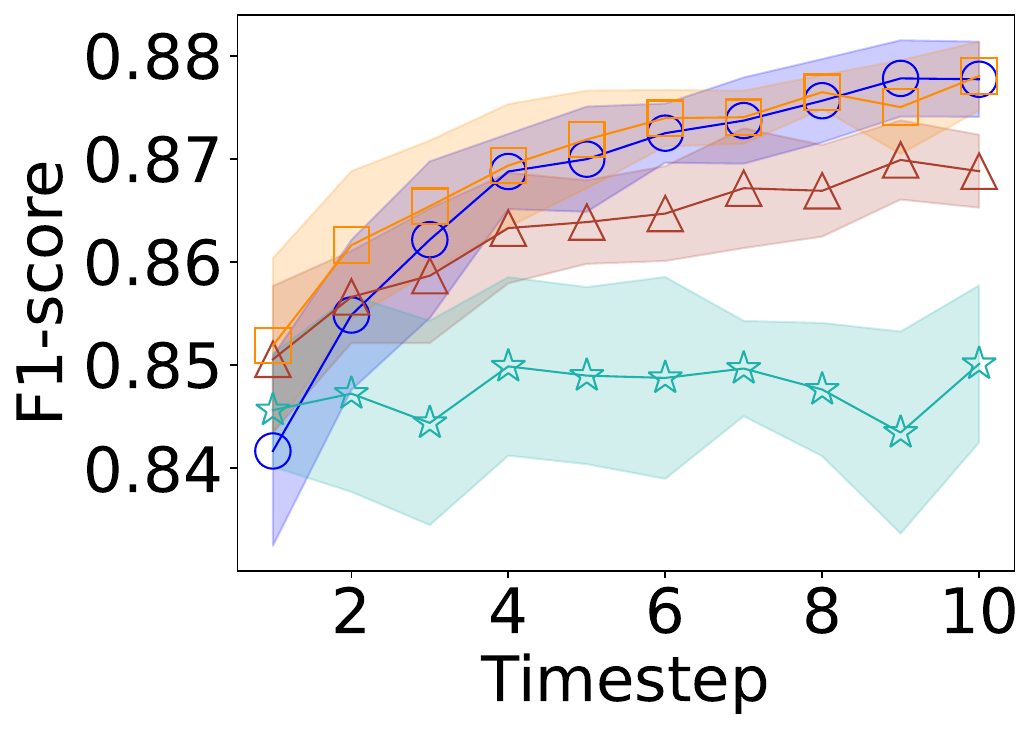}
  \subcaption{Query size 700}
\end{subfigure}

\begin{subfigure}[t]{0.3\columnwidth}
  \includegraphics[width=\columnwidth]{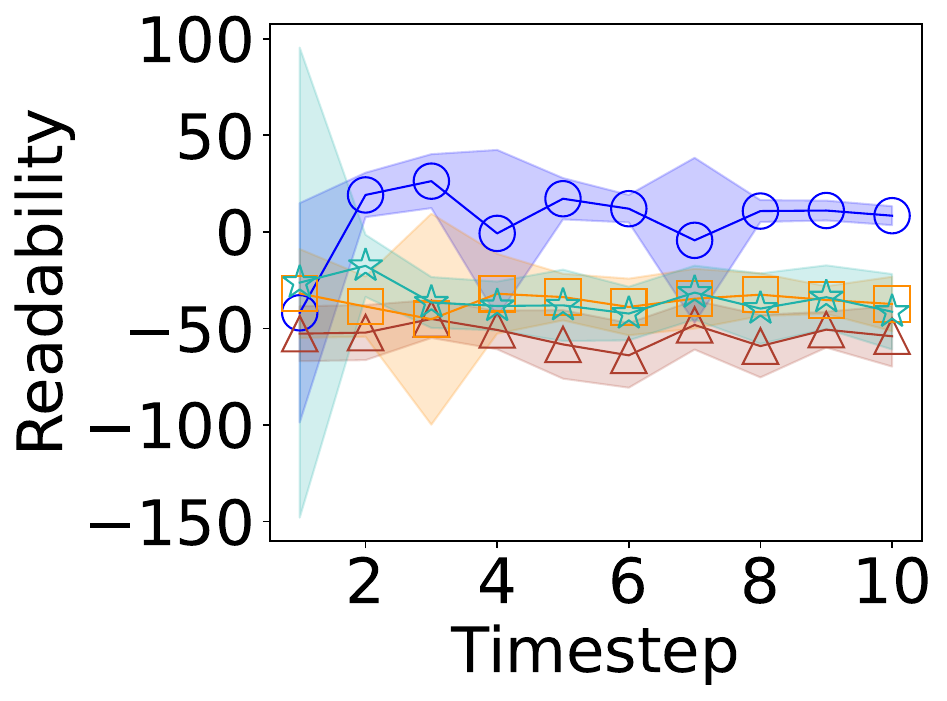}
  \subcaption{Query size 300}
\end{subfigure}%
\hspace{0.02\columnwidth}
\begin{subfigure}[t]{0.3\columnwidth}
  \includegraphics[width=\columnwidth]{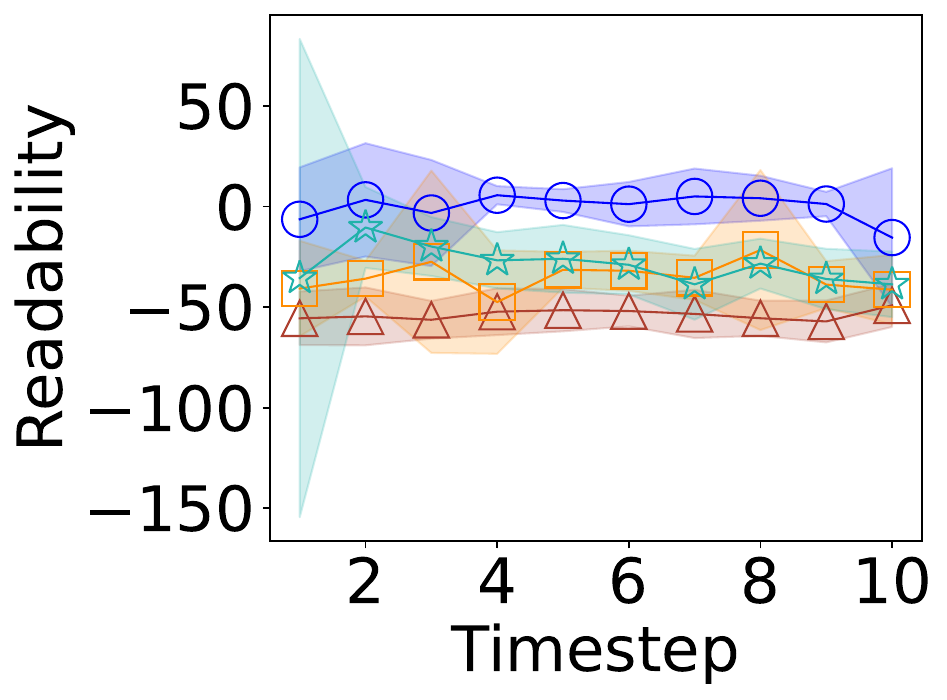}
  \subcaption{Query size 500}
\end{subfigure}%
\hspace{0.02\columnwidth}
\begin{subfigure}[t]{0.3\columnwidth}
  \includegraphics[width=\columnwidth]{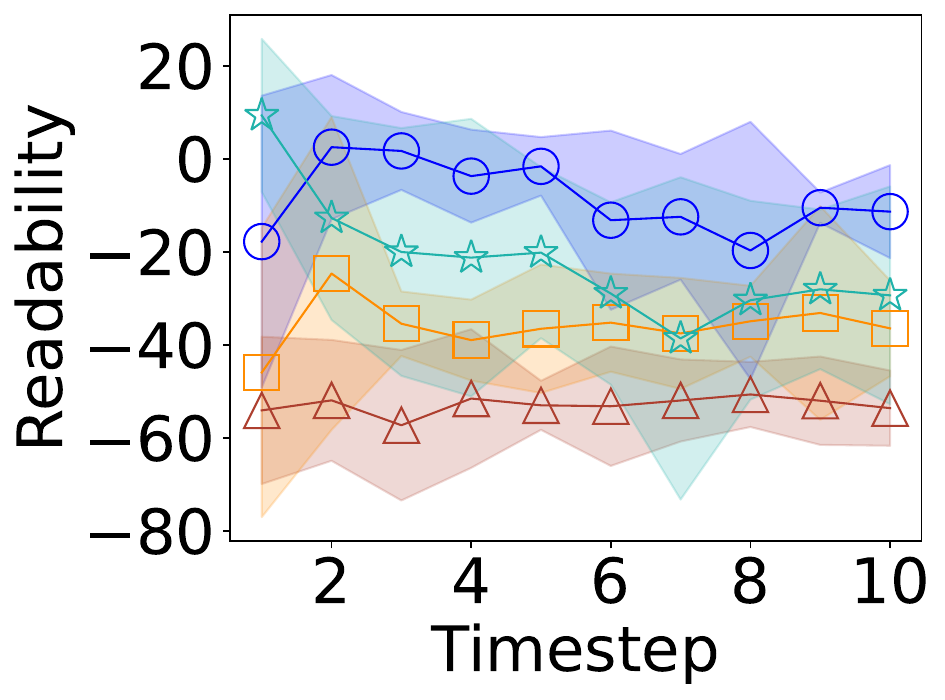}
  \subcaption{Query size 700}
\end{subfigure}

\begin{subfigure}[t]{0.3\columnwidth}
  \includegraphics[width=\columnwidth]{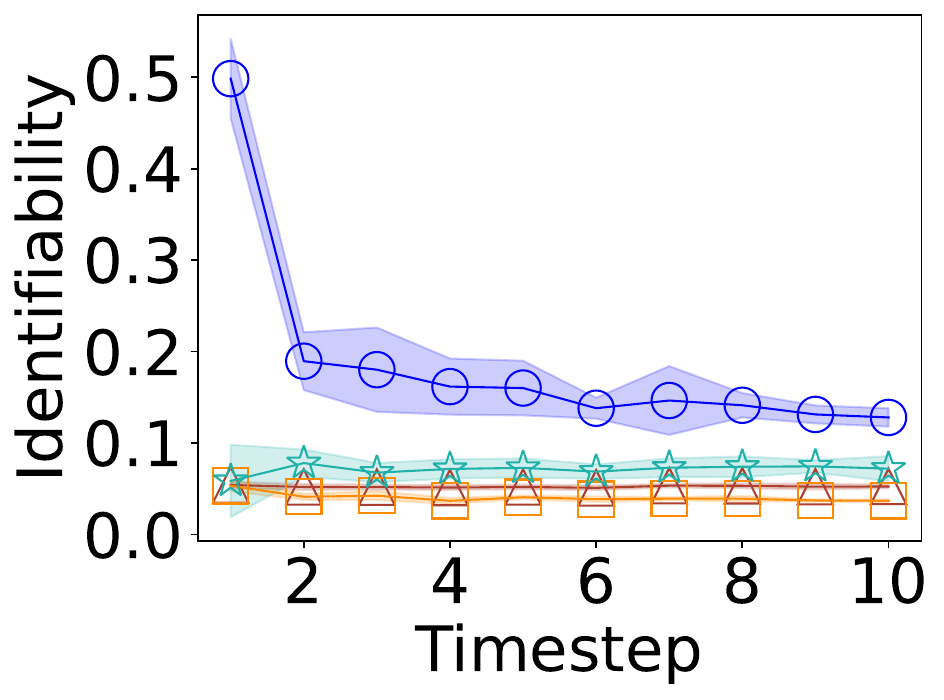}
  \subcaption{Query size 300}
\end{subfigure}%
\hspace{0.02\columnwidth}
\begin{subfigure}[t]{0.3\columnwidth}
  \includegraphics[width=\columnwidth]{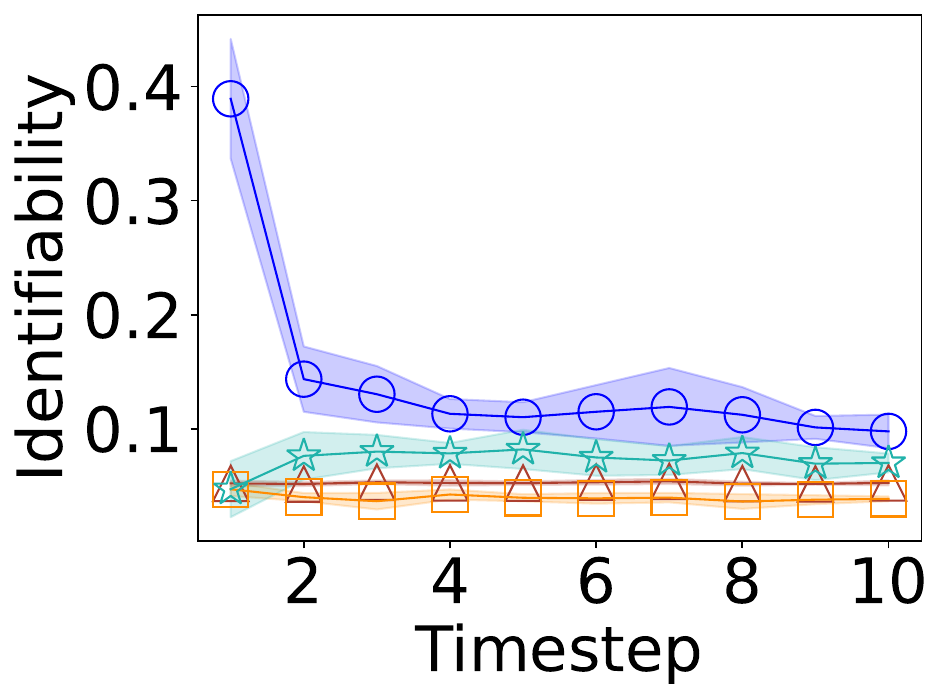}
  \subcaption{Query size 500}
\end{subfigure}%
\hspace{0.02\columnwidth}
\begin{subfigure}[t]{0.3\columnwidth}
  \includegraphics[width=\columnwidth]{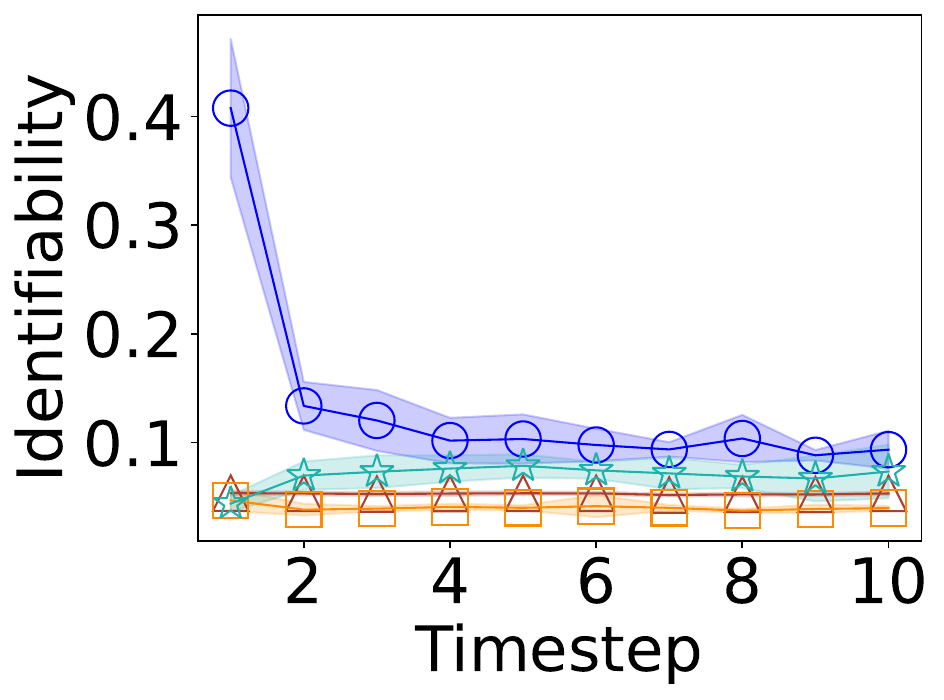}
  \subcaption{Query size 700}
\end{subfigure}

\caption{Comparing the effort-aware uncertainty sampling in \approach~with the state-of-the-art and baseline samplings over all 10 timesteps (10 runs each). The plots show the mean and standard deviation. \approach$_{ran}$,\approach$_{un}$ and \approach$_{conf}$ represent the variants that use random, uncertainty and confidence-based sampling, respectively.}
\label{fig:rq1-1}
\end{figure}

\subsubsection{Operationalization}

To answer \textbf{RQ1}, we conduct experiments to compare \approach~with three of its variants that use a different sampling scheme below (BERT is used as the default): 

\begin{itemize}
    \item \textbf{\approach$_{ran}$}: This is the baseline such that at each timestep, the $k$ unlabeled reports are randomly sampled from the unlabeled pool. It emulates the case where the neural language model is directly used and updated without active learning. We realize \approach$_{ran}$ by replacing the effort-aware uncertainty sampling in \approach~as a random sampling.
    
    \item \textbf{\approach$_{un}$}: This conducts the state-of-the-art sampling scheme in active learning based on uncertainty only \cite{DBLP:conf/iwpc/ThungLL15,DBLP:journals/infsof/WuZCZYM21,DBLP:journals/infsof/GeFQGQ22}. All other components are identical to the \approach. 
    \item \textbf{\approach$_{conf}$}: This implements the confidence-based sampling scheme from active learning, as described in \cite{DBLP:journals/infsof/WuZCZYM21,DBLP:journals/csur/RenXCHLGCW22}. All other components are identical to the \approach. 
 

\end{itemize}

The \approach$_{un}$ variant employs a state-of-the-art active learning sampling scheme based solely on uncertainty; \approach$_{ran}$ is one that randomly chooses reports for human labeling, which is a baseline. All other components are identical to those in \approach. We cannot simply rule out this component as in active learning, there has to be some mechanism to choose which reports are sent to the human for labeling, and \approach$_{ran}$ is the simplest and naive baseline.

Since there are multiple comparisons, we use the Scott-Knott test introduced in Section~\ref{sec:sta} to verify the statistical significance of the results.


\subsubsection{Findings}

We plot the traces of the key metrics for all 10 timesteps in Figure~\ref{fig:rq1-1}. Table~\ref{tab:rq1-1} also reports on the overall results (on all timesteps) of more performance metrics and the effort metrics, together with the ranks from the Scott-Knott test. As can be seen, \approach~obtains significantly better results than \approach$_{ran}$ and \approach$_{conf}$ on all performance metrics across the cases since the random sampling and confidence-based always fails to select representative reports for human labeling, especially when the query size is larger. The effectiveness of our effort-aware uncertainty sampling is quantitatively remarkable. When compared with a random sampling baseline (\approach$_{ran}$), \approach~achieves a 98.1\% improvement in readability and a 194.7\% improvement in identifiability. Against the state-of-the-art uncertainty-only sampling (\approach$_{un}$), \approach~ still shows a significant advantage with a 97.2\% improvement in readability and a 283.6\% improvement in identifiability. These figures, supported by the Scott-Knott test results in Table 2, demonstrate that the reports that \approach~ selects for human labeling are statistically and substantially easier for developers to process. 

This quantitative improvement translates to a significant reduction in human cognitive load. The reports selected by \approach~ are not only more likely to improve model performance but are also structured in a way that is more ``human-friendly,'' making the labeling task faster and less prone to fatigue. While there is a slight trade-off, resulting in marginally lower F1-scores compared to a pure uncertainty approach (\approach$_{un}$), the immense gain in labeling efficiency validates the core principle of our mutualistic framework: benefiting both the machine model and the human developer. The fluctuating results in readability and identifiability across timesteps are due to the evolving nature of the model; as the model learns, the uncertainty landscape shifts, causing different types of reports to become the most informative.

Compared with \approach$_{un}$, \approach~is still competitive but tends to be less effective for improving performance. This is understandable since the most useful reports for improving the performance of the neural language model might not be readable or identifiable. However, we see that \approach~achieves considerably better results than \approach$_{un}$ on readability and identifiability (97.2\% and 283.6\% improvement), thanks to the effort-aware uncertainty sampling. This indicates that \approach~is much more human-friendly which makes the labeling process effortless and time-saving for software developers, significantly benefiting one end of the mutualistic relation. Notably, in contrast to \approach$_{un}$, \approach~leads to a dramatic gain on the effort metrics while suffering a relatively marginal loss on the effectiveness of prediction. What we found is that there is a clear conflicting relationship between uncertainty and readability/identifiability. The key reason is that the most uncertain report which can be most helpful to improve the accuracy might be difficult to read/identify the bug relevance. On the other hand, those that can enable a human to easily decide its bug relevance could be samples that are already well-learned by a model.

\input{tab/rq1_avg}

An interesting observation is that, while the performance of \approach~is improved steadily with more timesteps, it tends to experience fluctuated results for readability and identifiability. This is because, as the uncertainty of the unlabeled reports evolves due to the continuously updated neural language model, the sampling landscape with respect to uncertainty, readability, and identifiability of the unlabeled reports also changes, leading to a different set of selected reports for labeling according to the effort-quality score. In particular, we see that there is a significant drop in the identifiability (and improvement in readability) from \textit{Timestep 1} to \textit{Timestep 2}, this is still caused by the change in the uncertainty of the unlabeled reports. Since the neural language model has not been fine-tuned when calculating the uncertainty in \textit{Timestep 1}, the model has not converged to a stable state compared with the case in \textit{Timestep 2}. This has caused considerably different uncertainty results that led to a drastic shift in the landscape, hence the significant change in the identifiability metric. In contrast, \approach$_{ran}$ and \approach$_{un}$ are not affected as the former relies on random exploration while the latter is only guided by the uncertainty value. We will further elaborate this observation with more evidence in Section~\ref{sec:trade-off}.



We can conclude the response to \textbf{RQ1} as:

\begin{quotebox}
   \noindent
   \textit{\textbf{\underline{To RQ1:}} Compared with the others, the effort-aware uncertainty sampling in \approach~exhibits a competitive evolving performance while significantly reducing the effort of manually labeling the queried reports for the developers.}
\end{quotebox}





\subsection{Usefulness of the Pseudo-labeling}
\label{sec:rq2}

\subsubsection{Operationalization}

To assess the benefit of the pseudo-labeling for \textbf{RQ2}, we compare the following approaches:

\begin{itemize}
    \item \textbf{No modification:} This denotes the \approach~variant without any pseudo labeling.
    \item \textbf{With augmentation:} This is EDA~\cite{DBLP:conf/emnlp/WeiZ19}, a state-of-the-art data augmentation technique, that seeks to address similar problem as the proposed pseudo labeling.
    \item \textbf{With pseudo labeling:} This is the full version of \approach~with the proposed pseudo labeling.
\end{itemize}



In essence, EDA and pseudo-labeling are two distinct techniques used to enhance machine learning models but address different aspects of model training. EDA focuses on improving text classification performance by artificially increasing the size and variability of training data through operations such as synonym replacement, random insertion, random swap, and random deletion. This helps models generalize better, especially with smaller datasets. On the other hand, pseudo-labeling is a semi-supervised learning technique where the model, after initial training on labeled data, generates predictions for unlabeled data. These predictions are then used as additional training data, effectively leveraging both labeled and unlabeled examples to refine the model's performance. In particular, while EDA enriches the existing labeled data, pseudo-labeling expands the training dataset by incorporating predictions on unlabeled data.

Since we have more than two approaches in the comparisons, we use the Scott-Knott test to verify the statistical significance between the results. Again, BERT is used as the default neural language model.

\begin{figure}[!t]
\centering
\footnotesize

\begin{subfigure}{0.1\columnwidth}
  \centering
  \includegraphics[width=\linewidth]{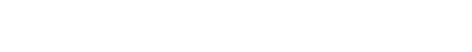} 
\end{subfigure}
\begin{subfigure}{0.6\columnwidth}
  \centering
  \includegraphics[width=0.9\linewidth]{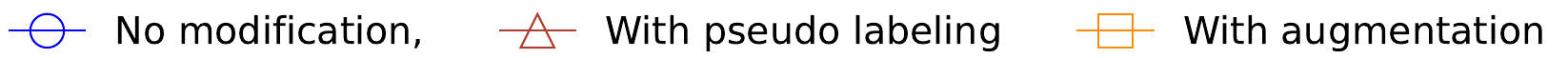} 
\end{subfigure}
\begin{subfigure}{0.1\columnwidth}
  \centering
  \includegraphics[width=\linewidth]{fig/blank_stripe.pdf} 
\end{subfigure}
\begin{subfigure}[t]{0.3\columnwidth}
\includegraphics[width=\columnwidth]{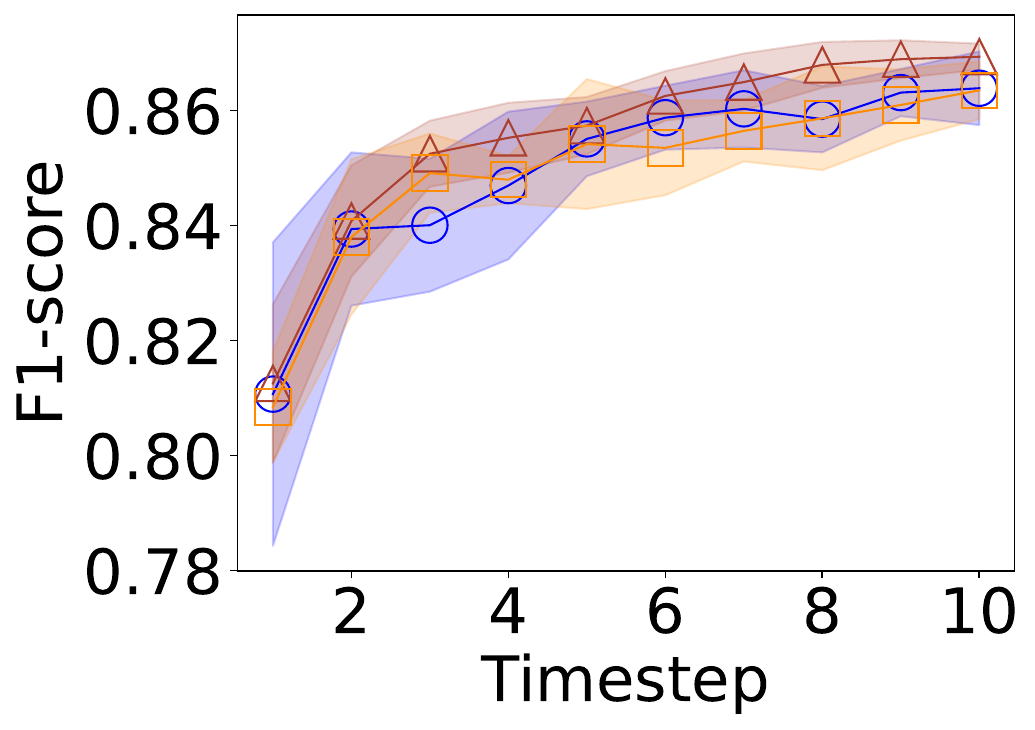}
    \subcaption{Query size 300}
    \end{subfigure}
\begin{subfigure}[t]{0.3\columnwidth}
\includegraphics[width=\columnwidth]{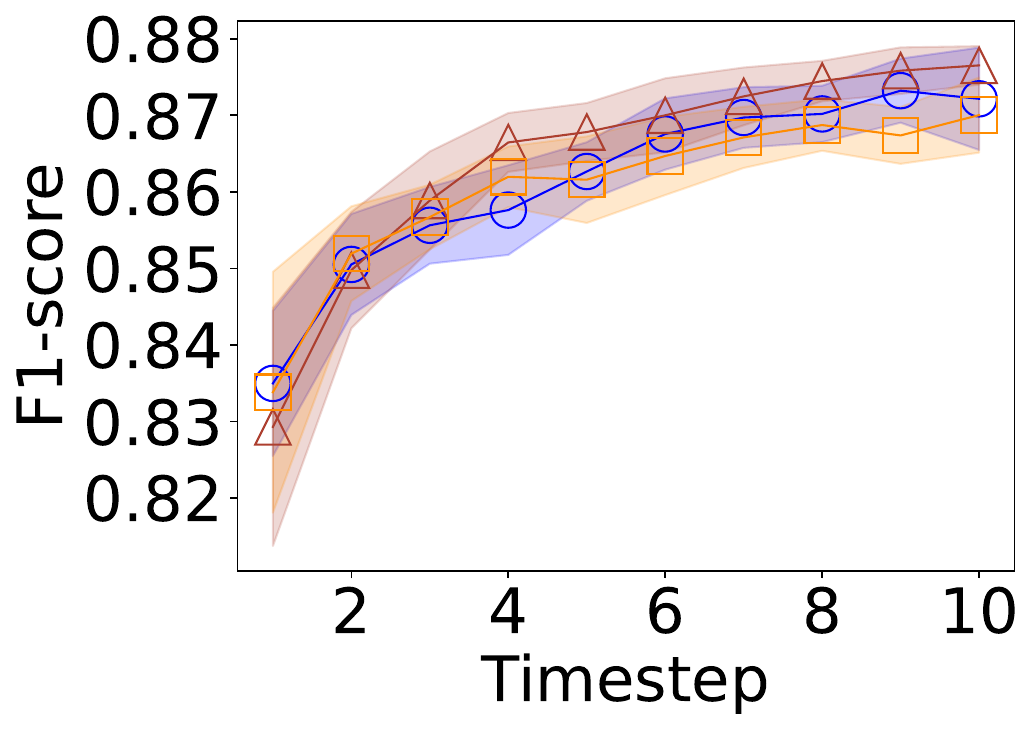}
   \subcaption{Query size 500}
    \end{subfigure}
\begin{subfigure}[t]{0.3\columnwidth}
\includegraphics[width=\columnwidth]{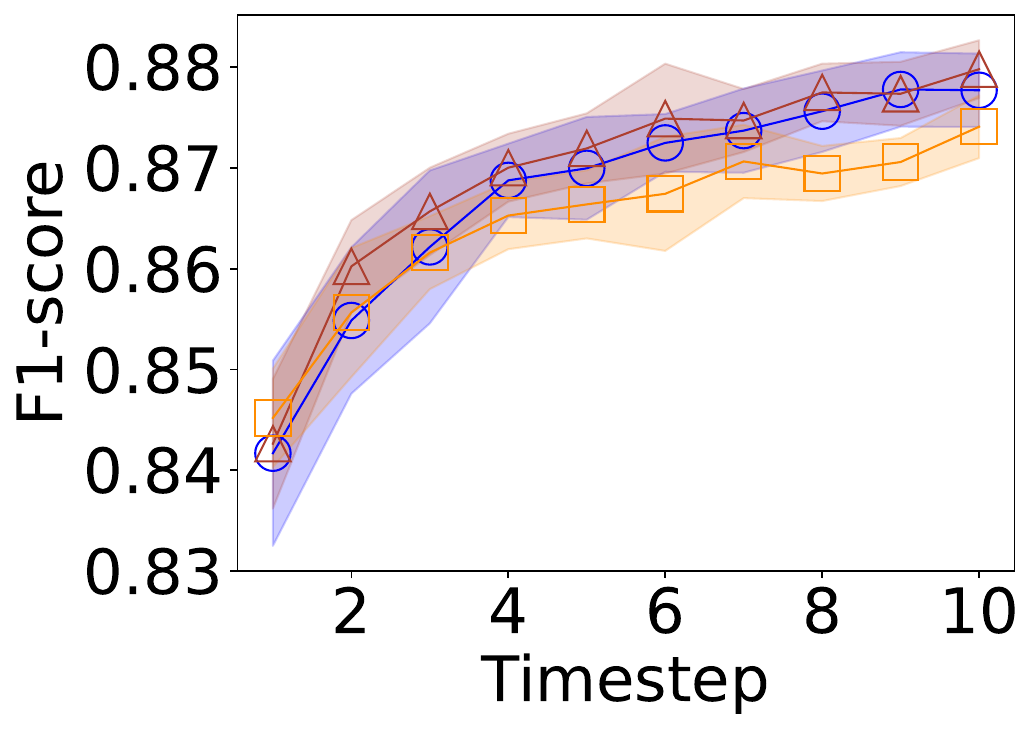}
   \subcaption{Query size 700}
    \end{subfigure}
\begin{subfigure}[t]{0.3\columnwidth}
\includegraphics[width=\columnwidth]{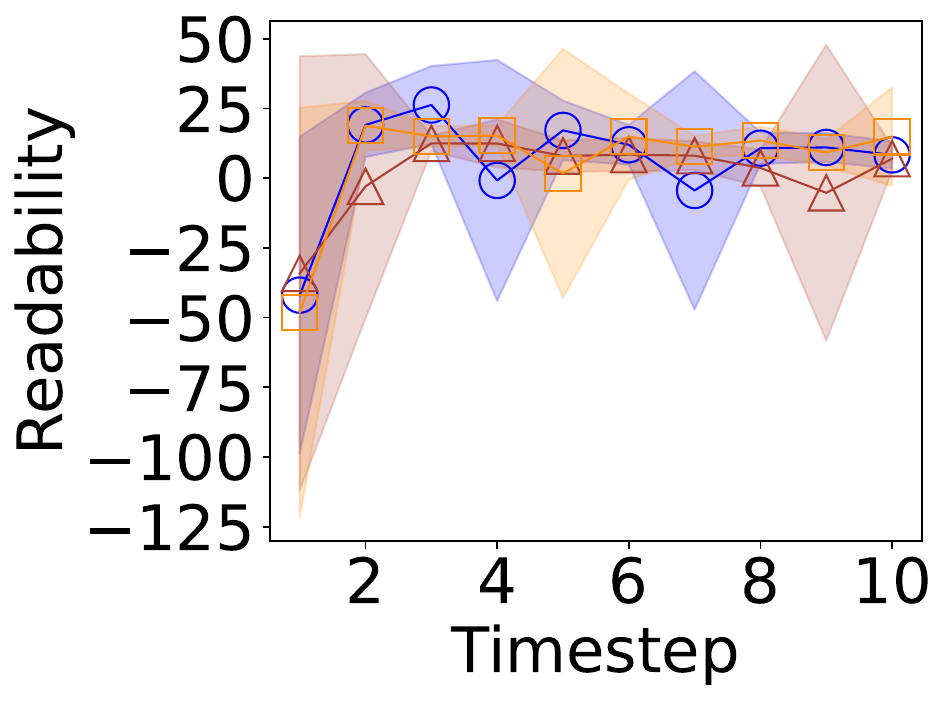}
    \subcaption{Query size 300}
    \end{subfigure}
\begin{subfigure}[t]{0.3\columnwidth}
\includegraphics[width=\columnwidth]{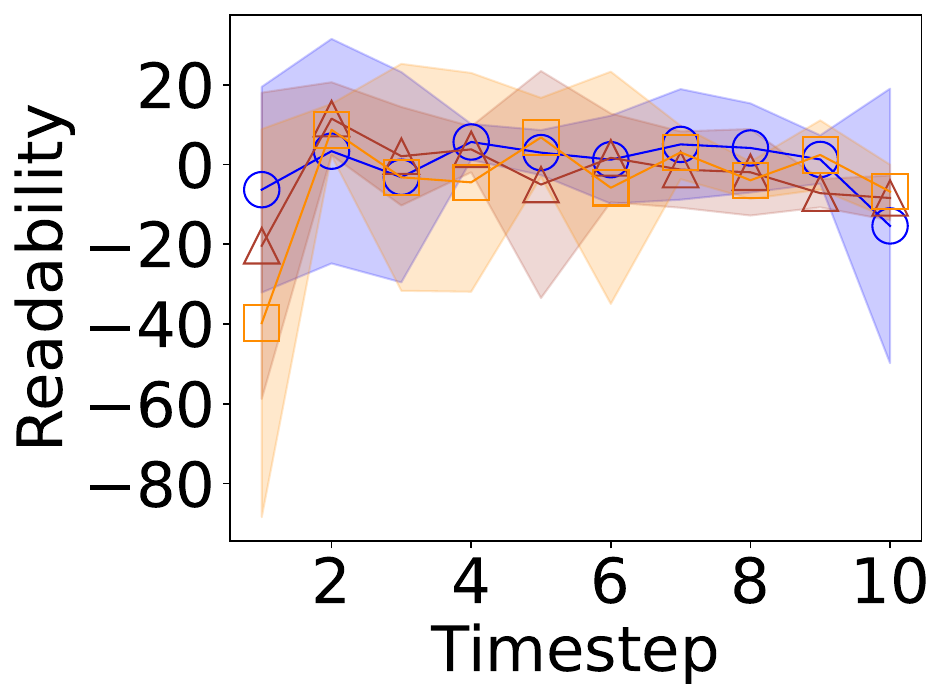}
   \subcaption{Query size 500}
    \end{subfigure}
\begin{subfigure}[t]{0.3\columnwidth}
\includegraphics[width=\columnwidth]{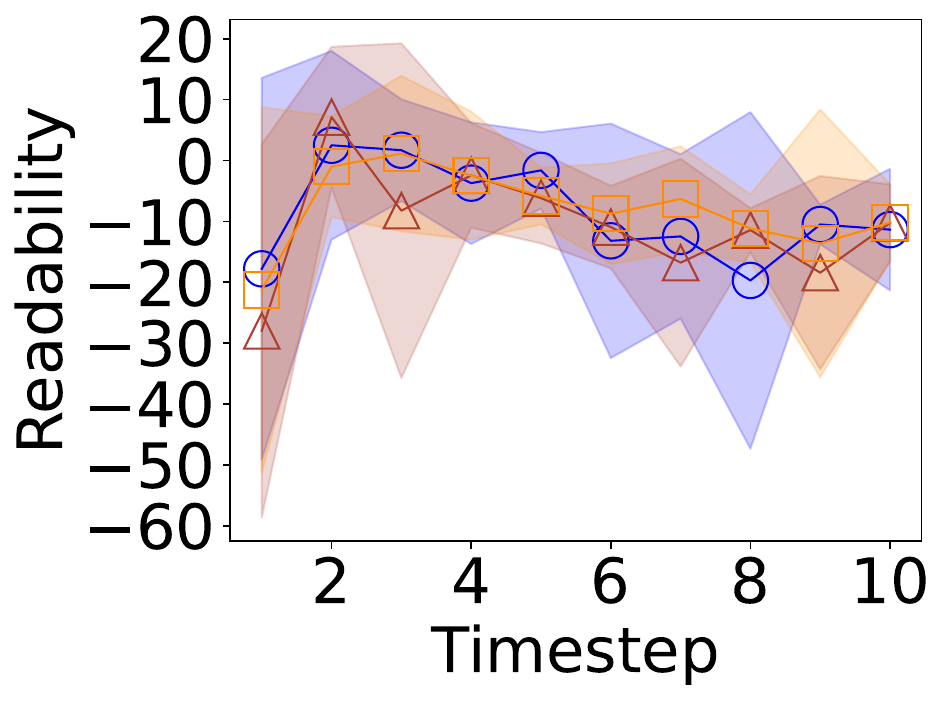}
   \subcaption{Query size 700}
    \end{subfigure}
\begin{subfigure}[t]{0.3\columnwidth}
\includegraphics[width=\columnwidth]{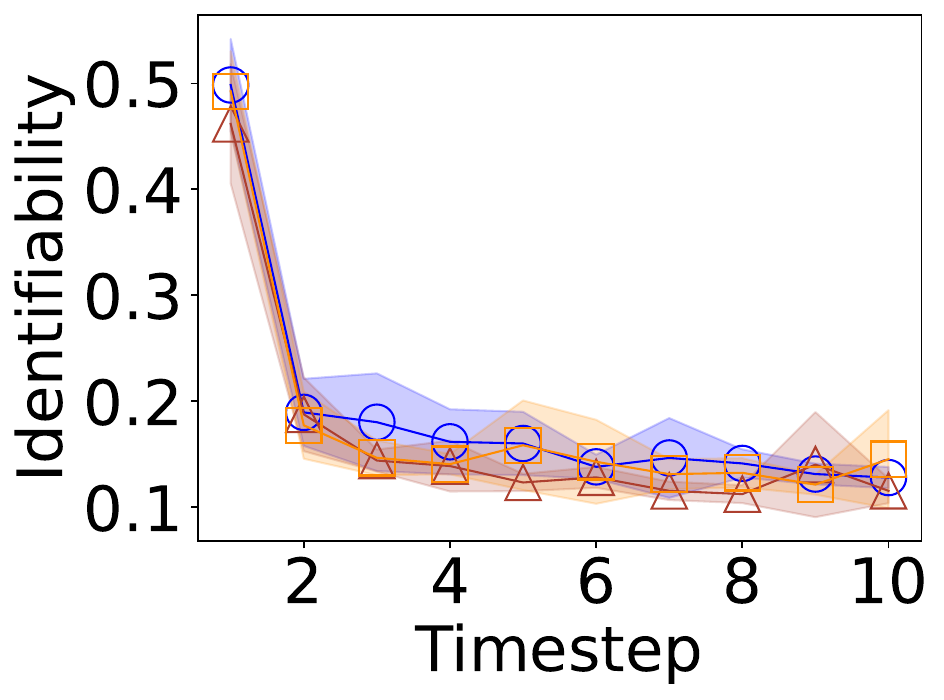}
    \subcaption{Query size 300}
    \end{subfigure}
\begin{subfigure}[t]{0.3\columnwidth}
\includegraphics[width=\columnwidth]{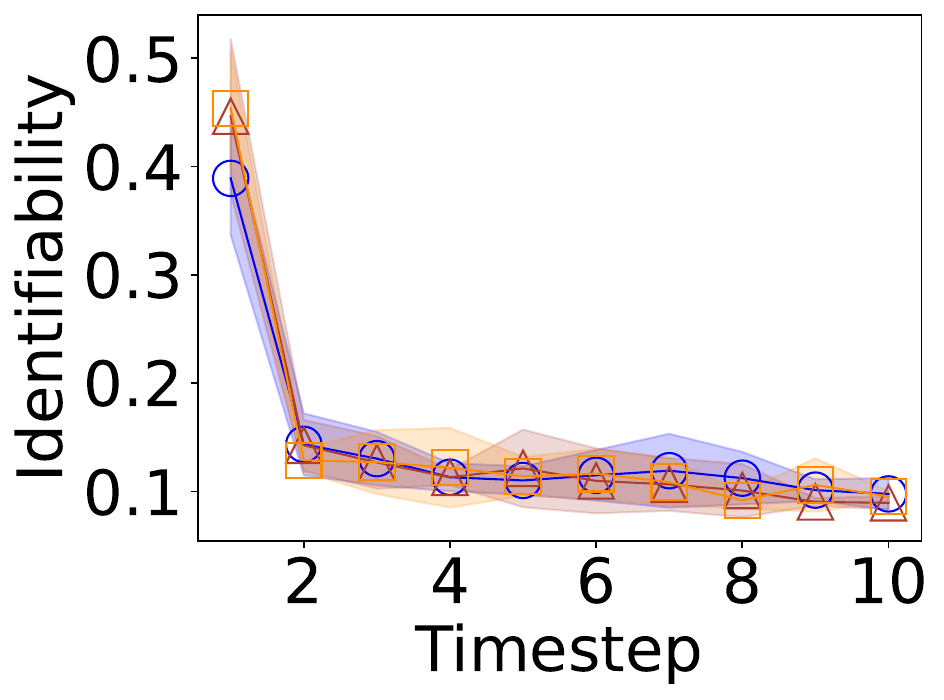}
   \subcaption{Query size 500}
    \end{subfigure}
\begin{subfigure}[t]{0.3\columnwidth}
\includegraphics[width=\columnwidth]{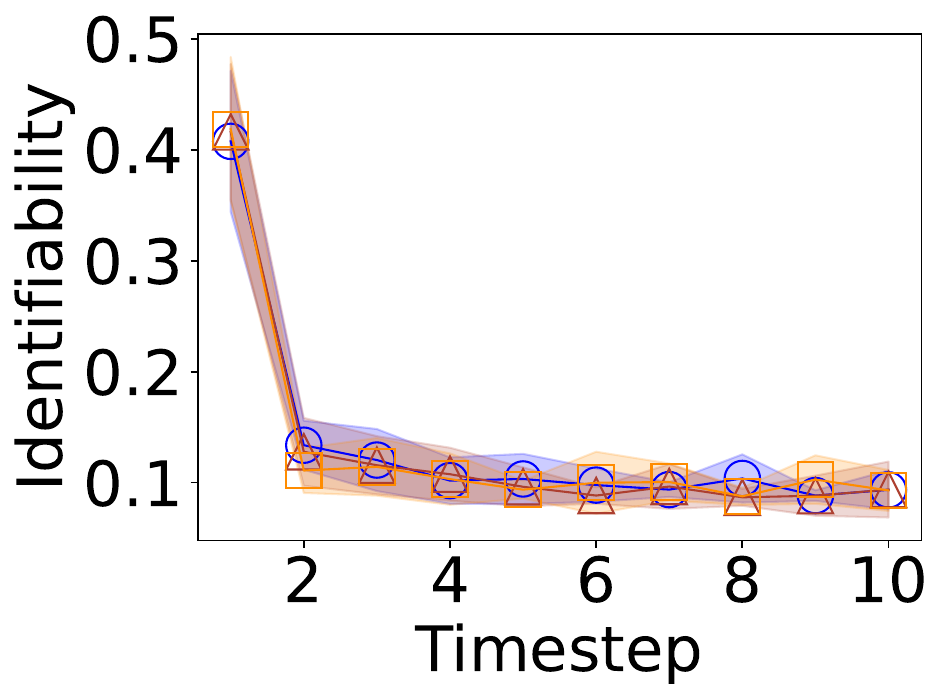}
   \subcaption{Query size 700}
    \end{subfigure}
    \caption{Comparing EDA (data augmentation) against \approach~with and without pseudo-labeling over all 10 timesteps (10 runs each). The plots show the mean and standard deviation.}
\label{fig:rq2-1}
\end{figure}

\input{tab/rq2_new}

\subsubsection{Findings}

Figure~\ref{fig:rq2-1} and Table~\ref{tab:rq2-1} show the trace along the timesteps and the overall statistical significance results over the cases, respectively. We see that the pseudo-labeling is indeed beneficial to the performance of the neural language model, even when compared with the EDA, as it achieves statistically better results on the majority of the performance metrics than the others. The inclusion of pseudo-labeling provides a clear, statistically significant boost to model performance. For instance, with a query size of 500, \approach with pseudo-labeling achieves an F1-score of {0.864}, surpassing both the no-modification version ({0.861}) and the EDA-augmented version ({0.860}). This shows the effectiveness of exploiting unlabeled data as opposed to perturbing labeled data (as in EDA). Pseudo-labeling is highly effective because it allows the model to learn from the vast, untapped pool of unlabeled reports without requiring extra human effort. By leveraging the model's own high-confidence predictions on unlabeled data that is semantically similar to human-labeled reports, we enrich the training set with diverse, real-world examples. This method is more powerful than data augmentation (EDA), which merely creates synthetic variations of existing labeled data.

There appears to be a slight degradation in readability and identifiability with pseudo-labeling, as the additional pseudo-labeled reports might lead to a model that produces better uncertainty for those reports of bad readability and identifiability in the subsequent timesteps. However, we see that the effect on readability and identifiability by pseudo-labeling is negligible with larger query sizes as the degradation is often statistically insignificant. This is because the more data samples used in the updating, the more reliable the neural language model becomes, hence helping the sampling to more accurately locate those reports with larger uncertainty while having better results on readability and identifiability in the landscape.




Therefore, we can conclude the following for \textbf{RQ2}: 

\begin{quotebox}
   \noindent
   \textit{\textbf{\underline{To RQ2:}} The pseudo-labeling in \approach~leads to a significant improvement in the performance of the neural language model with minimal effect on the readability and identifiability.}
\end{quotebox}

\subsection{Model-Agnostic Nature of \approach}
\label{sec:rq3}



\subsubsection{Operationalization}

To verify the model-agnostic nature of \approach, in \textbf{RQ3}, we equip it with  three alternative neural language models that replace BERT, these are:


\begin{itemize}


\item \textbf{CodeBERT~\cite{DBLP:conf/emnlp/FengGTDFGS0LJZ20}} A specifically designed model for code naturalness, in which the pre-training and the attention have been adopted to cater to the strong structure of code semantics. This fits our problem well since code is also very common in a report from GitHub.

\item \textbf{RoBERTa~\cite{DBLP:journals/corr/abs-1907-11692}} This is a robust and widely-used extension of BERT, known for its efficiency and accuracy in various natural language processing tasks. Its architecture and training methodology make it a strong contender in our study.

\item \textbf{RTA~\cite{DBLP:conf/icse/FangZTJXS23}} A pre-trained model tailored for analyzing bug reports via learning a universal representation. The pre-training is completed with two distinct learning objectives, including a masked language model and a contrastive learning objective. 

\end{itemize}

We selected CodeBERT, RoBERTa, and RTA for our experiments because these models represent the state-of-the-art in bug report identification tasks\cite{DBLP:journals/infsof/MeherBM24,DBLP:journals/corr/abs-2202-06149,DBLP:conf/icse/FangZTJXS23}. Our choice was based on their top performance and relevance to our task, ensuring a robust evaluation of \approach's model independence. Specifically, our reasons are demonstrated below:

\begin{itemize} \item CodeBERT and RoBERTa are optimized and widely recognized models based on the BERT architecture: 
\begin{itemize} \item CodeBERT is specifically designed for tasks in software engineering, such as code understanding and generation. It extends the BERT architecture to handle both natural language and programming language tasks, making it highly relevant for evaluating~\approach~ in code-related domains. CodeBERT has been demonstrated to outperform other models in various software engineering tasks, ensuring that our method is tested against a strong baseline in this field~\cite{DBLP:conf/emnlp/FengGTDFGS0LJZ20}. \item RoBERTa (Robustly optimized BERT approach) is known for its effectiveness in a wide range of NLP tasks. It improves upon BERT by utilizing a more extensive training dataset and better hyperparameter tuning, which has led to state-of-the-art results in many natural language understanding benchmarks. By including RoBERTa, we ensure that~\approach~ is evaluated using a highly optimized general-purpose text model, allowing us to assess its generalizability~\cite{DBLP:journals/corr/abs-1907-11692}. 
\end{itemize}
\item RTA is a recent and specialized model trained to generate a universal representation of bug reports. Unlike CodeBERT and RoBERTa, which are more general-purpose, RTA is tailored specifically for addressing multiple downstream software engineering problems, particularly bug report analysis. This model is designed to handle the nuances and specific requirements of software engineering tasks related to bug reports, such as prioritization and triage. By including RTA, we ensure that~\approach~ is tested against a model that is highly specialized, thereby verifying its effectiveness across both general and task-specific contexts~\cite{DBLP:conf/icse/FangZTJXS23}.

\end{itemize}

\begin{figure}[!t]
\centering

\begin{subfigure}{0.2\columnwidth}
  \centering
  \includegraphics[width=\linewidth]{fig/blank_stripe.pdf} 
\end{subfigure}
\begin{subfigure}{0.33\columnwidth}
  \centering
  \includegraphics[width=0.7\linewidth]{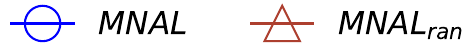} 
\end{subfigure}
\begin{subfigure}{0.2\columnwidth}
  \centering
  \includegraphics[width=\linewidth]{fig/blank_stripe.pdf} 
\end{subfigure}
\begin{subfigure}[t]{0.3\columnwidth}
\includegraphics[width=\columnwidth]{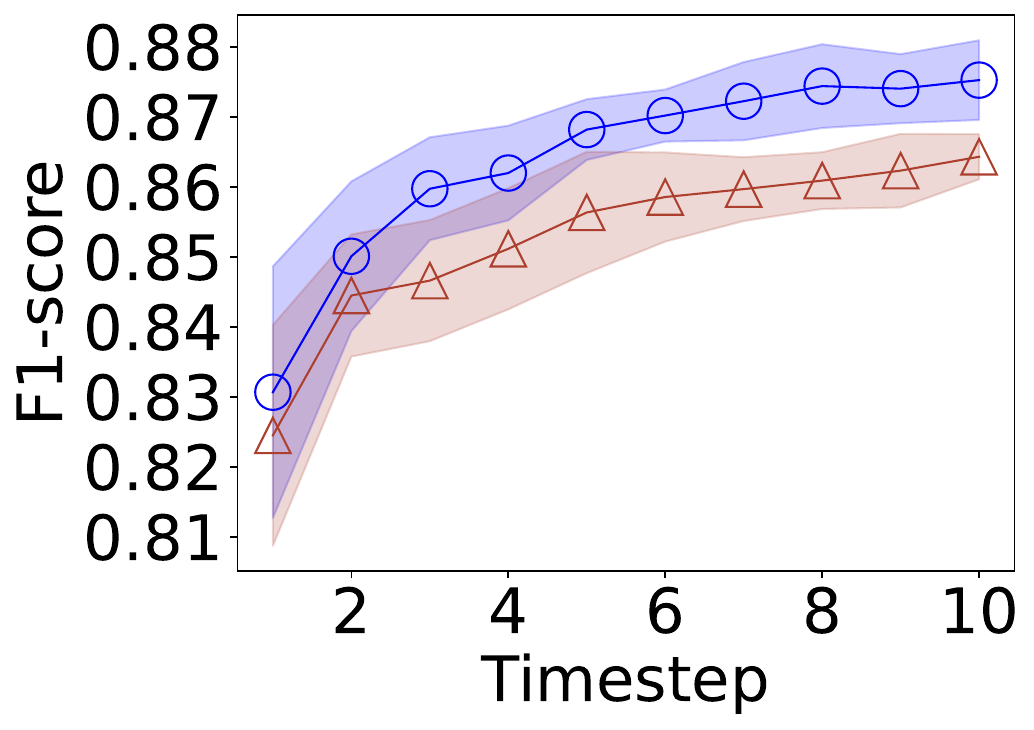}
    \subcaption{CodeBERT, Query size 300}
    \end{subfigure}
\begin{subfigure}[t]{0.3\columnwidth}
\includegraphics[width=\columnwidth]{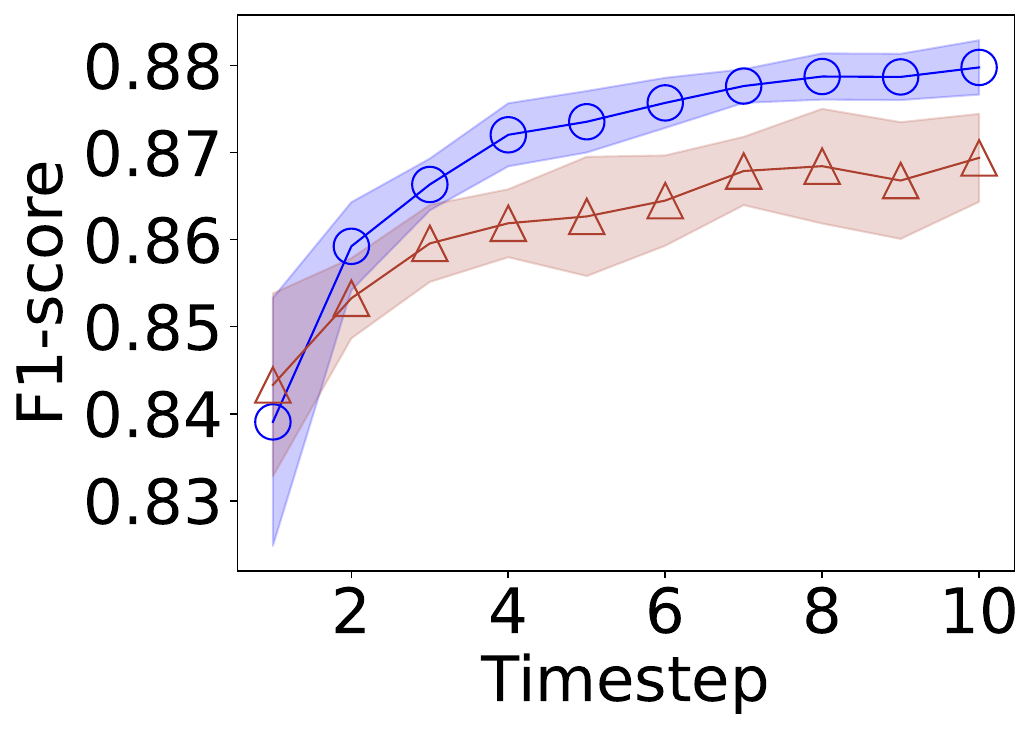}
   \subcaption{CodeBERT, Query size 500}
    \end{subfigure}
\begin{subfigure}[t]{0.3\columnwidth}
\includegraphics[width=\columnwidth]{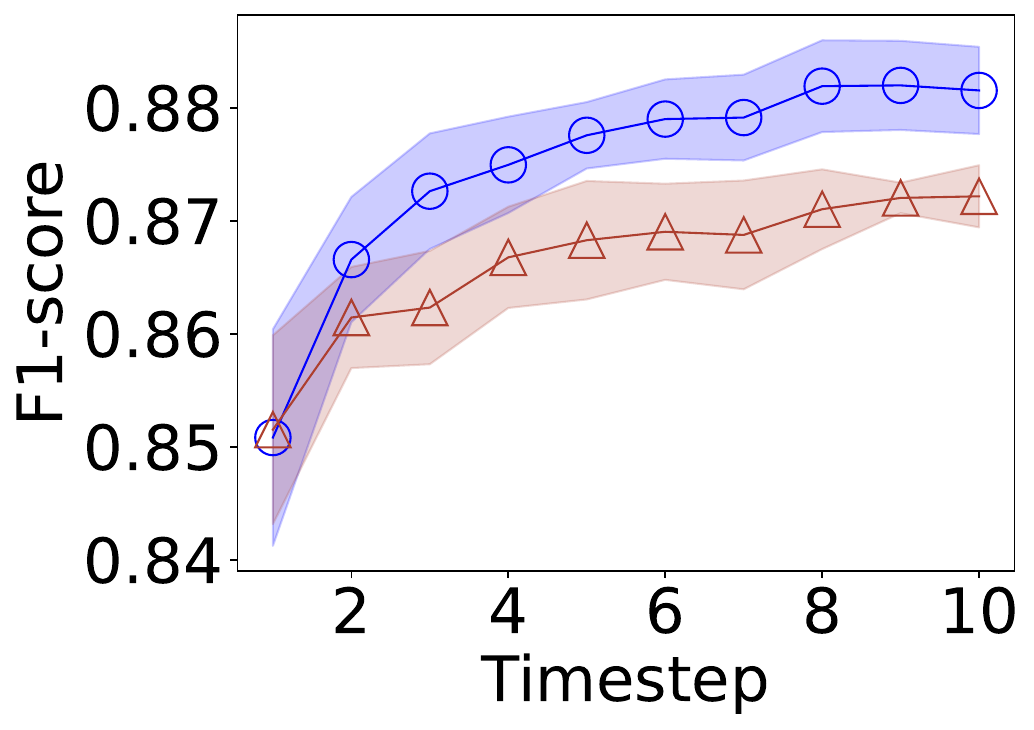}
   \subcaption{CodeBERT, Query size 700}
    \end{subfigure}
\begin{subfigure}[t]{0.3\columnwidth}
\includegraphics[width=\columnwidth]{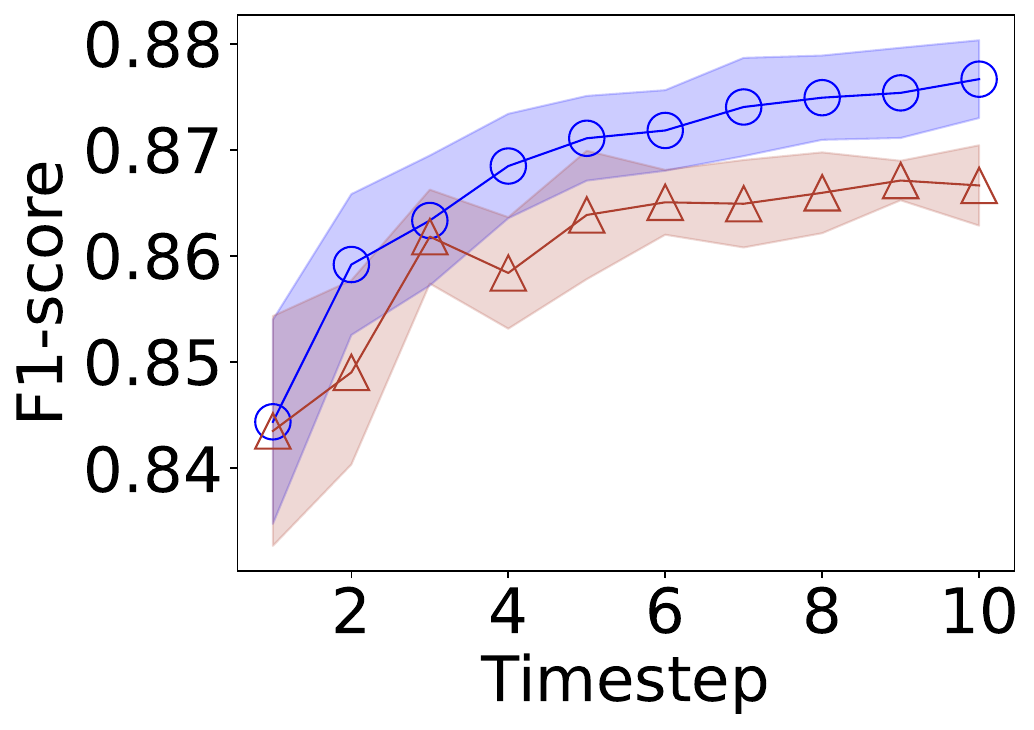}
    \subcaption{RoBERTa, Query size 300}
    \end{subfigure}
\begin{subfigure}[t]{0.3\columnwidth}
\includegraphics[width=\columnwidth]{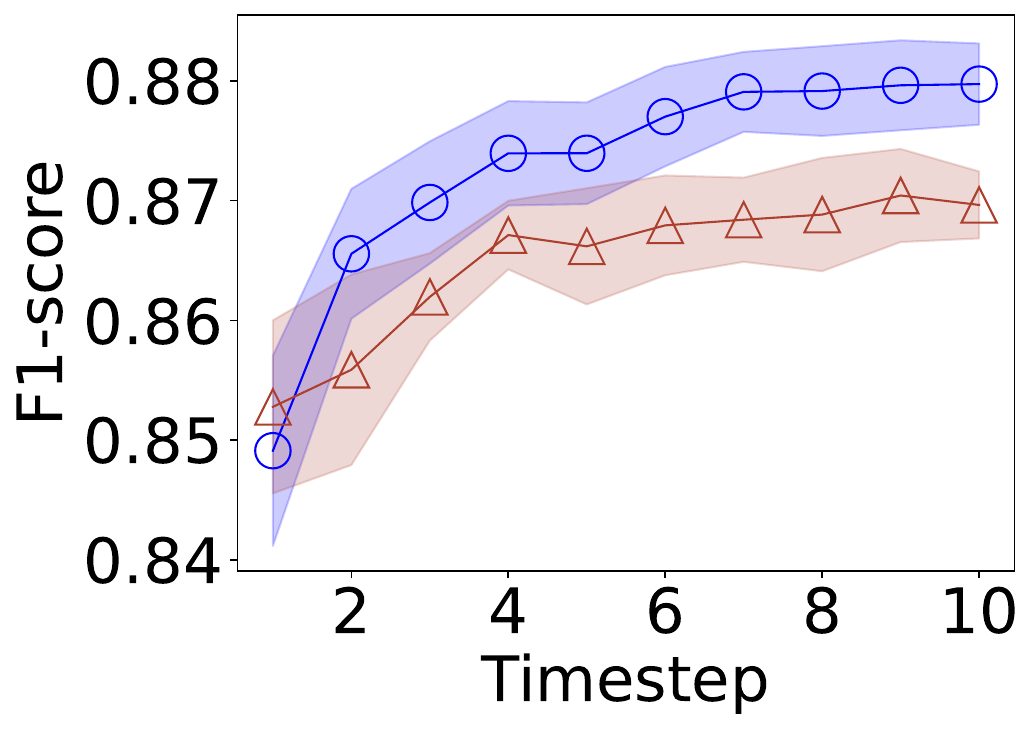}
   \subcaption{RoBERTa, Query size 500}
    \end{subfigure}
\begin{subfigure}[t]{0.3\columnwidth}
\includegraphics[width=\columnwidth]{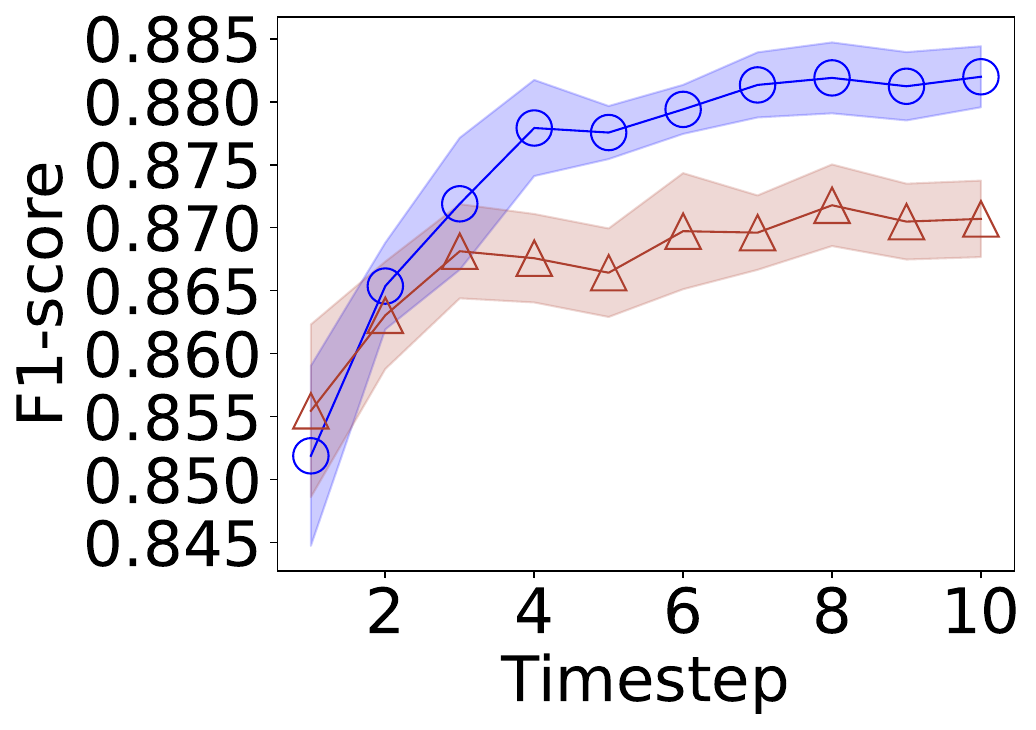}
   \subcaption{RoBERTa, Query size 700}
    \end{subfigure}
\begin{subfigure}[t]{0.3\columnwidth}
\includegraphics[width=\columnwidth]{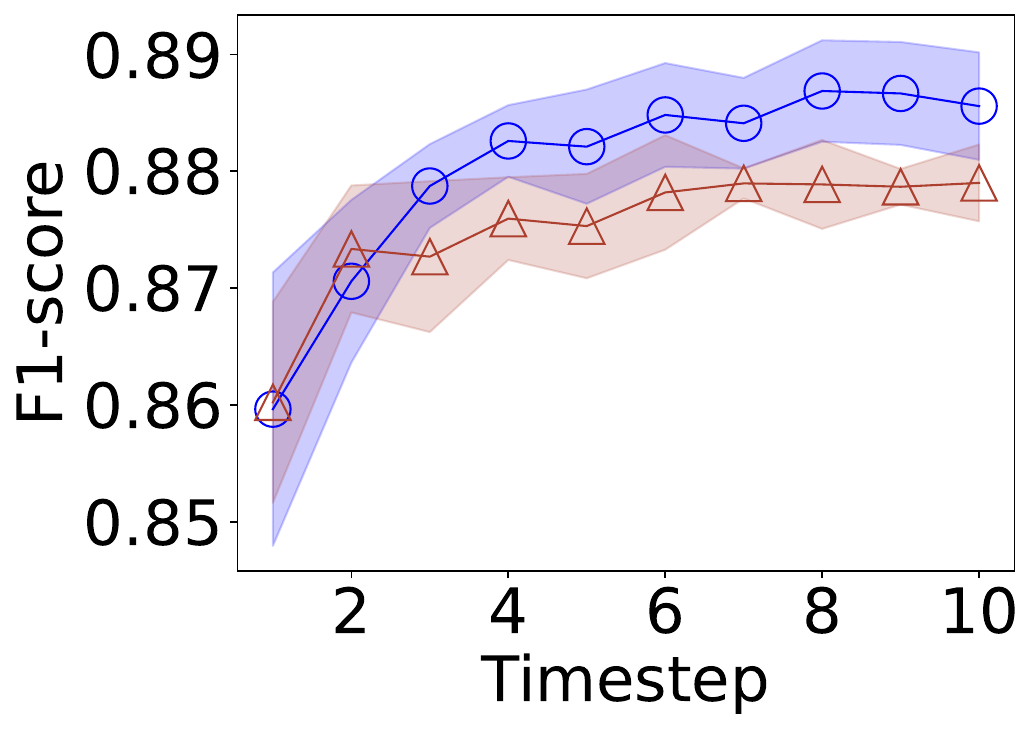}
    \subcaption{RTA, Query size 300}
    \end{subfigure}
\begin{subfigure}[t]{0.3\columnwidth}
\includegraphics[width=\columnwidth]{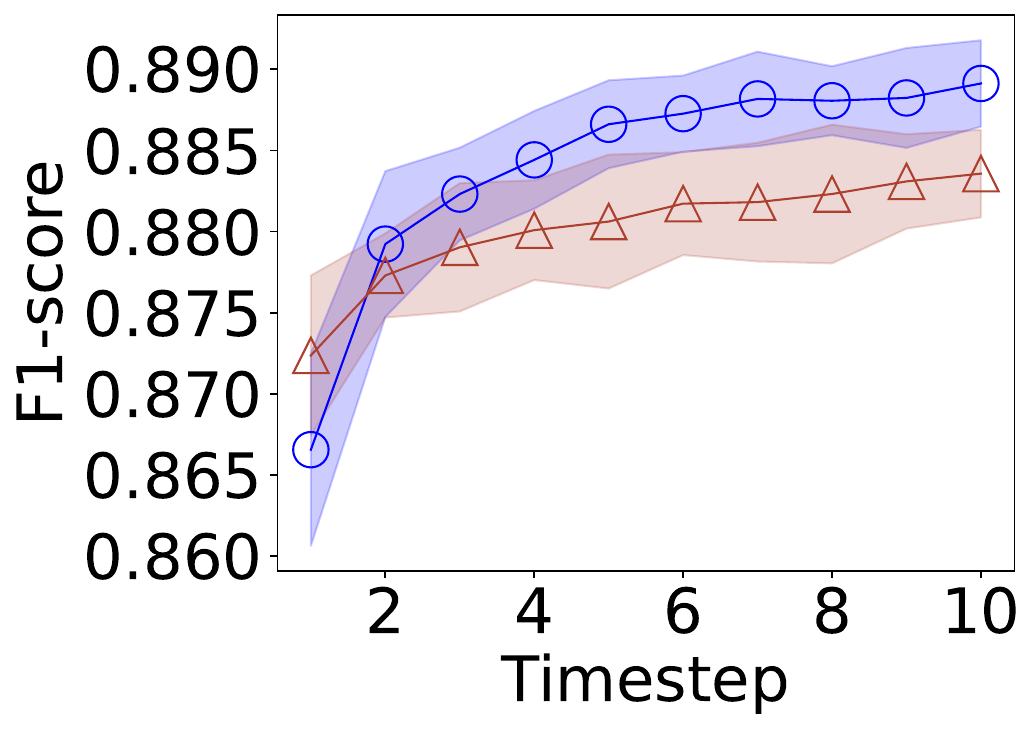}
   \subcaption{RTA, Query size 500}
    \end{subfigure}
\begin{subfigure}[t]{0.3\columnwidth}
\includegraphics[width=\columnwidth]{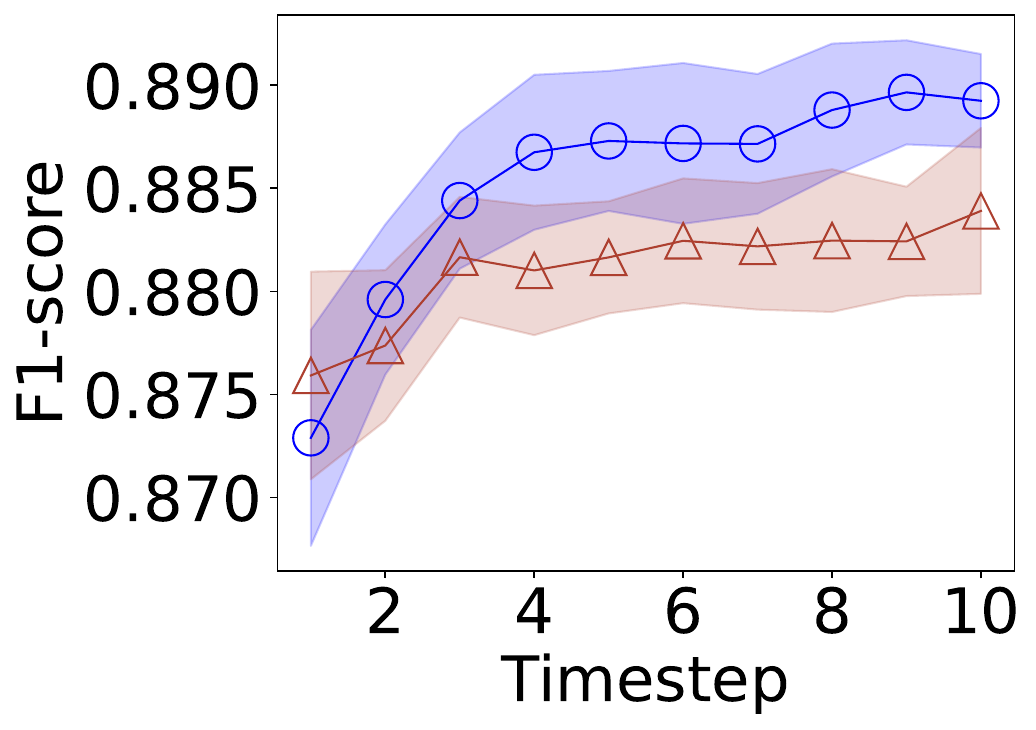}
   \subcaption{RTA, Query size 700}
    \end{subfigure}

    \caption{Comparing the F1-score of different neural language models with and without pairing \approach~over all 10 timesteps (10 runs each). The plots show the mean and standard deviation.}
\label{fig:rq3-1}
\end{figure}

Our selection process was guided by the need to evaluate \approach~ under a diverse set of conditions: a general-purpose code model (CodeBERT), a general-purpose text model (RoBERTa), and a task-specific model (RTA). This diversity helps to demonstrate the model independence of \approach~ by showing its effectiveness across different types of pre-trained language models. Additionally, these models were selected because they represent state-of-the-art techniques in their respective fields. By choosing widely recognized and effective models, we ensure that our findings are robust and generalizable. 


All the above neural language models can be pre-trained and fine-tuned, hence they can be triggered in a warm start manner under \approach. As such, we pertain those models using the same datasets as published by their original authors. To assess the benefits of \approach, we pair each of the above models with \approach~and \approach$_{ran}$, which uses random sampling to select reports to be labeled for updating the model without the notion of active learning. Again, since there are mainly pairwise comparisons, we use the Wilcoxon Sign-rank test for verifying statistical significance.

\begin{figure}[!t]
\centering

\begin{subfigure}{0.2\columnwidth}
  \centering
  \includegraphics[width=\linewidth]{fig/blank_stripe.pdf} 
\end{subfigure}
\begin{subfigure}{0.33\columnwidth}
  \centering
  \includegraphics[width=0.7\linewidth]{fig/rq3_legend.pdf} 
\end{subfigure}
\begin{subfigure}{0.2\columnwidth}
  \centering
  \includegraphics[width=\linewidth]{fig/blank_stripe.pdf} 
\end{subfigure}
\begin{subfigure}[t]{0.3\columnwidth}
\includegraphics[width=\columnwidth]{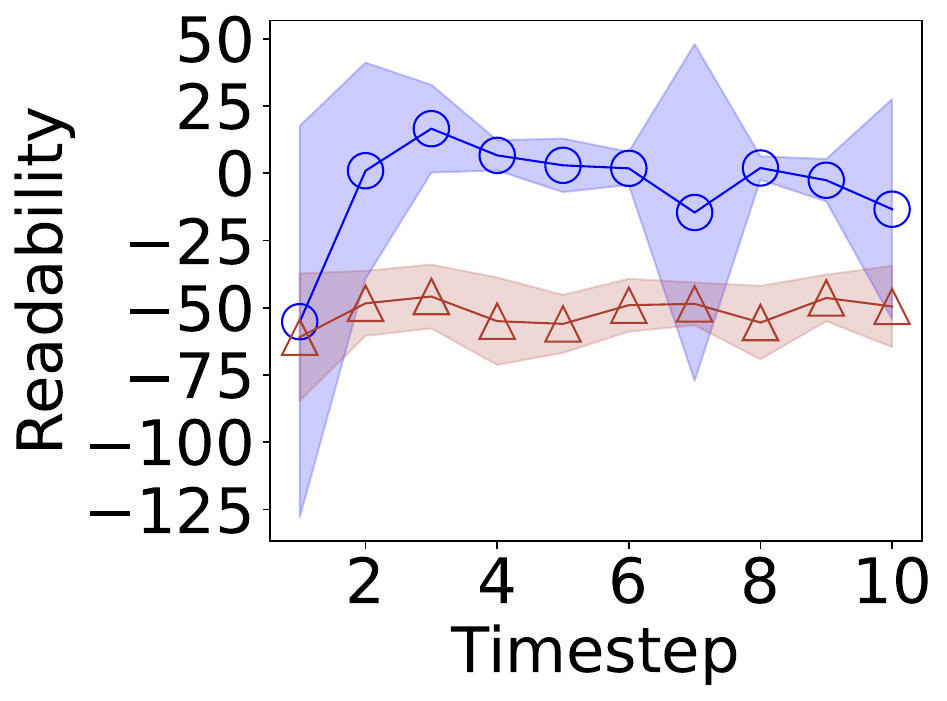}
    \subcaption{CodeBERT, Query size 300}
    \end{subfigure}
\begin{subfigure}[t]{0.3\columnwidth}
\includegraphics[width=\columnwidth]{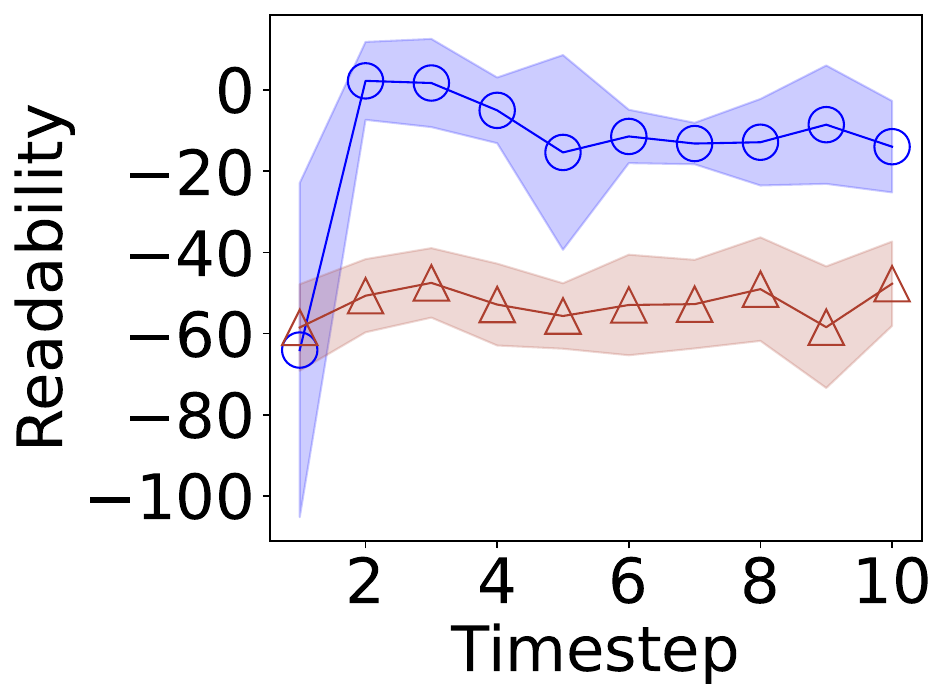}
   \subcaption{CodeBERT, Query size 500}
    \end{subfigure}
\begin{subfigure}[t]{0.3\columnwidth}
\includegraphics[width=\columnwidth]{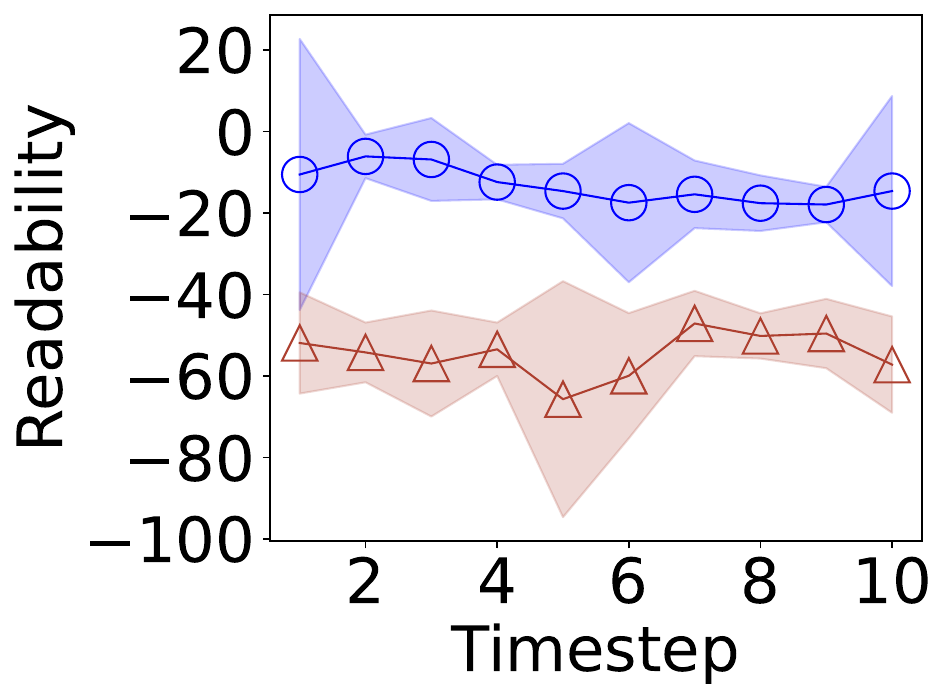}
   \subcaption{CodeBERT, Query size 700}
    \end{subfigure}
\begin{subfigure}[t]{0.3\columnwidth}
\includegraphics[width=\columnwidth]{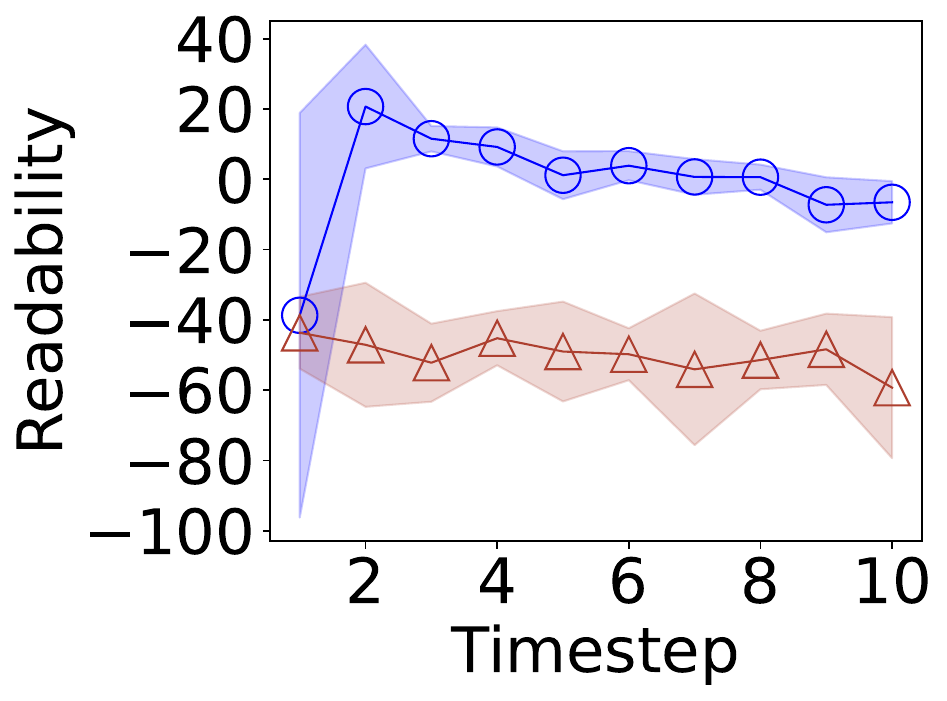}
    \subcaption{RoBERTa, Query size 300}
    \end{subfigure}
\begin{subfigure}[t]{0.3\columnwidth}
\includegraphics[width=\columnwidth]{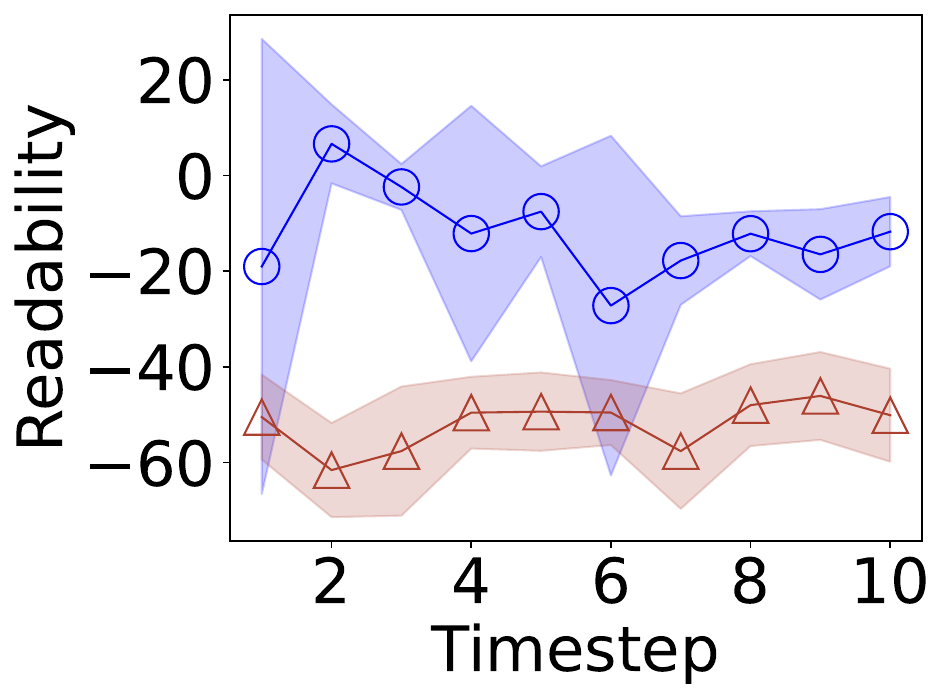}
   \subcaption{RoBERTa, Query size 500}
    \end{subfigure}
\begin{subfigure}[t]{0.3\columnwidth}
\includegraphics[width=\columnwidth]{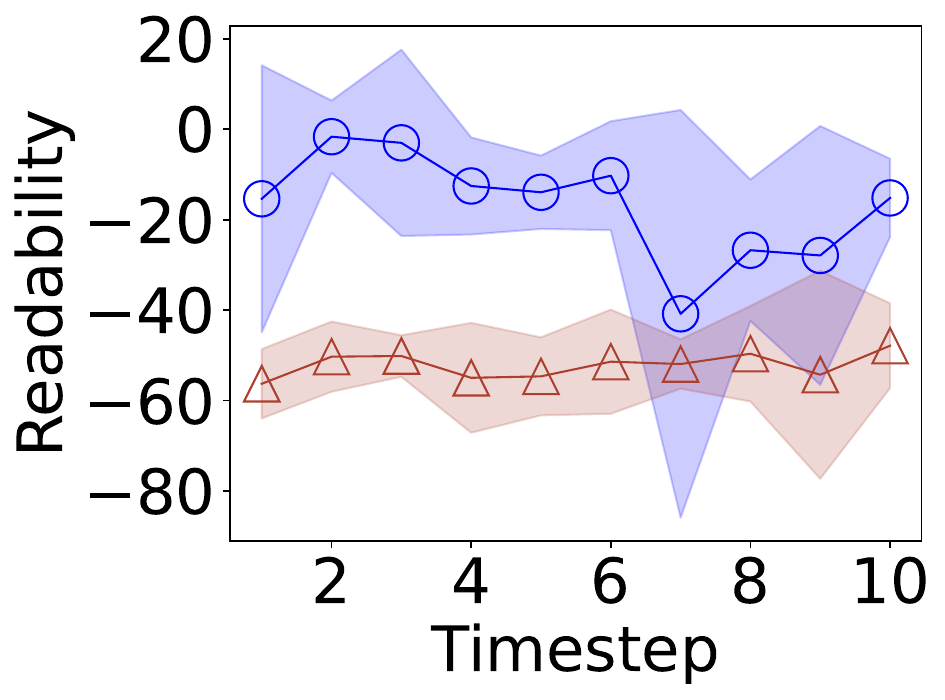}
   \subcaption{RoBERTa, Query size 700}
    \end{subfigure}
\begin{subfigure}[t]{0.3\columnwidth}
\includegraphics[width=\columnwidth]{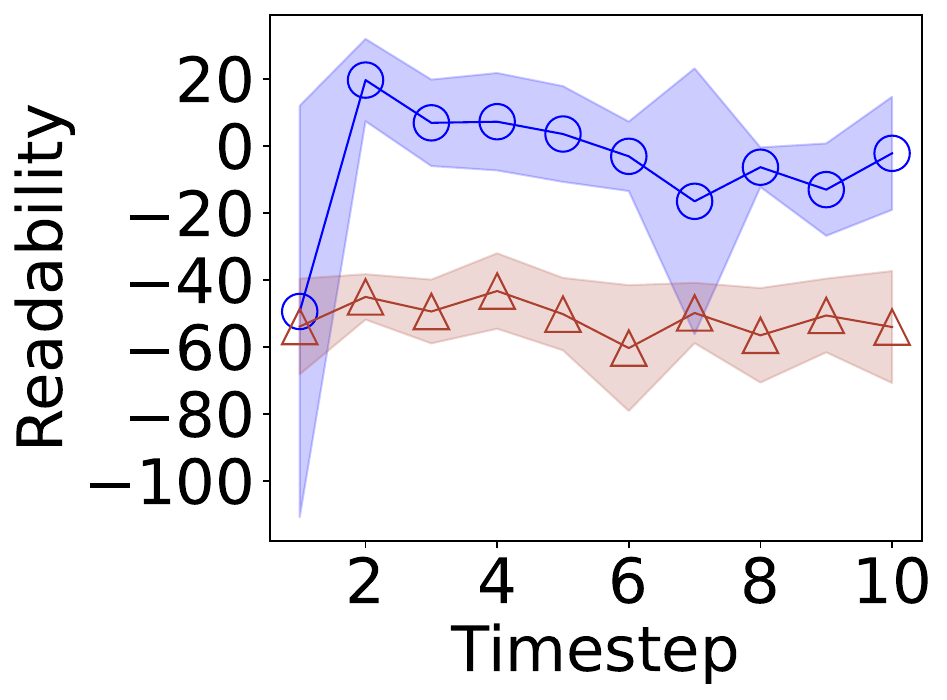}
    \subcaption{RTA, Query size 300}
    \end{subfigure}
\begin{subfigure}[t]{0.3\columnwidth}
\includegraphics[width=\columnwidth]{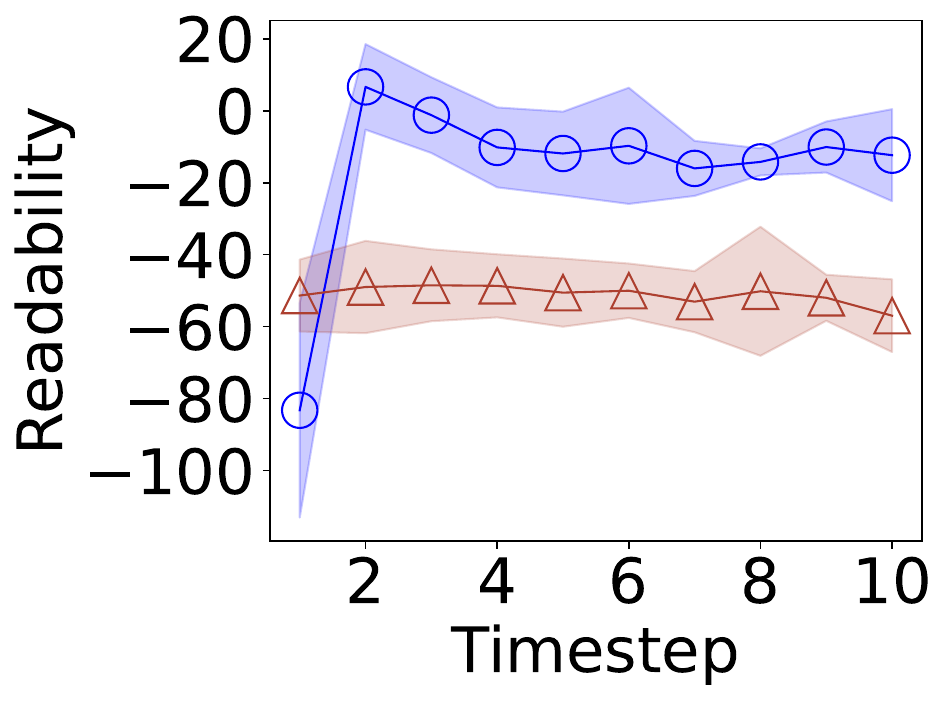}
   \subcaption{RTA, Query size 500}
    \end{subfigure}
\begin{subfigure}[t]{0.3\columnwidth}
\includegraphics[width=\columnwidth]{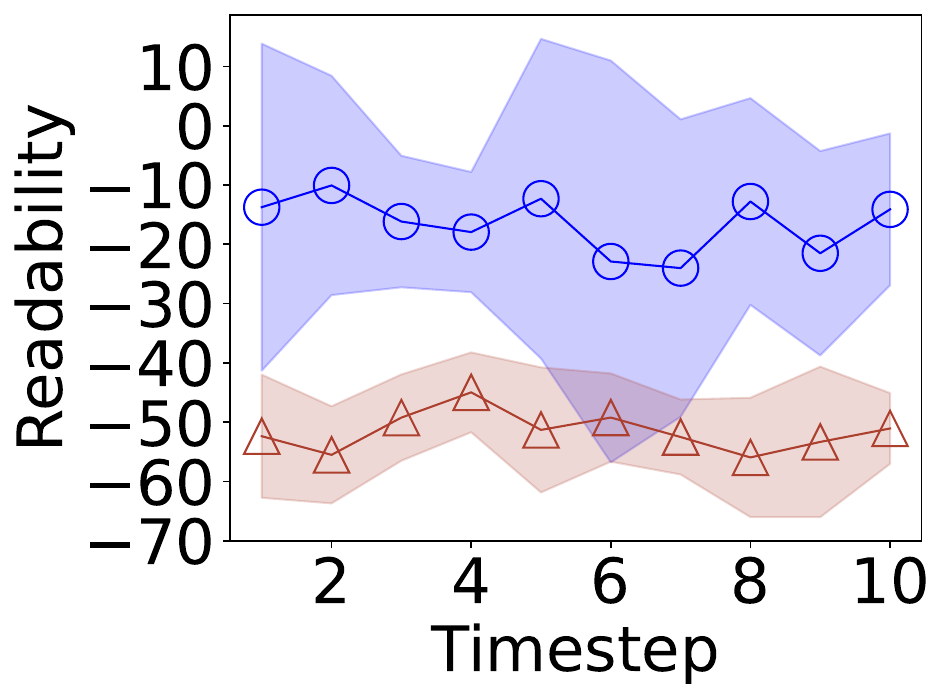}
   \subcaption{RTA, Query size 700}
    \end{subfigure}
      \caption{Comparing the readability of different neural language models with and without pairing \approach~over all 10 timesteps (10 runs each). The plots show the mean and standard deviation.}
 \label{fig:rq3-2}
\end{figure}

\begin{figure}[!t]
\centering

\begin{subfigure}{0.2\columnwidth}
  \centering
  \includegraphics[width=\linewidth]{fig/blank_stripe.pdf} 
\end{subfigure}
\begin{subfigure}{0.33\columnwidth}
  \centering
  \includegraphics[width=0.7\linewidth]{fig/rq3_legend.pdf} 
\end{subfigure}
\begin{subfigure}{0.2\columnwidth}
  \centering
  \includegraphics[width=\linewidth]{fig/blank_stripe.pdf} 
\end{subfigure}
\begin{subfigure}[t]{0.3\columnwidth}
\includegraphics[width=\columnwidth]{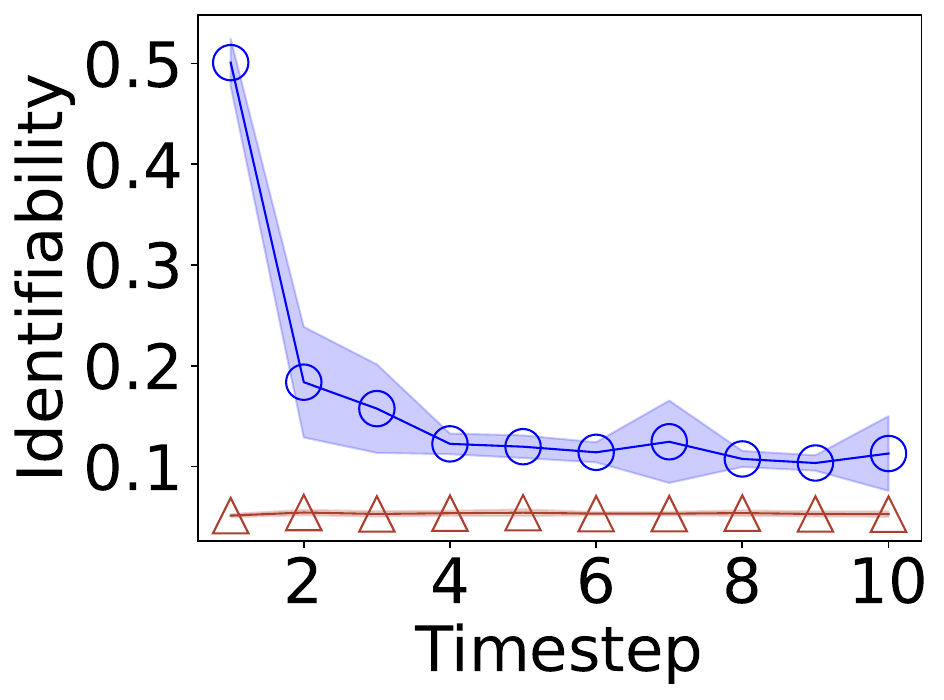}
    \subcaption{CodeBERT, Query size 300}
    \end{subfigure}
\begin{subfigure}[t]{0.3\columnwidth}
\includegraphics[width=\columnwidth]{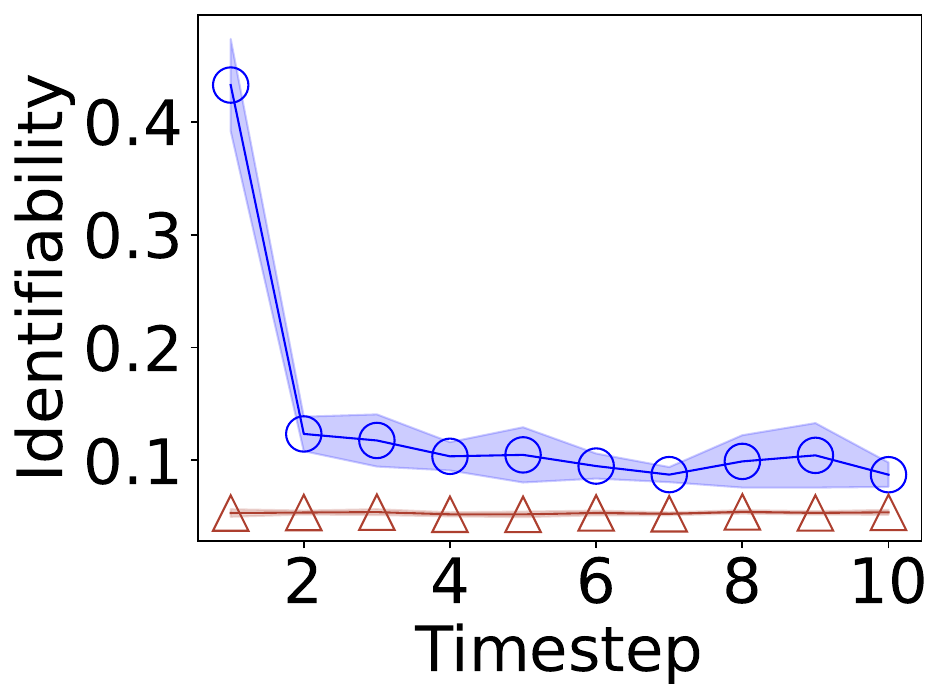}
   \subcaption{CodeBERT, Query size 500}
    \end{subfigure}
\begin{subfigure}[t]{0.3\columnwidth}
\includegraphics[width=\columnwidth]{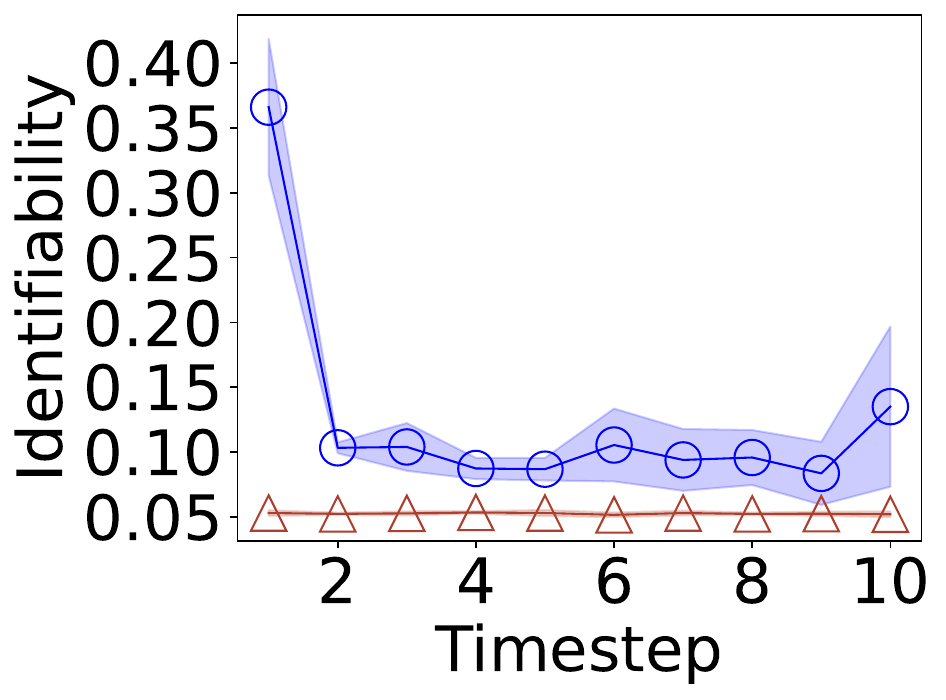}
   \subcaption{CodeBERT, Query size 700}
    \end{subfigure}
\begin{subfigure}[t]{0.3\columnwidth}
\includegraphics[width=\columnwidth]{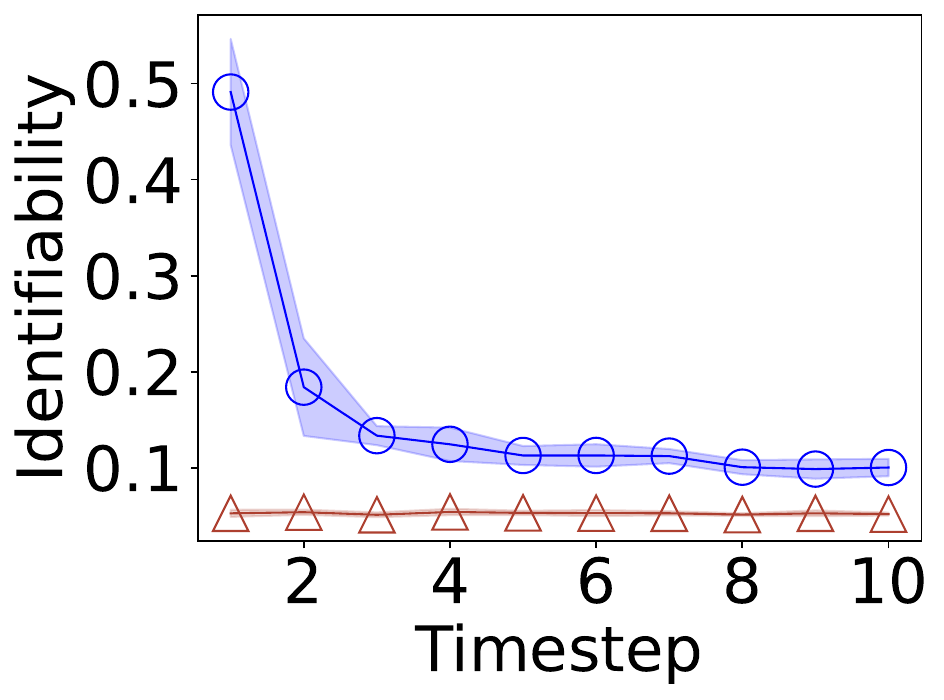}
    \subcaption{RoBERTa, Query size 300}
    \end{subfigure}
\begin{subfigure}[t]{0.3\columnwidth}
\includegraphics[width=\columnwidth]{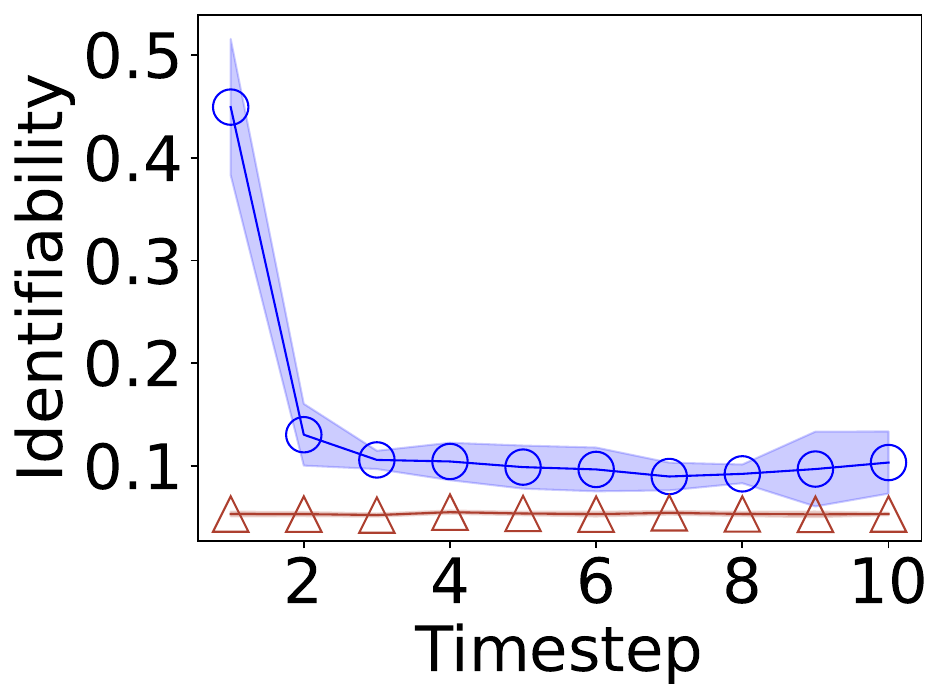}
   \subcaption{RoBERTa, Query size 500}
    \end{subfigure}
\begin{subfigure}[t]{0.3\columnwidth}
\includegraphics[width=\columnwidth]{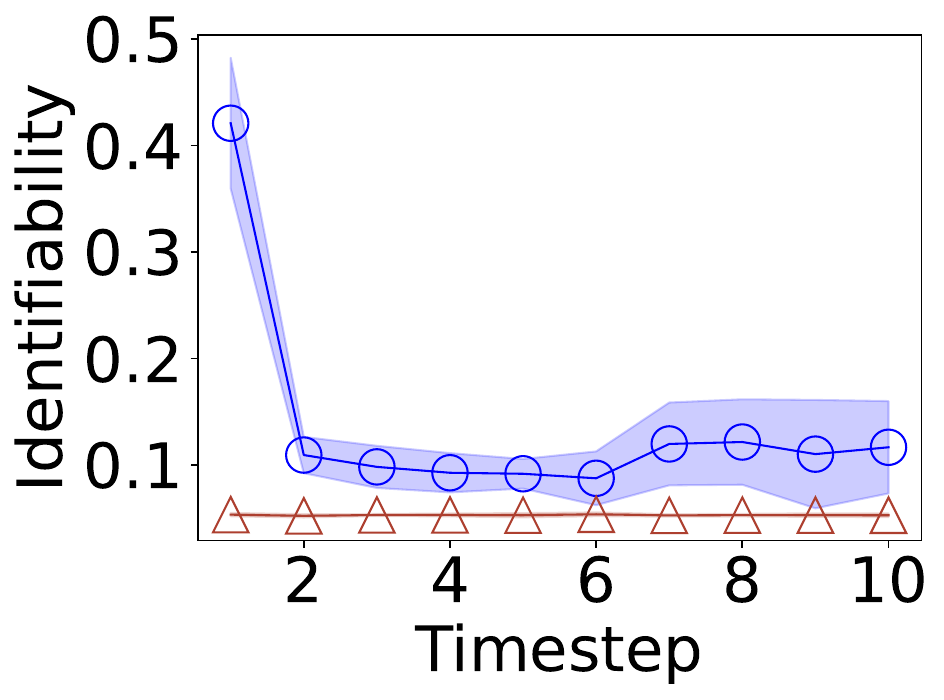}
   \subcaption{RoBERTa, Query size 700}
    \end{subfigure}
\begin{subfigure}[t]{0.3\columnwidth}
\includegraphics[width=\columnwidth]{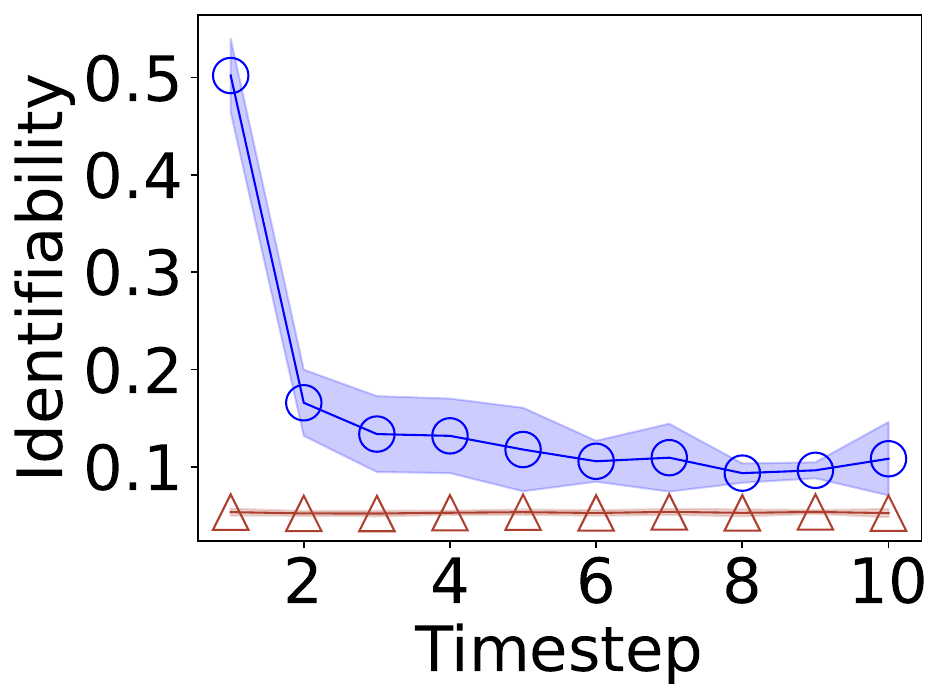}
    \subcaption{RTA, Query size 300}
    \end{subfigure}
\begin{subfigure}[t]{0.3\columnwidth}
\includegraphics[width=\columnwidth]{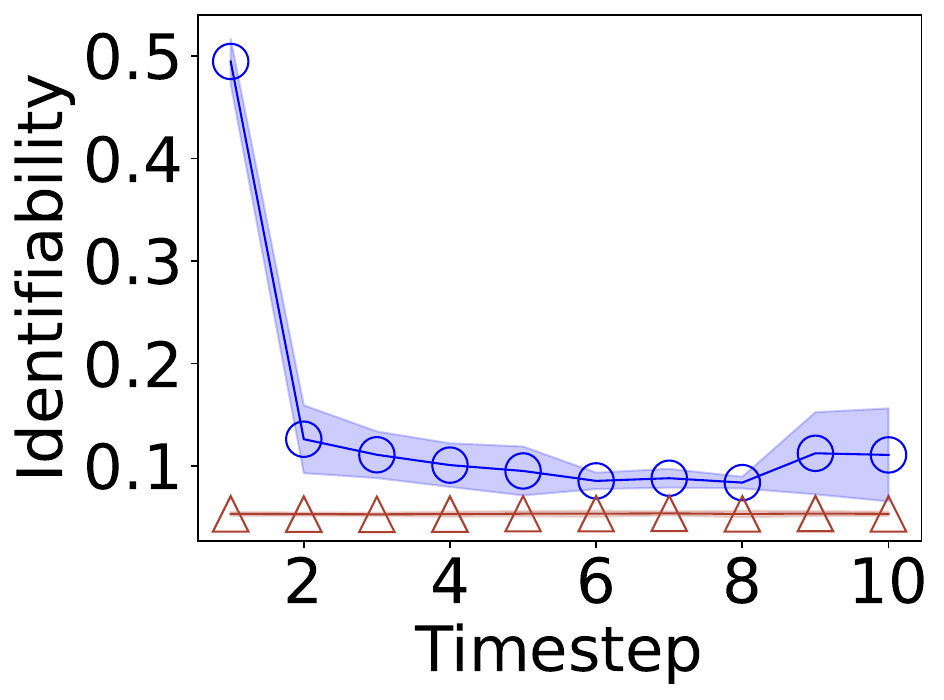}
   \subcaption{RTA, Query size 500}
    \end{subfigure}
\begin{subfigure}[t]{0.3\columnwidth}
\includegraphics[width=\columnwidth]{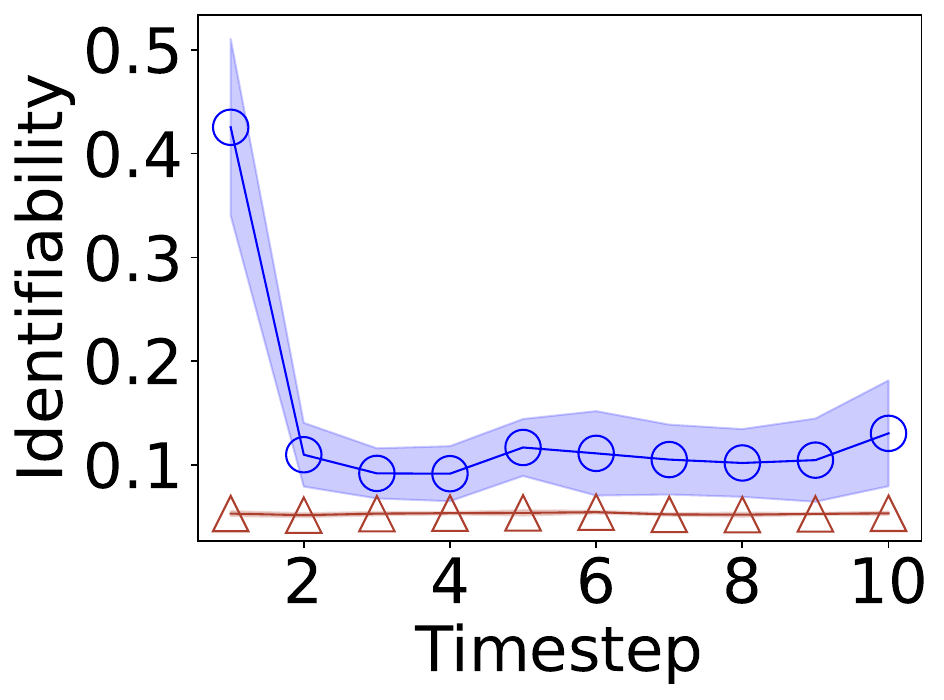}
   \subcaption{RTA, Query size 700}
    \end{subfigure}

      \caption{Comparing the Identifiability of different neural language models with and without pairing \approach~over all 10 timesteps (10 runs each). The plots show the mean and standard deviation.}
\label{fig:rq3-3}
\end{figure}




\subsubsection{Findings}


Figures~\ref{fig:rq3-1},~\ref{fig:rq3-2}, and~\ref{fig:rq3-3}, together with Table~\ref{tab:rq3-1}, demonstrate the traces for 10 timesteps and the overall results with the statistical test. Our results confirm that \approach is model-agnostic. When paired with various neural language models (CodeBERT, RoBERTa, RTA), \approach consistently and significantly outperforms the baseline where the same models are updated with randomly sampled data (i.e., $p < 0.001$). For example, when applied to RoBERTa (query size 700), \approach improves readability from {-52.139 to -16.749} and identifiability from {0.053 to 0.137}. The superiority is even more obvious in readability and identifiability, where the average improvements are {78.6\% and 171.5\%} respectively across all models. This proves that the benefits of \approach are not tied to a specific model architecture but stem from the framework itself. By intelligently selecting which data to label (effort-aware sampling) and augmenting it with pseudo-labeled reports, \approach provides a superior training signal to any underlying neural language model. This allows organizations to apply \approach to their preferred or existing models, enhancing both performance and human efficiency without being locked into a single technology.

We still observe that there are fluctuations in the readability and identifiability due to the same reason mentioned for \textbf{RQ1}. The performance in terms of the F1-score, on the other hand, is improved in a steady state.


In conclusion, we say that:

\begin{quotebox}
   \noindent
   \textit{\textbf{\underline{To RQ3:}} \approach~is indeed model-agnostic: no matter which underlying neural language model is employed, the \approach~framework can still considerably improve their performance with significantly reduced efforts required for labeling.}
\end{quotebox}

\input{tab/rq3_new}

\input{tab/sota}
\input{tab/sota_cross}
\subsection{Comparing \approach~with State-of-the-art Approaches}
\label{sec:rq4}

\subsubsection{Operationalization}

To confirm the overall benefit of \approach, we compare it against state-of-the-art approaches that rely on active learning for bug report identification. Table~\ref{tab:sota} illustrates the compared approaches published since 2015. As can be seen, those approaches leverage mainly TF-IDF to quantify the reports, which are paired with different machine learning models. In particular, all of them mainly rely on uncertainty for the sampling with active learning. \texttt{EMBLEM} is a unique example in which the certainty metric is used to further extract the reports from the set of highly uncertain samples. However, none of them have exploited pseudo-labeling to enrich the amount of labeled reports. 

In addition, we also conducted a comparison between \approach~and the state-of-the-art cross-project methods. Table~\ref{tab:sota_cross} presents the state-of-the-art methods we used for comparison. Since those methods follow a typical supervised learning paradigm, we used 3,000, 5,000 and 7,000 training data sizes to train the state-of-the-art cross-project method, while \approach was trained using the 10 time step with 300, 500 and 700 query sizes.

Furthermore, we also compare \approach~against a state-of-the-art Large Language Model (LLM) - GPT-4o-mini\footnote{https://openai.com/index/gpt-4o-mini-advancing-cost-efficient-intelligence/}. Our implementation leverages the GPT-4o-mini model to automatically classify bug reports using few-shot learning, varying the number of provided examples (n-shot) from 0 to 5 to analyze the performance improvement with additional labeled examples, where 1 shot here equivalent to a pair of bug report and non-bug report. The implementation includes several steps. We structured the prompts to be concise yet informative, asking the model directly if a given text was a bug report, to which it should respond with a binary answer (0 for no, 1 for yes). Specifically, we designed the basic prompt for zero-shot learning, where the model was asked directly if a given text was a bug report. The prompt was formulated as:  ``\textit{``Is this a bug report? Respond with only 1 for yes or 0 for no: $T$''} followed by the descriptions in the bug report $T$ to be predicted. For few-shot in context learning, we include labeled examples in the prompt to provide context and improve the model's understanding. A few-shot prompt with a pair of positive and negative examples was structured as: ``Here are some examples of bug reports and non-bug reports: \textit{``Here are some examples of bug reports and non-bug reports: $E_1$, $E_1$, ..., $E_n$. Now, is this a bug report? Respond with only 1 for yes or 0 for no: T"}, where $E_1$, $E_1$, ..., $E_n$ are description of different non-bug relevant reports and $T$ is the bug report to be predicted. To ensure the validity of our few-shot prompting testing, for each test example, we randomly select an example text from the labeled training data to use as a prompt. Finally, we prepared the same test data used for \approach~to evaluate the performance of GPT-4o-mini. The observation that 1-shot learning outperforms 5-shot learning is indeed counterintuitive, but a reasonable explanation can be: the examples used for the few-shot prompts were selected randomly from the labeled training data for each report being tested. The superior performance of the 1-shot setting suggests that a single, randomly selected example was often sufficient and, on average, provided a clearer, more direct signal to the model for this binary classification task.

Conversely, when providing five randomly selected examples, the probability of introducing noise, ambiguity, or even contradictory information into the prompt increases significantly. The bug reports within the dataset are noted to be diverse in format and complexity. A random set of five reports could easily include examples that are atypical, poorly written, or present conflicting patterns. This noisy and more complex context may have confused the model, hindering its ability to identify the core task, rather than refining its understanding. This phenomenon, where performance does not monotonically increase with the number of examples, can occur in few-shot learning when the quality and direct relevance of the examples are more critical than the sheer quantity. Essentially, the focused context of a single good example (selected by chance) proved more effective than the potentially confusing context of five random examples.

To reduce unnecessary noise when running the state-of-the-art approaches, we use the publicly available code published by their authors. Since there are more than two comparisons, we use the Scott-Knott test and Wilcoxon sign-rank test to verify the statistical significance.

\begin{figure}[!t]
\centering

\begin{subfigure}{0.01\columnwidth}
  \centering
  \includegraphics[width=\linewidth]{fig/blank_stripe.pdf} 
\end{subfigure}
\begin{subfigure}{0.9\columnwidth}
  \centering
  \includegraphics[width=0.9\linewidth]{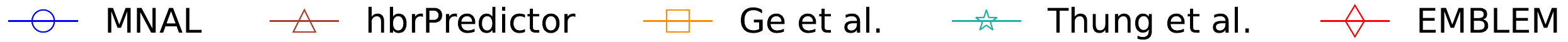} 
\end{subfigure}
\begin{subfigure}{0.01\columnwidth}
  \centering
  \includegraphics[width=\linewidth]{fig/blank_stripe.pdf} 
\end{subfigure}
\begin{subfigure}[t]{0.3\columnwidth}
\includegraphics[width=\columnwidth]{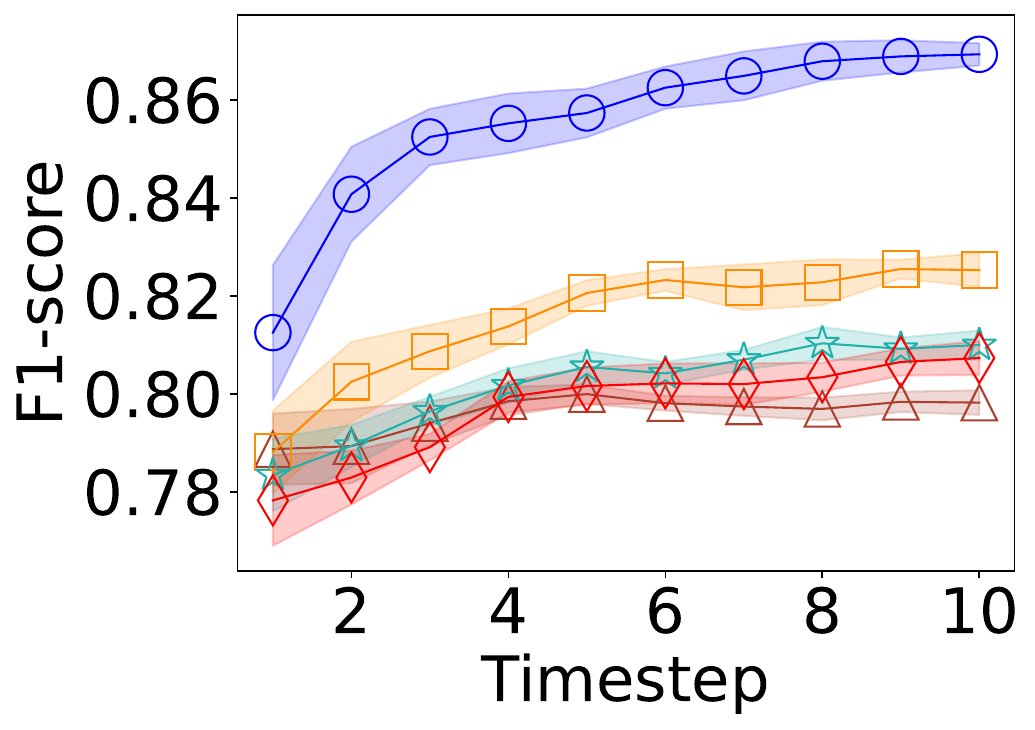}
    \subcaption{Query size 300}
    \end{subfigure}
\begin{subfigure}[t]{0.3\columnwidth}
\includegraphics[width=\columnwidth]{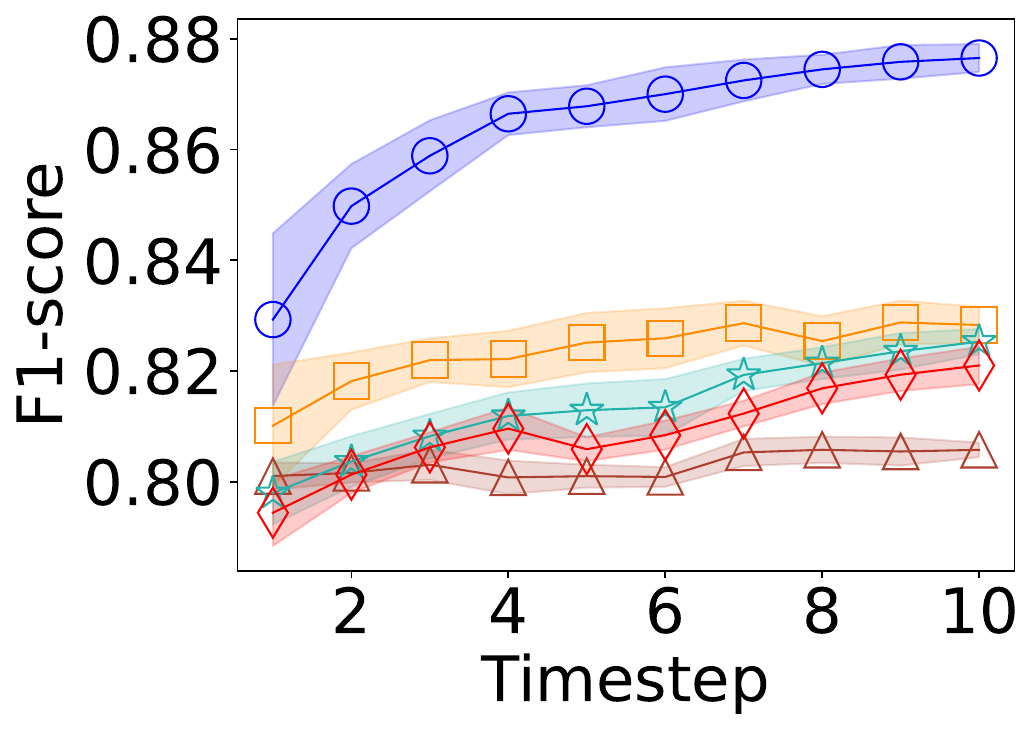}
   \subcaption{Query size 500}
    \end{subfigure}
\begin{subfigure}[t]{0.3\columnwidth}
\includegraphics[width=\columnwidth]{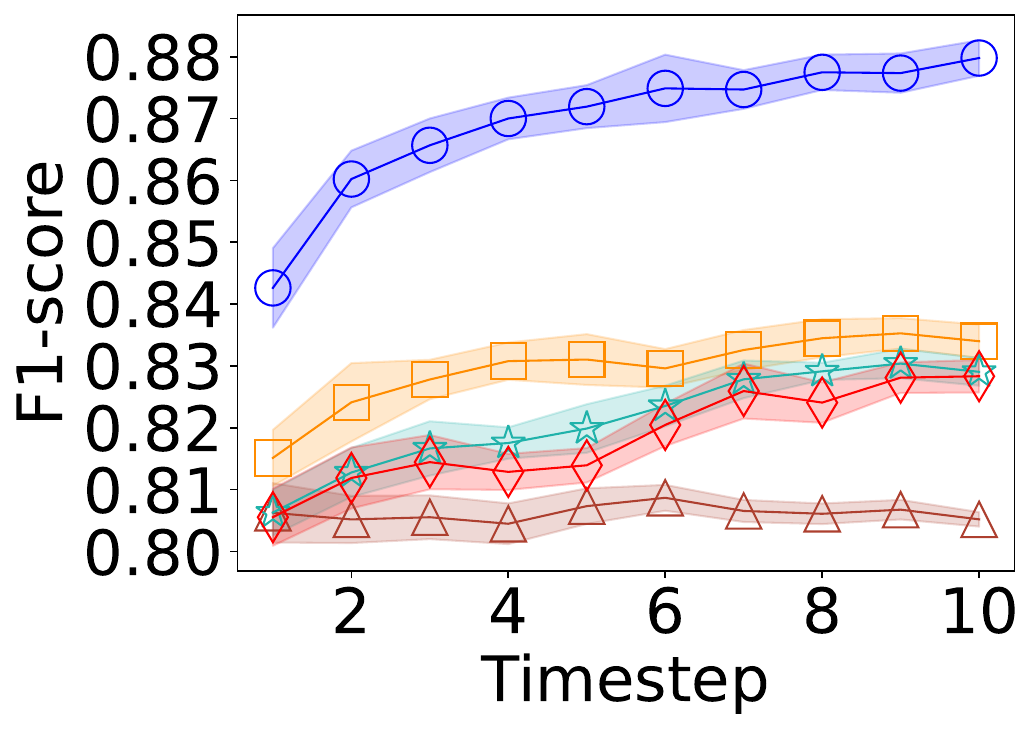}
   \subcaption{Query size 700}
    \end{subfigure}
\begin{subfigure}[t]{0.3\columnwidth}
\includegraphics[width=\columnwidth]{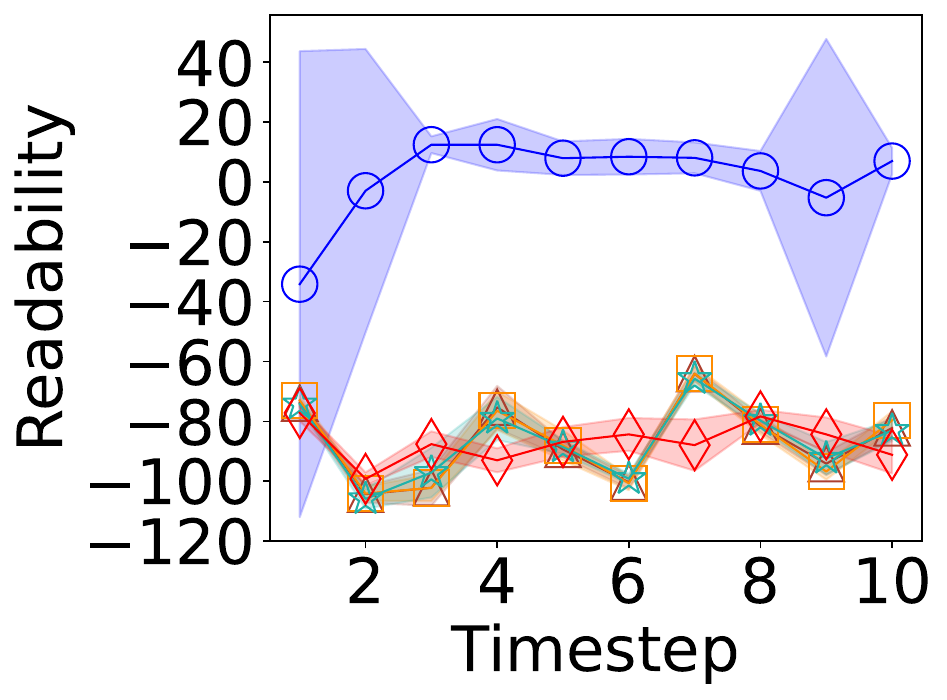}
    \subcaption{Query size 300}
    \end{subfigure}
\begin{subfigure}[t]{0.3\columnwidth}
\includegraphics[width=\columnwidth]{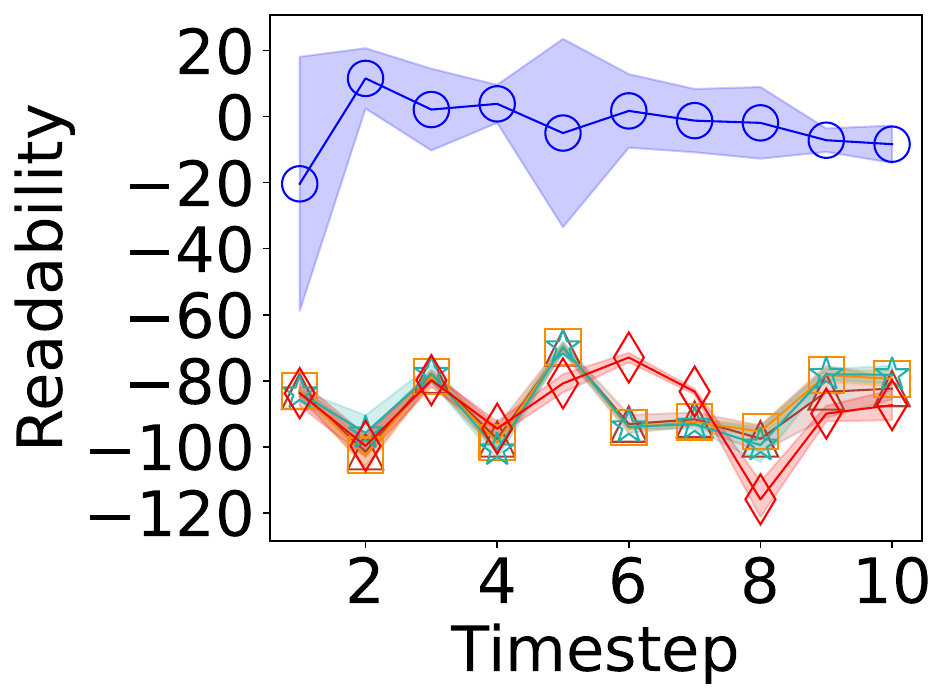}
   \subcaption{Query size 500}
    \end{subfigure}
\begin{subfigure}[t]{0.3\columnwidth}
\includegraphics[width=\columnwidth]{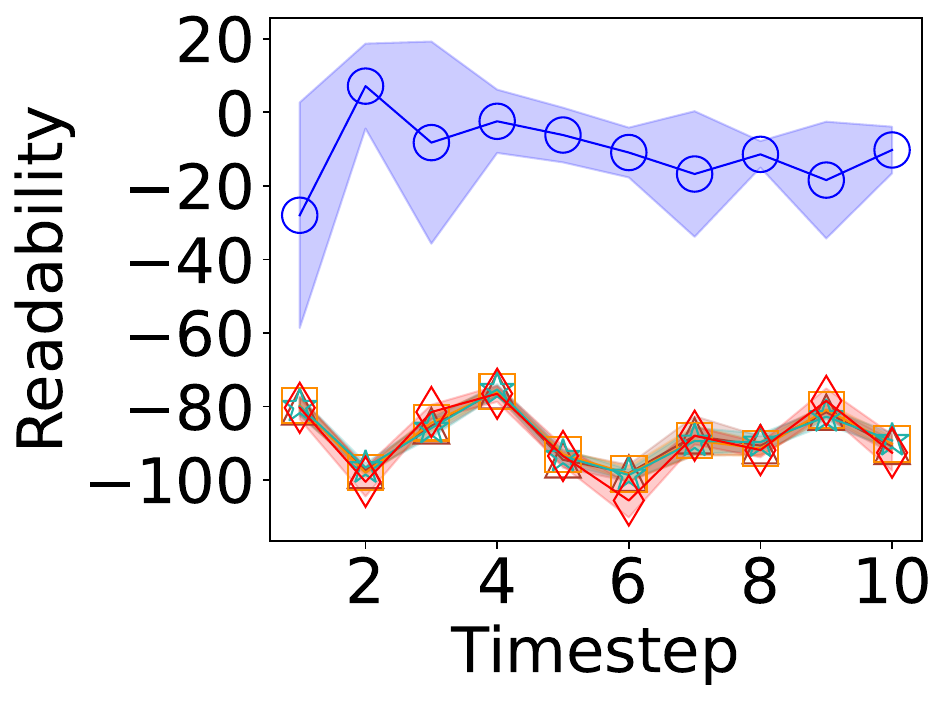}
   \subcaption{Query size 700}
    \end{subfigure}
\begin{subfigure}[t]{0.3\columnwidth}
\includegraphics[width=\columnwidth]{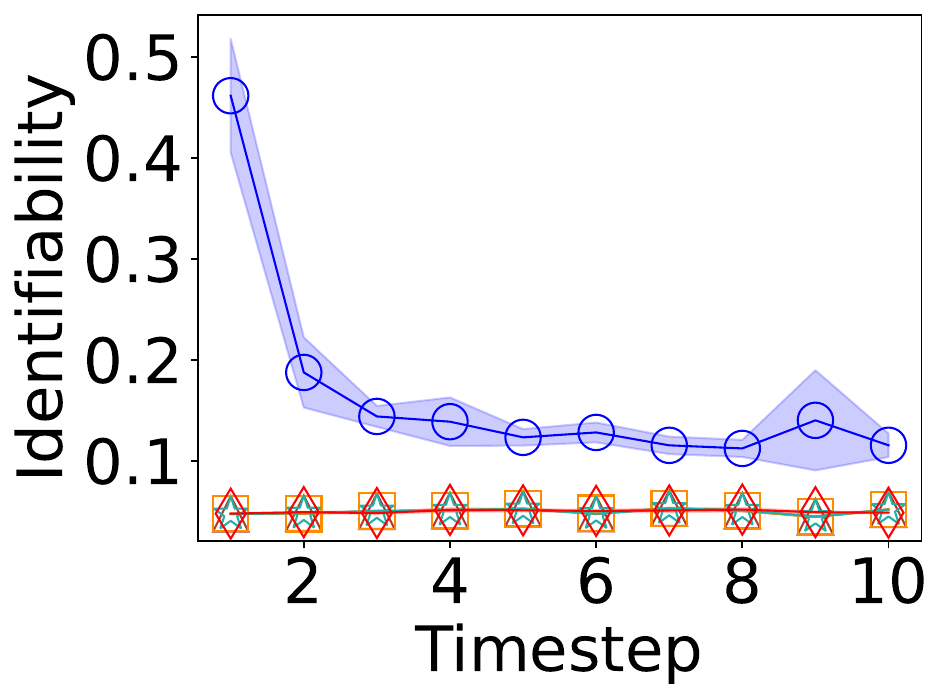}
    \subcaption{Query size 300}
    \end{subfigure}
\begin{subfigure}[t]{0.3\columnwidth}
\includegraphics[width=\columnwidth]{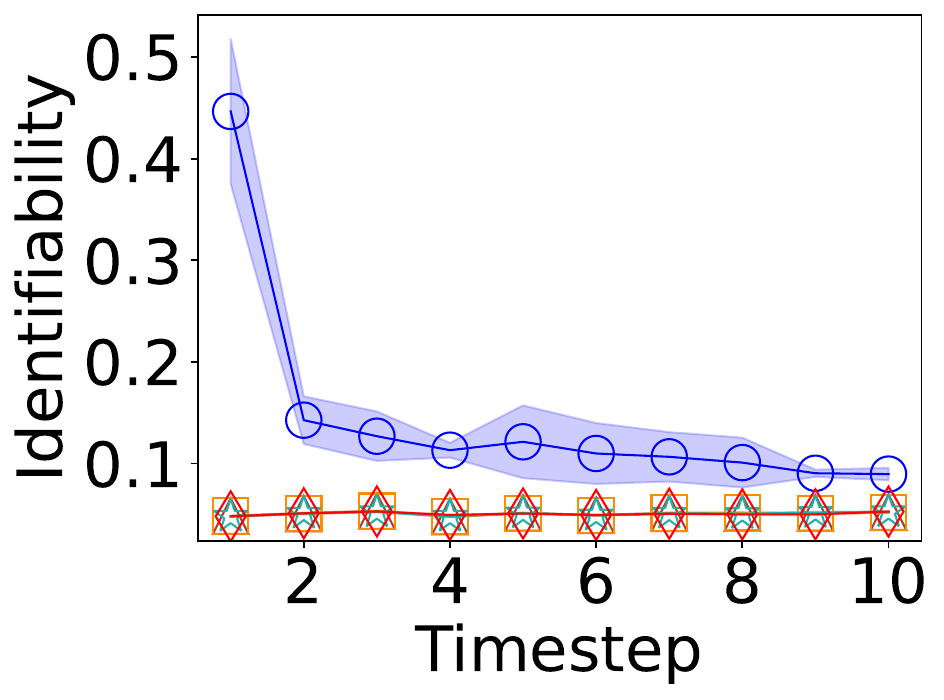}
   \subcaption{Query size 500}
    \end{subfigure}
\begin{subfigure}[t]{0.3\columnwidth}
\includegraphics[width=\columnwidth]{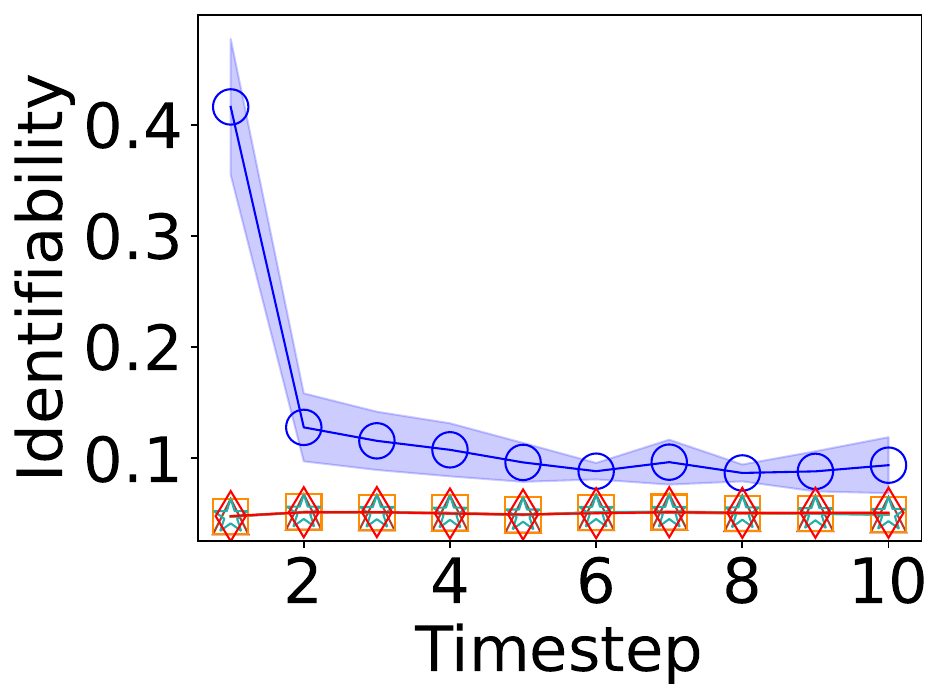}
   \subcaption{Query size 700}
    \end{subfigure}

    \caption{Comparing \approach~with state-of-the-art active learning approaches for bug report identification over all 10 timesteps (10 runs each). The plots show the mean and standard deviation. }
 \label{fig:rq4-1}
\end{figure}

\subsubsection{Findings}

As illustrated in Figure~\ref{fig:rq4-1}, remarkably, \approach~significantly outperforms the state-of-the-art active learning approaches in all cases starting from the first timestep. At a query size of 700, \approach's F1-score is {0.869}, while the next best approach (Ge et al.) scores only {0.829}, as shown in Table 7. This advantage stems from its holistic design; unlike prior active learning methods that rely on simpler models (e.g., TF-IDF) and focus only on uncertainty, \approach~combines the semantic power of neural networks with a dual focus on model accuracy and human efficiency. The combined effect of pseudo-labeling and the neural language model within active learning leads to superior F1-scores. Apart from the performance in identifying bug reports, the effort-aware uncertain sampling results in drastically improved readability and identifiability for the queried reports, with up to {95.8\% and 196.0\% improvements}, respectively. The statistical test from Table~\ref{tab:rq4-1} further confirms that \approach~achieves superior results on nearly all performance metrics. Although compared state-of-the-art approaches also rely on active learning, they do not consider identifiability, reliability, or the pseudo-labeling process. Consequently, they incur much larger human efforts, as seen from the identifiability and reliability values in Table~\ref{tab:rq4-1}, without achieving better accuracy than \approach.

Similarly, as shown in Table~\ref{tab:rq4-cross}, we also compare \approach~against the state-of-the-art cross-project approaches for bug report identification. Remarkably, \approach~demonstrates significant improvements over other cross-project methods, not only in terms of accuracy and F1-score but also readability and identifiability. Unlike traditional approaches, which mainly rely on a passive learning approach, the integration of neural active learning and effort-aware uncertainty sampling in \approach~has led to a considerable reduction in human labeling efforts. This is further supported by the superior performance metrics across all evaluation criteria compared to the SOTA cross-project approach~\cite{DBLP:journals/infsof/MeherBM24} using the advanced neural language model BERT, while maintaining a competitive recall. Overall, \approach~effectively improves labeling efficiency and reduces labeling effort, providing a more reliable solution for cross-project bug report identification.

\input{tab/rq4_avg}

\input{tab/rq4_cross}

\input{tab/llms}

In comparison with LLMs, \approach’s performance is also impressive. As shown in Table~\ref{tab:llms}, \approach’s F1-score is 16.2\% higher than GPT-4o-mini’s zero-shot baseline and 8.7\% higher than the LLM’s best one-shot performance. This demonstrates that a smaller, specialized model fine-tuned with the intelligent, human-in-the-loop \approach framework can achieve better performance on a specific task than a massive, general-purpose model. Furthermore, we compared \approach with HINT, another state-of-the-art approach that utilizes pseudo-labeling. As shown in Table 9, HINT outperforms \approach on the Precision and AUC metrics. However, \approach demonstrates superior performance in F1-score, Accuracy, and Recall. This is significant because, fundamentally, HINT does not incorporate the human-machine collaborative aspect that is central to our work. \approach's ability to actively reduce annotator fatigue by selecting for readable and identifiable reports represents a key practical advantage not captured by these metrics alone.



Therefore, for \textbf{RQ4}, we conclude that:
\begin{quotebox}
   \noindent
   \textit{\textbf{\underline{To RQ4:}} \approach significantly outperforms other state-of-the-art approaches on different performance metrics, readability and identifiability. }
\end{quotebox}

\subsubsection{Analysis of \approach's Superiority over Active Learning Counterparts}

The experimental results above demonstrate that \approach consistently and significantly outperforms state-of-the-art active learning approaches for bug report identification across performance and effort metrics. The underlying reasons for this superiority can be attributed to several design choices that fundamentally advance beyond traditional active learning frameworks.

\begin{enumerate}
\item \approach is built on Neural Active Learning, which pairs a powerful neural language model with the active learning. In contrast, the compared state-of-the-art methods rely on statistical machine learning models like SVM, NBM, or Random Forest, combined with TF-IDF for text representation. TF-IDF can capture keyword frequency but fails to understand the deeper semantic and contextual nuances of language found in bug reports. \approach, by using models like BERT, leverages contextualized embeddings that provide a much richer, more meaningful representation of the reports. 

\item \approach introduces a more comprehensive, human-centric sampling strategy. The compared active learning methods primarily use uncertainty sampling to select reports for labeling. While this can improve model accuracy, it often queries complex and hard-to-understand reports, leading to cognitive fatigue for developers. \approach's effort-aware uncertainty sampling explicitly addresses this by creating a mutualistic relationship between the human and the machine. It balances the model's need for informative samples (uncertainty) with the developer's need for reports that are easier to process (readability and identifiability). This multi-objective approach not only reduces human labeling effort significantly  but also ensures the quality and timeliness of the labels provided, indirectly benefiting the model.

\item \approach enriches the training data at each step through pseudo-labeling, a feature absent in the other active learning approaches. After developers label a small set of queried reports, \approach leverages the neural model's rich embedding space to find the most similar unlabeled reports and assigns them a ``pseudo-label''. This process effectively multiplies the value of each human-provided label, expanding the training set with high-confidence samples without incurring additional human labeling cost. 
\end{enumerate}



\subsection{Human-centric Case Study}
\label{sec:rq5}

\subsubsection{Operationalization}

To fully verify the effort reduction of active learning achieved by the mutualistic relation introduced in \approach and demonstrate the practical values of \approach, we conduct a case study involving human developers who actually label the reports during the training/updating.


We recruit 10 software engineering research students who have the necessary understanding of software development and have identified Github reports before, but their years of experience vary between one and four years. 
{Given that bug identification is a relatively straightforward task based on objective evidence (e.g., crash logs, reproduction steps), existing literature suggests that such students are qualified proxies for industrial developers in this context.}

In particular, the participants are asked to label two sets of 300 reports queried by \approach~and \approach$_{ran}$ (from \textbf{RQ1}) in one timestep, respectively. To avoid perception bias, each participant labels 60 randomly selected reports of which half are from \approach, but the participants have no knowledge about the source of their assigned reports. Upon completing the labeling of a report, the participants are asked to rate how they feel about its readability and identifiability using a Likert scale ranging from 0 to 4, where 0 denotes the most readable/identifiable level. This enables us to qualitatively assess \approach~in terms of human cognition. In addition, the participants are paid £10.42/hour in compliance with the national minimum wage legislation in the UK, thus serving as a basis to evaluate the real cost savings when applying \approach.

{To ensure the quality of the study, we implemented two key procedures:}

\begin{itemize}
\item {\textbf{Shuffled order:} The 60 reports assigned to each participant were presented in a randomly shuffled order to prevent ordering effects and perception bias.}

\item {\textbf{Mandatory breaks:} To avoid fatigue-induced errors, participants were required to take a mandatory break every 20 reports or were allowed to complete the task across multiple sessions.}
\end{itemize}

{Furthermore, to enhance the reliability of our findings, we invited an expert an expert to conduct a post-hoc spot check by re-evaluating 10\% of the human-labeled results with equal split from both two sets (i.e., 60 reports). The consistency/inter-rater reliability of labels between the expert and students on the same set of reports reaches 93.33\% with a Cohen's $\kappa=0.867$, confirming the reliability of the results given by the participants.}

\input{tab/rq5}

\subsubsection{Findings}

The human case study provides compelling real-world evidence of \approach's benefits. From Table~\ref{tab:rq5-1}, we see that reports selected by \approach~were significantly easier to handle. Compared with the random baseline (\approach$_{ran}$), participants' qualitative ratings showed a {74.7\% improvement in perceived readability and a 64.8\% improvement in perceived identifiability}. This subjective feedback was mirrored by objective measurements. When comparing the saving of labeling efforts, \approach$_{ran}$ needs {128 minutes} to label all 300 reports while \approach~only needs {40 minutes}---a remarkable reduction of 68.8\%. This implies that \approach~can enable humans to label 460 reports per hour, in contrast to the 140 reports by \approach$_{ran}$. The benefits provided by \approach~are even more clear when considering the monetary cost; to label 300 reports, \approach~costs only {\pounds6.67}, which is more than {3x cheaper} than the \pounds21.91 incurred by \approach$_{ran}$. These results directly validate the practical utility of \approach, confirming it provides a tangible, efficient, and cost-effective solution for the human-intensive task of bug report labeling in practice.

\begin{figure}[!t]
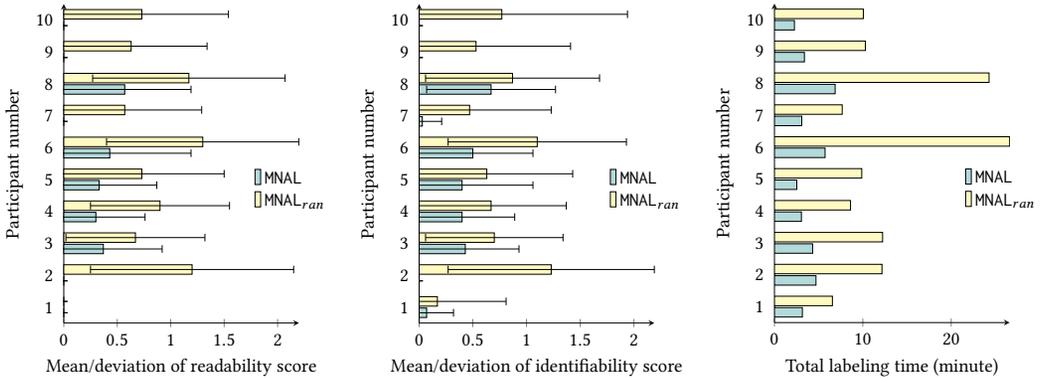

\centering
\begin{subfigure}[t]{0.32\columnwidth}
\includestandalone[width=\columnwidth]{fig/rq5-1-new}
    \end{subfigure}
~\hfill
\begin{subfigure}[t]{0.32\columnwidth}
\includestandalone[width=\columnwidth]{fig/rq5-2-new}
\end{subfigure}
~\hfill
\begin{subfigure}[t]{0.32\columnwidth}
\includestandalone[width=\columnwidth]{fig/rq5-3-new}
\end{subfigure}

    \caption{Comparing \approach~and \approach$_{ran}$ with respect to every individual of the 10 human participants (each labels 30 reports). For each participant, the pairwise comparison shows $p< 0.001$.}
\label{fig:rq5-1}%
\end{figure}

To take a closer look at each individual participant, we observe similar results, as shown in Figure~\ref{fig:rq5-1}. The significant reduction in time and cost is a direct result of the effort-aware sampling, which successfully filters for reports that are less ambiguous and require less cognitive effort to classify. As can be seen, this holds true across all 10 participants, regardless of their years of experience; almost all participants rated \approach~better than its counterpart in terms of readability and identifiability (i.e., a lower score means better). In particular, a few participants even considered all 30 reports queried by \approach~as ``highly readable'' or ``highly identifiable'' (an average rating of 0). Similarly, when comparing the total labeling time of 30 reports for each participant, \approach~needs considerably shorter time on each case. For example, \textit{Participant 6} used 5.73 minutes to label the reports queried by \approach, but this became 26.62 minutes for those from \approach$_{ran}$.

Therefore, in response to \textbf{RQ5}, we conclude that: 
\begin{quotebox}
   \noindent
   \textit{\textbf{\underline{To RQ5:}} Comparing to \approach$_{ran}$, \approach~significantly reduces the effort for labeling the reports, within the same time frame, $3\times$ monetary saving while having 74.7\% and 64.8\% improvement in the qualitative readability and identifiability score.}
\end{quotebox}



%% file: tab/rq1_avg.tex
  	

\begin{table}[t!]
	\centering
        \footnotesize
	\setlength{\tabcolsep}{0.8mm}
	\caption{Comparing the mean and standard deviation (SD) of the metrics values for the effort-aware uncertainty sampling in \approach~and its variant counterparts that use existing samplings over all timesteps on 10 runs. $r$ denotes the rank produced by Scott-Knott test and the statistically best results are highlighted in \setlength{\fboxsep}{1.5pt}\colorbox{teal!20}{green}.}
	\label{tab:rq1-1}%
\begin{adjustbox}{width=\textwidth}
\begin{tabular}{ll|llllllll|llllll}
\toprule
 \multirow{2}{*}{\textbf{Query size}} &
  \multirow{2}{*}{\textbf{Approach}} &
  \multicolumn{2}{c}{\textbf{F1-score}} &
  \multicolumn{2}{c}{\textbf{Accuracy}} &
  \multicolumn{2}{c}{\textbf{Recall}} &
  \multicolumn{2}{c|}{\textbf{Precision}}&
  \multicolumn{2}{c}{\textbf{Readability}} &
  \multicolumn{2}{c}{\textbf{Identifiability}}\\ \cline{3-14}

 &
   &
  \textbf{$r$} &
  \textbf{Mean (SD)} &
  \textbf{$r$} &
  \textbf{Mean (SD)} &
  \textbf{$r$} &
  \textbf{Mean (SD)} &
  \textbf{$r$} &
  \textbf{Mean (SD)} &
  \textbf{$r$} &
  \textbf{Mean (SD)} &
  \textbf{$r$} &
  \textbf{Mean (SD)} \\ \midrule
 &
  \texttt{\approach$_{ran}$}  &2&0.849 ($\pm0.013$)&2&0.842 ($\pm0.011$)&2&0.837 ($\pm0.026$)&2&0.862 ($\pm0.011$)&3&-53.530 ($\pm 14.955$)&3&0.052 ($\pm0.003$)\\
 &
  \texttt{\approach$_{un}$} &\cellcolor{teal!20}1&\cellcolor{teal!20}0.857 ($\pm 0.016$)&\cellcolor{teal!20}1&\cellcolor{teal!20}0.849 ($\pm 0.015$)&\cellcolor{teal!20}1&\cellcolor{teal!20}0.849 ($\pm 0.030$)&\cellcolor{teal!20}1&\cellcolor{teal!20}0.865 ($\pm 0.014$)&2&-36.049 ($\pm 23.043$)&4&0.041 ($\pm 0.006$)\\
&
  \texttt{\approach$_{conf}$} &3&0.827($\pm0.016$)&3&0.819($\pm0.013$)&3&0.812($\pm0.033$)&4&0.842($\pm0.015$)&2&-34.575($\pm42.019$)&2&0.071($\pm0.017$)\\
\multirow{-4}{*}{300} &
  \texttt{\approach} &2&0.850 ($\pm 0.019$)&2&0.841 ($\pm 0.017$)&\cellcolor{teal!20}1&\cellcolor{teal!20}0.848 ($\pm 0.038$)&3&0.852 ($\pm 0.016$)&\cellcolor{teal!20}1&\cellcolor{teal!20}5.750 ($\pm 32.774$)&\cellcolor{teal!20}1&\cellcolor{teal!20}0.188 ($\pm 0.109$)\\  \hline
  
 &
  \texttt{\approach$_{ran}$}  &2&0.86 ($\pm 0.009$)&2&0.851 ($\pm 0.008$)&2&0.855 ($\pm 0.016$)&2&0.864 ($\pm 0.009$)&4&-53.724 ($\pm 11.302$)&3&0.053 ($\pm 0.002$)\\
 &
  \texttt{\approach$_{un}$} &\cellcolor{teal!20}1&\cellcolor{teal!20}0.864 ($\pm 0.014$)&\cellcolor{teal!20}1&\cellcolor{teal!20}0.856 ($\pm 0.013$)&\cellcolor{teal!20}1&\cellcolor{teal!20}0.859 ($\pm 0.024$)&\cellcolor{teal!20}1&\cellcolor{teal!20}0.869 ($\pm 0.011$)&3&-35.195 ($\pm 24.862$)&4&0.040 ($\pm 0.006$)\\
&
  \texttt{\approach$_{conf}$} &3&0.837($\pm0.012$)&3&0.829($\pm0.01$)&3&0.829($\pm0.031$)&4&0.847($\pm0.015$)&2&-28.908($\pm41.391$)&2&0.073($\pm0.018$)\\
\multirow{-4}{*}{500} &
  \texttt{\approach} &2&0.861 ($\pm 0.013$)&2&0.853 ($\pm 0.012$)&\cellcolor{teal!20}1&\cellcolor{teal!20}0.861 ($\pm 0.022$)&3&0.862 ($\pm 0.012$)&\cellcolor{teal!20}1&\cellcolor{teal!20}-0.119 ($\pm 20.631$)&\cellcolor{teal!20}1&\cellcolor{teal!20}0.143 ($\pm 0.087$)\\  \hline
  
 &
  \texttt{\approach$_{ran}$}  &3&0.863 ($\pm 0.008$)&3&0.855 ($\pm 0.007$)&2&0.858 ($\pm 0.016$)&2&0.868 ($\pm 0.008$)&4&-52.903 ($\pm 11.866$)&3&0.053 ($\pm 0.002$)\\
 &
  \texttt{\approach$_{un}$} &\cellcolor{teal!20}1&\cellcolor{teal!20}0.870 ($\pm 0.009$)&\cellcolor{teal!20}1&\cellcolor{teal!20}0.861 ($\pm 0.009$)&\cellcolor{teal!20}1&\cellcolor{teal!20}0.869 ($\pm 0.016$)&\cellcolor{teal!20}1&\cellcolor{teal!20}0.871 ($\pm 0.009$)&3&-35.872 ($\pm 18.985$)&4&0.040 ($\pm 0.006$)\\
&
  \texttt{\approach$_{conf}$} &4&0.848($\pm0.009$)&4&0.839($\pm0.008$)&3&0.839($\pm0.02$)&4&0.856($\pm0.012$)&2&-22.015($\pm26.695$)&2&0.07($\pm0.017$)\\
\multirow{-4}{*}{700} &
  \texttt{\approach} &2&0.867 ($\pm 0.012$)&2&0.859 ($\pm 0.012$)&\cellcolor{teal!20}1&\cellcolor{teal!20}0.869 ($\pm 0.023$)&3&0.867 ($\pm 0.011$)&\cellcolor{teal!20}1&\cellcolor{teal!20}-8.617($\pm18.488$)&\cellcolor{teal!20}1&\cellcolor{teal!20}0.134($\pm 0.096$)\\\bottomrule

\end{tabular}
\end{adjustbox}
\end{table}%

%% file: tab/rq2_new.tex
\begin{table}[t!]
	\centering
	\setlength{\tabcolsep}{0.8mm}
	\caption{Comparing the mean and standard deviation (SD) of the metrics values for \approach~, the version without pseudo-labeling, and EDA for bug report identification. The format is the same as Table~\ref{tab:rq1-1}.}
	\label{tab:rq2-1}%
\begin{adjustbox}{width=\textwidth}

\begin{tabular}{ll|llllllll|llllll}
\toprule
 \multirow{2}{*}{\textbf{Query Size}} &
  \multirow{2}{*}{\textbf{Approach}} &
  \multicolumn{2}{c}{\textbf{F1-score}} &
  \multicolumn{2}{c}{\textbf{Accuracy}} &
  \multicolumn{2}{c}{\textbf{Recall}} &
  \multicolumn{2}{c|}{\textbf{Precision}} &
  \multicolumn{2}{c}{\textbf{Readability}} &
  \multicolumn{2}{c}{\textbf{Identifiability}}\\ \cline{3-14}

 &
   &
  \textbf{$r$} &
  \textbf{Mean (SD)} &
  \textbf{$r$} &
  \textbf{Mean (SD)} &
  \textbf{$r$} &
  \textbf{Mean (SD)} &
  \textbf{$r$} &
  \textbf{Mean (SD)} &
  \textbf{$r$} &
  \textbf{Mean (SD)} &
  \textbf{$r$} &
  \textbf{Mean (SD)} \\ \midrule
  
 &
  \texttt{No modification} &2&0.850 ($\pm0.019$)&2&0.841 ($\pm0.017$)&2&0.848 ($\pm0.038$)&2&0.852 ($\pm0.016$)&2&5.750 ($\pm32.774$)&\cellcolor{teal!20}1&\cellcolor{teal!20}0.188 ($\pm0.109$)\\ &
  \texttt{With augmentation} &2&0.849 ($\pm0.014$)&2&0.841 ($\pm0.013$)&\cellcolor{teal!20}1&\cellcolor{teal!20}0.853 ($\pm0.028$)&3&0.849 ($\pm0.014$)&\cellcolor{teal!20}1&\cellcolor{teal!20}6.632 ($\pm34.131$)&2&0.179 ($\pm0.11$)\\ 
  \multirow{-3}{*}{300} &
  \texttt{With pseudo-labeling} &\cellcolor{teal!20}1&\cellcolor{teal!20}0.855 ($\pm0.018$)&\cellcolor{teal!20}1&\cellcolor{teal!20}0.847 ($\pm0.017$)&\cellcolor{teal!20}1&\cellcolor{teal!20}0.852 ($\pm0.031$)&\cellcolor{teal!20}1&\cellcolor{teal!20}0.859 ($\pm0.016$)&3&1.071 ($\pm36.207$)&3&0.155 ($\pm0.104$)\\ \hline
  
 &
  \texttt{No modification}  &2&0.861 ($\pm0.013$)&2&0.853 ($\pm0.012$&3&0.861 ($\pm0.022$)&2&0.862 ($\pm0.012$)&\cellcolor{teal!20}1&\cellcolor{teal!20}-0.119 ($\pm20.631$)&2&0.143 ($\pm0.087$)\\ &
  \texttt{With augmentation} &2&0.860 ($\pm0.012$)&3&0.851 ($\pm0.011$)&\cellcolor{teal!20}1&\cellcolor{teal!20}0.865 ($\pm0.024$)&3&0.856 ($\pm0.01$)&3&-4.309 ($\pm26.025$)&\cellcolor{teal!20}1&\cellcolor{teal!20}0.146 ($\pm0.107$)\\   
\multirow{-3}{*}{500} &
  \texttt{With pseudo-labeling} &\cellcolor{teal!20}1&\cellcolor{teal!20}0.864 ($\pm0.015$)&\cellcolor{teal!20}1&\cellcolor{teal!20}0.856 ($\pm0.014$)&2&0.863 ($\pm0.026$)&\cellcolor{teal!20}1&\cellcolor{teal!20}0.865 ($\pm0.010$)&2&-1.671 ($\pm18.956$)&3&0.136 ($\pm0.106$)\\ \hline
  
 &
  \texttt{No modification}  &2&0.867 ($\pm0.012$)&2&0.859 ($\pm0.012$)&1&0.869 ($\pm0.023$)&\cellcolor{teal!20}1&\cellcolor{teal!20}0.867 ($\pm0.011$)&2&-8.617 ($\pm18.488$)&\cellcolor{teal!20}1&\cellcolor{teal!20}0.134 ($\pm0.096$)\\ &
  \texttt{With augmentation} &3&0.865 ($\pm0.09$)&3&0.855 ($\pm0.009$)&1&0.87 ($\pm0.017$)&2&0.86 ($\pm0.01$)&\cellcolor{teal!20}1&\cellcolor{teal!20}-7.948 ($\pm15.341$)&\cellcolor{teal!20}1&\cellcolor{teal!20}0.132 ($\pm0.1$)\\  
\multirow{-3}{*}{700} &
  \texttt{With pseudo-labeling} &\cellcolor{teal!20}1&\cellcolor{teal!20}0.869 ($\pm0.011$)&\cellcolor{teal!20}1&\cellcolor{teal!20}0.861 ($\pm0.011$)&1&0.871 ($\pm0.019$)&\cellcolor{teal!20}1&\cellcolor{teal!20}0.868 ($\pm0.009$)&3&-9.530 ($\pm18.465$)&2&0.122 ($\pm0.100$)\\\bottomrule

\end{tabular}
\end{adjustbox}
\end{table}%

%% file: tab/rq3_new.tex
\begin{table}[t!]
	\centering
	\setlength{\tabcolsep}{0.8mm}
	\caption{Comparing the mean and standard deviation (SD) of the metrics values for \approach~under different neural language models against updating them directly. Where $^\ddagger$ indicates the value is higher than the counterpart with high statistical significance (i.e., $p < 0.001$).}
\label{tab:rq3-1}
\begin{adjustbox}{width=\textwidth}
\begin{tabular}{lll|llll|lll}

\toprule
\textbf{Query Size} &\textbf{Model} &\textbf{Approach} &
\textbf{F1-score} &
\textbf{Accuracy} &
\textbf{Recall} &
\textbf{Precision} &
\textbf{Readability} &
\textbf{Identifiability}\\ 

\midrule
&&
  \texttt{\approach$_{ran}$}  &0.853 ($\pm0.014$)&0.846 ($\pm0.013$)&0.841 ($\pm0.025$)&0.866 ($\pm0.011$)&-51.532 ($\pm14.501$)&0.053 ($\pm0.003$)\\
&\multirow{-2}{*}{CodeBERT}&
  \texttt{\approach} &\cellcolor{teal!20}0.864 ($\pm0.016$)$^\ddagger$&\cellcolor{teal!20}0.855 ($\pm0.015$)$^\ddagger$&\cellcolor{teal!20}0.863 ($\pm0.027$)$^\ddagger$&0.865 ($\pm0.01$)&\cellcolor{teal!20}-5.500 ($\pm40.631$)$^\ddagger$&\cellcolor{teal!20}0.165 ($\pm0.118$)$^\ddagger$\\\cline{2-9}
&&
  \texttt{\approach$_{ran}$}  &0.861 ($\pm0.010$)&0.852 ($\pm0.009$)&0.86 ($\pm0.021$)&0.861 ($\pm0.010$)&-50.017 ($\pm14.424$)&0.053 ($\pm0.003$)\\
&\multirow{-2}{*}{RoBERTa}&
  \texttt{\approach}  &\cellcolor{teal!20}0.868 ($\pm0.011$)$^\ddagger$&\cellcolor{teal!20}0.858 ($\pm0.011$)$^\ddagger$&\cellcolor{teal!20}0.874 ($\pm0.017$)$^\ddagger$&0.862 ($\pm0.012$)&\cellcolor{teal!20}-0.473 ($\pm24.745$)$^\ddagger$&\cellcolor{teal!20}0.157 ($\pm0.117$)$^\ddagger$\\\cline{2-9}
&&
  \texttt{\approach$_{ran}$} &0.875 ($\pm0.007$)&0.866 ($\pm0.007$)&0.880 ($\pm0.015$)&0.870 ($\pm0.010$)&-51.317 ($\pm13.598$)&0.053 ($\pm0.003$)\\
\multirow{-6}{*}{300} &\multirow{-2}{*}{RTA} &
  \texttt{\approach}  &\cellcolor{teal!20}0.880 ($\pm0.010$)$^\ddagger$&\cellcolor{teal!20}0.871 ($\pm0.010$)$^\ddagger$&\cellcolor{teal!20}0.891 ($\pm0.017$)$^\ddagger$&0.870 ($\pm0.011$)&\cellcolor{teal!20}-5.331 ($\pm31.372$)$^\ddagger$&\cellcolor{teal!20}0.156 ($\pm0.121$)$^\ddagger$\\ \hline
  
&&
  \texttt{\approach$_{ran}$}  &0.862 ($\pm0.010$)&0.854 ($\pm0.009$)&0.855 ($\pm0.019$)&0.869 ($\pm0.008$)&-52.605 ($\pm11.605$)&0.053 ($\pm0.002$)\\
&\multirow{-2}{*}{CodeBERT}&
  \texttt{\approach} &\cellcolor{teal!20}0.870 ($\pm0.013$)$^\ddagger$&\cellcolor{teal!20}0.861 ($\pm0.012$)$^\ddagger$&\cellcolor{teal!20}0.872 ($\pm0.024$)$^\ddagger$&0.868 ($\pm0.009$)&\cellcolor{teal!20}-14.026 ($\pm24.921$)$^\ddagger$&\cellcolor{teal!20}0.135 ($\pm0.102$)$^\ddagger$\\\cline{2-9}
&&
  \texttt{\approach$_{ran}$}  &0.865 ($\pm0.008$)&0.856 ($\pm0.007$)&0.865 ($\pm0.015$)&0.865 ($\pm0.007$)&-52.010 ($\pm10.750$)&0.053 ($\pm0.002$)\\
&\multirow{-2}{*}{RoBERTa}&
  \texttt{\approach}  &\cellcolor{teal!20}0.873 ($\pm0.010$)$^\ddagger$&\cellcolor{teal!20}0.864 ($\pm0.010$)$^\ddagger$&\cellcolor{teal!20}0.878 ($\pm0.019$)$^\ddagger$&0.867 ($\pm0.008$)&\cellcolor{teal!20}-12.000 ($\pm23.367$)$^\ddagger$&\cellcolor{teal!20}0.137 ($\pm0.109$)$^\ddagger$\\\cline{2-9}
&&
  \texttt{\approach$_{ran}$} &0.880 ($\pm0.005$)&0.871 ($\pm0.005$)&0.887 ($\pm0.011$)&0.873 ($\pm0.008$)&-51.033 ($\pm10.880$)&0.053 ($\pm0.002$)\\
\multirow{-6}{*}{500} &\multirow{-2}{*}{RTA} &
  \texttt{\approach}  &\cellcolor{teal!20}0.884 ($\pm0.007$)$^\ddagger$&\cellcolor{teal!20}0.875 ($\pm0.007$)$^\ddagger$&\cellcolor{teal!20}0.893 ($\pm0.012$)$^\ddagger$&0.875 ($\pm0.008$)&\cellcolor{teal!20}-16.241 ($\pm27.093$)$^\ddagger$&\cellcolor{teal!20}0.141 ($\pm0.122$)$^\ddagger$\\ \hline

&&
  \texttt{\approach$_{ran}$}  &0.866 ($\pm0.008$)&0.858 ($\pm0.007$)&0.862 ($\pm0.015$)&0.871 ($\pm0.008$)&-54.596 ($\pm14.400$)&0.052 ($\pm0.002$)\\
&\multirow{-2}{*}{CodeBERT}&
  \texttt{\approach} &\cellcolor{teal!20}0.875 ($\pm0.010$)$^\ddagger$&\cellcolor{teal!20}0.866 ($\pm0.010$)$^\ddagger$&\cellcolor{teal!20}0.878 ($\pm0.018$)$^\ddagger$&0.871 ($\pm0.007$)&\cellcolor{teal!20}-13.356 ($\pm15.890$)$^\ddagger$&\cellcolor{teal!20}0.126 ($\pm0.087$)$^\ddagger$\\\cline{2-9}
&&
  \texttt{\approach$_{ran}$}  &0.867 ($\pm0.006$)&0.859 ($\pm0.006$)&0.867 ($\pm0.013$)&0.868 ($\pm0.006$)&-52.139 ($\pm11.494$)&0.053 ($\pm0.002$)\\ 
&\multirow{-2}{*}{RoBERTa}&
  \texttt{\approach}  &\cellcolor{teal!20}0.875 ($\pm0.010$)$^\ddagger$&\cellcolor{teal!20}0.866 ($\pm0.010$)$^\ddagger$&\cellcolor{teal!20}0.882 ($\pm0.015$)$^\ddagger$&0.869 ($\pm0.008$)&\cellcolor{teal!20}-16.749 ($\pm24.792$)$^\ddagger$&\cellcolor{teal!20}0.137 ($\pm0.102$)$^\ddagger$\\\cline{2-9}
&&
  \texttt{\approach$_{ran}$} &0.881 ($\pm0.004$)&0.873 ($\pm0.004$)&0.887 ($\pm0.010$)&0.875 ($\pm0.007$)&-51.531 ($\pm9.335$)&0.053 ($\pm0.002$)\\
\multirow{-6}{*}{700} &\multirow{-2}{*}{RTA} &
  \texttt{\approach}  &\cellcolor{teal!20}0.885 ($\pm0.006$)$^\ddagger$&\cellcolor{teal!20}0.877 ($\pm0.006$)$^\ddagger$&\cellcolor{teal!20}0.894 ($\pm0.011$)$^\ddagger$&0.877 ($\pm0.006$)&\cellcolor{teal!20}-16.551 ($\pm21.921$)$^\ddagger$&\cellcolor{teal!20}0.139 ($\pm0.105$)$^\ddagger$\\\bottomrule

\end{tabular}
\end{adjustbox}
\label{tab:rq3-1}
\end{table}%

%% file: tab/sota.tex
\begin{table}
  \caption{The state-of-the-art active learning approaches for bug report identification or a similar problem. (TF-IDF, NBM, RF, and SVM stand for Term Frequency–Inverse Document Frequency, Multinomial Naive Bayes, Random Forest, and Support Vector Machine, respectively.)}
  \label{tab:sota}
  \begin{adjustbox}{width=0.8\textwidth}
  \begin{tabular}{lllll}
    \toprule
    \textbf{Name}&\textbf{Model}&\textbf{Sampling}&\textbf{Pseudo Labeling}&\textbf{Year}\\
    \midrule

\citeauthor{DBLP:conf/iwpc/ThungLL15}~\cite{DBLP:conf/iwpc/ThungLL15}&TF-IDF and SVM&Uncertainty&No&2015\\

    \texttt{hbrPredictor}~\cite{DBLP:journals/infsof/WuZCZYM21}&TF-IDF and NBM&Uncertainty&No&2021\\

    \citeauthor{DBLP:journals/infsof/GeFQGQ22}~\cite{DBLP:journals/infsof/GeFQGQ22}&TF-IDF and RF&Uncertainty&No&2021\\

    \texttt{EMBLEM}~\cite{DBLP:journals/tse/TuYM22}&TF-IDF and SVM&Uncertainty and Certainty&No&2022\\




  \bottomrule
\end{tabular}
\end{adjustbox}
\end{table}

%% file: tab/sota_cross.tex
\begin{table}
  \caption{The state-of-the-art cross project approaches for bug report identification or a similar problem. (MLP and BERT stands for Multilayer Perceptron and Bidirectional encoder representations from transformers, respectively.)}
  \label{tab:sota_cross}
  \begin{adjustbox}{width=0.5\textwidth}

  \begin{tabular}{lllll}
    \toprule
    \textbf{Name}&\textbf{Model}&\textbf{Year}\\
    \midrule

\citeauthor{DBLP:journals/ese/HerboldTT20}~\cite{DBLP:journals/ese/HerboldTT20}&auto-fastText&2020\\

\citeauthor{DBLP:conf/iwpc/PerezJUV21}~\cite{DBLP:conf/iwpc/PerezJUV21}&TF-IDF and MLP&2021\\

    \texttt{MHNurf}~\cite{DBLP:conf/apsec/LongCC22}&Attention Network&2022\\

    \citeauthor{DBLP:journals/infsof/MeherBM24}~\cite{DBLP:journals/infsof/MeherBM24}&BERT&2024\\

  \bottomrule
\end{tabular}
\end{adjustbox}
\end{table}

%% file: tab/rq4_avg.tex
\begin{table}[t!]
	\centering
	\setlength{\tabcolsep}{0.8mm}
	\caption{Comparing the mean and standard deviation (SD) of the metrics values for \approach~and the state-of-the-art approaches for bug report identification. The format is the same as Table~\ref{tab:rq1-1}.}
	\label{tab:rq4-1}%
\begin{adjustbox}{width=\textwidth}

\begin{tabular}{ll|llllllll|llllll}
\toprule
 \multirow{2}{*}{\textbf{Query Size}} &
  \multirow{2}{*}{\textbf{Approach}} &
  \multicolumn{2}{c}{\textbf{F1-score}} &
  \multicolumn{2}{c}{\textbf{Accuracy}} &
  \multicolumn{2}{c}{\textbf{Recall}} &
  \multicolumn{2}{c|}{\textbf{Precision}} &
  \multicolumn{2}{c}{\textbf{Readability}} &
  \multicolumn{2}{c}{\textbf{Identifiability}}\\ \cline{3-14}

 &
   &
  \textbf{$r$} &
  \textbf{Mean (SD)} &
  \textbf{$r$} &
  \textbf{Mean (SD)} &
  \textbf{$r$} &
  \textbf{Mean (SD)} &
  \textbf{$r$} &
  \textbf{Mean (SD)} &
  \textbf{$r$} &
  \textbf{Mean (SD)} &
  \textbf{$r$} &
  \textbf{Mean (SD)} \\ \midrule

 &
  \texttt{EMBLEM} &4&0.797 ($\pm0.011$)&4&0.775 ($\pm0.010$)&4&0.831 ($\pm0.025$)&3&0.767 ($\pm0.014$)&2&-87.015 ($\pm13.525$)&2&0.050 ($\pm0.002$)\\
 &
  \texttt{hbrPredictor} &4&0.796 ($\pm0.006$)&5&0.756 ($\pm0.005$)&\cellcolor{teal!20}1&\cellcolor{teal!20}0.894 ($\pm0.021$)&4&0.718 ($\pm0.009$)&2&-86.748 ($\pm13.417$)&2&0.050 ($\pm0.003$)\\
&
  Ge et al. &2&0.815 ($\pm0.013$)&2&0.803 ($\pm0.014$)&5&0.819 ($\pm0.031$)&2&0.813 ($\pm0.024$)&2&-86.625 ($\pm13.521$)&2&0.050 ($\pm0.003$)\\  
&
  Thung et al. &3&0.802 ($\pm0.009$)&3&0.778 ($\pm0.010$)&3&0.842 ($\pm0.021$)&3&0.766 ($\pm0.014$)&2&-86.505 ($\pm12.561$)&2&0.050 ($\pm0.002$)\\
\multirow{-5}{*}{300}  &
  \approach  &\cellcolor{teal!20}1&\cellcolor{teal!20}0.855 ($\pm0.018$)&\cellcolor{teal!20}1&\cellcolor{teal!20}0.847 ($\pm0.017$)&2&0.852 ($\pm0.031$)&\cellcolor{teal!20}1&\cellcolor{teal!20}0.859 ($\pm0.016$)&\cellcolor{teal!20}1&\cellcolor{teal!20}1.770 ($\pm36.207$)&\cellcolor{teal!20}1&\cellcolor{teal!20}0.167 ($\pm0.104$)\\\hline
  
 &
  \texttt{EMBLEM} &4&0.810 ($\pm0.009$)&4&0.789 ($\pm0.009$)&5&0.842 ($\pm0.015$)&4&0.779 ($\pm0.011$)&2&-88.767 ($\pm12.004$)&2&0.050 ($\pm0.001$)\\  
 &
  \texttt{hbrPredictor} &5&0.803 ($\pm0.003$)&5&0.764 ($\pm0.005$)&\cellcolor{teal!20}1&\cellcolor{teal!20}0.902 ($\pm0.011$)&5&0.724 ($\pm0.008$)&2&-87.837 ($\pm10.198$)&2&0.051 ($\pm0.002$)\\
&
  Ge et al. &2&0.823 ($\pm0.008$)&2&0.805 ($\pm0.010$)&3&0.853 ($\pm0.023$)&2&0.797 ($\pm0.021$)&2&-87.202 ($\pm10.415$)&2&0.050 ($\pm0.002$)\\
&
  Thung et al. &3&0.814 ($\pm0.009$)&3&0.794 ($\pm0.010$)&4&0.846 ($\pm0.018$)&3&0.784 ($\pm0.013$)&2&-87.023 ($\pm10.714$)&2&0.051 ($\pm0.002$)\\
\multirow{-5}{*}{500}  &
  \approach  &\cellcolor{teal!20}1&\cellcolor{teal!20}0.864 ($\pm0.015$)&\cellcolor{teal!20}1&\cellcolor{teal!20}0.856 ($\pm0.014$)&2&0.863 ($\pm0.026$)&\cellcolor{teal!20}1&\cellcolor{teal!20}0.865 ($\pm0.010$)&\cellcolor{teal!20}1&\cellcolor{teal!20}-2.511 ($\pm18.956$)&\cellcolor{teal!20}1&\cellcolor{teal!20}0.145 ($\pm0.106$)\\\hline
  
 &
  \texttt{EMBLEM} &4&0.819 ($\pm0.008$)&4&0.801 ($\pm0.009$)&4&0.844 ($\pm0.015$)&4&0.795 ($\pm0.009$)&2&-88.860 ($\pm9.759$)&2&0.051 ($\pm0.001$)\\
 &
  \texttt{hbrPredictor} &5&0.806 ($\pm0.003$)&5&0.770 ($\pm0.005$)&\cellcolor{teal!20}1&\cellcolor{teal!20}0.899 ($\pm0.010$)&5&0.731 ($\pm0.007$)&2&-88.194 ($\pm7.588$)&2&0.050 ($\pm0.001$)\\
&
  Ge et al. &2&0.829 ($\pm0.007$)&2&0.817 ($\pm0.006$)&5&0.835 ($\pm0.017$)&2&0.825 ($\pm0.010$)&2&-88.113 ($\pm7.541$)&2&0.050 ($\pm0.001$)\\
&
  Thung et al. &3&0.821 ($\pm0.008$)&3&0.804 ($\pm0.008$)&3&0.848 ($\pm0.015$)&3&0.797 ($\pm0.009$)&2&-88.102 ($\pm7.290$)&2&0.050 ($\pm0.001$)\\
\multirow{-5}{*}{700}  &
  \approach  &\cellcolor{teal!20}1&\cellcolor{teal!20}0.869 ($\pm0.011$)&\cellcolor{teal!20}1&\cellcolor{teal!20}0.861 ($\pm0.011$)&2&0.871 ($\pm0.019$)&\cellcolor{teal!20}1&\cellcolor{teal!20}0.868 ($\pm0.009$)&\cellcolor{teal!20}1&\cellcolor{teal!20}-10.555 ($\pm18.465$)&\cellcolor{teal!20}1&\cellcolor{teal!20}0.132 ($\pm0.100$)\\\bottomrule

\end{tabular}
\end{adjustbox}
\end{table}%

%% file: tab/rq4_cross.tex
\begin{table}[t!]
	\centering
	\setlength{\tabcolsep}{0.8mm}
	\caption{Comparing the mean and standard deviation of the metrics values for \approach~and the state-of-the-art cross project approaches for bug report identification. The format is the same as Table~\ref{tab:rq1-1}.}
	\label{tab:rq4-cross}%
\begin{adjustbox}{width=\textwidth}

\begin{tabular}{ll|llllllll|llllll}
\toprule
 \multirow{2}{*}{\textbf{Training Size}} &
  \multirow{2}{*}{\textbf{Approach}} &
  \multicolumn{2}{c}{\textbf{F1-score}} &
  \multicolumn{2}{c}{\textbf{Accuracy}} &
  \multicolumn{2}{c}{\textbf{Recall}} &
  \multicolumn{2}{c|}{\textbf{Precision}} &
  \multicolumn{2}{c}{\textbf{Readability}} &
  \multicolumn{2}{c}{\textbf{Identifiability}}\\ \cline{3-14}

 &
   &
  \textbf{$r$} &
  \textbf{Mean (SD)} &
  \textbf{$r$} &
  \textbf{Mean (SD)} &
  \textbf{$r$} &
  \textbf{Mean (SD)} &
  \textbf{$r$} &
  \textbf{Mean (SD)} &
  \textbf{$r$} &
  \textbf{Mean (SD)} &
  \textbf{$r$} &
  \textbf{Mean (SD)} \\ \midrule

&
  Herbold et al.&5&0.700 ($\pm0.016$)&5&0.696 ($\pm0.012$)&5&0.708 ($\pm0.005$)&5&0.706 ($\pm0.021$)&2&-86.896 ($\pm12.999$)&2&0.051 ($\pm0.001$)\\
 &
  Perez et al. &4&0.712 ($\pm0.015$)&4&0.715 ($\pm0.012$)&4&0.721 ($\pm0.022$)&4&0.722 ($\pm0.024$)&2&-87.5678 ($\pm12.152$)&2&0.050 ($\pm0.003$)\\
&
  \texttt{MHNurf} &3&0.804 ($\pm0.043$)&3&0.795 ($\pm0.015$)&3&0.814 ($\pm0.011$)&3&0.806 ($\pm0.035$)&2&-86.650 ($\pm13.658$)&2&0.050 ($\pm0.002$)\\  
&
  Meher et al. &2&0.847 ($\pm0.004$)&2&0.838 ($\pm0.008$)&2&0.836 ($\pm0.021$)&2&0.853 ($\pm0.014$)&2&-85.432 ($\pm13.431$)&2&0.051 ($\pm0.001$)\\
\multirow{-5}{*}{3000}  &
  \approach  &\cellcolor{teal!20}1&\cellcolor{teal!20}0.855 ($\pm0.018$)&\cellcolor{teal!20}1&\cellcolor{teal!20}0.847 ($\pm0.017$)&\cellcolor{teal!20}1&\cellcolor{teal!20}0.852 ($\pm0.031$)&\cellcolor{teal!20}1&\cellcolor{teal!20}0.859 ($\pm0.016$)&\cellcolor{teal!20}1&\cellcolor{teal!20}1.770 ($\pm36.207$)&\cellcolor{teal!20}1&\cellcolor{teal!20}0.167 ($\pm0.104$)\\\hline
  
 &
  Herbold et al. &5&0.711 ($\pm0.006$)&5&0.704 ($\pm0.004$)&5&0.700 ($\pm0.002$)&4&0.713 ($\pm0.06$)&2&-86.958 ($\pm11.576$)&2&0.050 ($\pm0.003$)\\  
 &
  Perez et al. &4&0.724 ($\pm0.007$)&4&0.734 ($\pm0.008$)&4&0.744 ($\pm0.011$)&4&0.736 ($\pm0.012$)&2&-86.388 ($\pm12.547$)&2&0.051 ($\pm0.001$)\\
&
  \texttt{MHNurf} &3&0.816 ($\pm0.006$)&3&0.807 ($\pm0.018$)&3&0.821 ($\pm0.015$)&3&0.815 ($\pm0.018$)&2&-88.494 ($\pm11.979$)&2&0.051 ($\pm0.003$)\\
&
  Meher et al. &2&0.857 ($\pm0.009$)&2&0.848 ($\pm0.007$)&2&0.858 ($\pm0.017$)&2&0.861 ($\pm0.013$)&2&-86.754 ($\pm12.567$)&2&0.051 ($\pm0.002$)\\
\multirow{-5}{*}{5000}  &
  \approach  &\cellcolor{teal!20}1&\cellcolor{teal!20}0.864 ($\pm0.015$)&\cellcolor{teal!20}1&\cellcolor{teal!20}0.856 ($\pm0.014$)&\cellcolor{teal!20}1&\cellcolor{teal!20}0.863 ($\pm0.026$)&\cellcolor{teal!20}1&\cellcolor{teal!20}0.865 ($\pm0.010$)&\cellcolor{teal!20}1&\cellcolor{teal!20}-2.511 ($\pm18.956$)&\cellcolor{teal!20}1&\cellcolor{teal!20}0.145 ($\pm0.106$)\\\hline
  
 &
  Herbold et al. &5&0.721 ($\pm0.015$)&5&0.715 ($\pm0.021$)&5&0.709 ($\pm0.021$)&5&0.719 ($\pm0.007$)&2&-88.861 ($\pm8.875$)&2&0.051 ($\pm0.002$)\\
 &
  Perez et al. &4&0.732 ($\pm0.004$)&4&0.741 ($\pm0.007$)&4&0.752 ($\pm0.008$)&4&0.741 ($\pm0.013$)&2&-88.056 ($\pm6.994$)&2&0.050 ($\pm0.003$)\\
&
  \texttt{MHNurf} &3&0.824 ($\pm0.013$)&3&0.812 ($\pm0.012$)&3&0.828 ($\pm0.008$)&3&0.821 ($\pm0.004$)&2&-88.339 ($\pm6.856$)&2&0.050 ($\pm0.001$)\\
&
  Meher et al. &2&0.864 ($\pm0.015$)&2&0.852 ($\pm0.011$)&2&0.859 ($\pm0.023$)&2&0.866 ($\pm0.013$)&2&-87.875 ($\pm7.067$)&2&0.051 ($\pm0.006$)\\
\multirow{-5}{*}{7000}  &
  \approach  &\cellcolor{teal!20}1&\cellcolor{teal!20}0.869 ($\pm0.011$)&\cellcolor{teal!20}1&\cellcolor{teal!20}0.861 ($\pm0.011$)&\cellcolor{teal!20}1&\cellcolor{teal!20}0.871 ($\pm0.019$)&\cellcolor{teal!20}1&\cellcolor{teal!20}0.868 ($\pm0.015$)&\cellcolor{teal!20}1&\cellcolor{teal!20}-10.555 ($\pm18.413$)&\cellcolor{teal!20}1&\cellcolor{teal!20}0.132 ($\pm0.100$)\\\bottomrule

\end{tabular}
\end{adjustbox}
\end{table}%

%% file: tab/llms.tex
\begin{table}[t!]
	\centering
	\caption{Comparing the performance of \approach~and the state-of-the-art pseudo-labeling model and LLMs in bug report identification task. The format is the same as Table~\ref{tab:rq3-1}.}
	\label{tab:llms}%
\begin{adjustbox}{width=0.7\textwidth}

\begin{tabular}{cc|ccccc}
\toprule
 \textbf{Approach} &
  \textbf{Prompting} &
  \textbf{F1-score} &
  \textbf{Accuracy} &
  \textbf{Recall} &
  \textbf{Precision} &
  \textbf{AUC}  \\ \midrule
  
\multirow{6}{*}{GPT-4o-mini} &
  \texttt{0-shot}&0.707&0.563&$^\ddagger$\cellcolor{teal!20}0.989&0.550&0.533\\ &
  \texttt{1-shot} &0.782&0.732&0.901&0.690&0.720\\ &
  \texttt{2-shot} &0.653&0.563&0.773&0.566&0.548\\ &
  \texttt{3-shot} &0.648&0.516&0.835&0.529&0.494\\ &
  \texttt{4-shot} &0.689&0.534&0.968&0.534&0.504\\ &
  \texttt{5-shot} &0.668&0.526&0.894&0.533&0.500\\  
  \hline
HINT &
  \texttt{-} &0.848&0.845&0.821&\cellcolor{teal!20}0.877$^\ddagger$&\cellcolor{teal!20}0.889$^\ddagger$\\
  \hline
\approach &
  \texttt{-} &\cellcolor{teal!20}0.869$^\ddagger$&\cellcolor{teal!20}0.861$^\ddagger$&0.871&0.868&0.878\\ \bottomrule

\end{tabular}
\end{adjustbox}
\end{table}%

%% file: tab/rq5.tex


\begin{table}[t!]
	\centering
	\caption{Comparing \approach~and \approach$_{ran}$ in the qualitative study involving humans. The format is the same as Table~\ref{tab:rq3-1}.}
	\label{tab:rq5-1}%
    \begin{adjustbox}{width=0.8\textwidth}
	\begin{tabular}{llllll}\toprule
	\textbf{Type}&\textbf{Metric}&\textbf{\approach$_{ran}$}&\textbf{\approach}\\\midrule 

   \multirow{2}{*}{Effort} &Mean/deviation of readability score&0.79 ($\pm 0.83$)&\cellcolor{teal!20}0.20 ($\pm 0.47$)$^\ddagger$\\
   &Mean/deviation of identifiability score&0.71 ($\pm 0.88$)&\cellcolor{teal!20}0.25 ($\pm 0.48$)$^\ddagger$\\
\hline
    \multirow{2}{*}{Time} &Total labeling time for a timestep (300 reports)&128 mins&\cellcolor{teal!20}40 mins$^\ddagger$\\
    &Expected number of reports labeled per hour&140&\cellcolor{teal!20}460$^\ddagger$\\
\hline
    \multirow{1}{*}{Money} &Total monetary cost for labeling in a timestep (300 reports)&\pounds21.91&\cellcolor{teal!20}\pounds6.67$^\ddagger$\\
    \bottomrule
	\end{tabular}
 \end{adjustbox}
\end{table}%

%% file: fig/rq5-1-new.tex
\begin{tikzpicture}
    \begin{axis}[
    height=7.5cm, width=6cm,
        xbar=1pt,
        bar width=5pt,
        ylabel=Participant number,
        xlabel=Mean/deviation of readability score,
        enlarge y limits=0.5,
        xmin=0,
        ymin=0.5,
        ymax=10.5,
        ytick style={
            /pgfplots/major tick length=0pt,
        },
        ytick={1,2,...,10},
      axis x line  = bottom,
    axis y line  = left  ,
legend cell align={left},
  legend style={nodes={scale=1, transform shape},draw=none,fill=none,at={(1.15,0.5)}}
    ]
           \addplot+ [fill=teal!30,draw=black,
            error bars/.cd,
                x dir=both,
                x explicit,
                error bar style={color=black}
        ] coordinates {
          (0,1) +- (0,0)
                    (0,2) +- (0,0)
                        (0.37,3) +- (0.55,0)
                            (0.3,4) +- (0.46,0)
                                (0.33,5) +- (0.54,0)
                                    (0.43,6) +- (0.76,0)
           (0,7) +- (0,0)
          (0.57,8) +- (0.62,0)
            (0,9) +- (0,0)
            (0,10) +- (0,0)

        };

           \addplot+ [fill=yellow!30,draw=black,
            error bars/.cd,
                x dir=both,
                x explicit,
                 error bar style={color=black}
        ] coordinates {

        (0,1) +- (0,0)
               (1.2,2) +- (0.95,0)
                (0.67,3) +- (0.65,0)
                 (0.9,4) +- (0.65,0)
                  (0.73,5) +- (0.77,0)
                   (1.3,6) +- (0.9,0)
            (0.57,7) +- (0.72,0)
              (1.17,8) +- (0.9,0)
             (0.63,9) +- (0.71,0)
             (0.73,10) +- (0.81,0)

        };

        \legend{
            \approach,
           \approach$_{ran}$
        }
    \end{axis}
\end{tikzpicture}

%% file: fig/rq5-2-new.tex
\begin{tikzpicture}
    \begin{axis}[
    height=7.5cm, width=6cm,
        xbar=1pt,
        bar width=5pt,
         ylabel=Participant number,
        xlabel=Mean/deviation of identifiability score,
        enlarge y limits=0.5,
        xmin=0,
        ymin=0.5,
        ymax=10.5,
        ytick style={
            /pgfplots/major tick length=0pt,
        },
        ytick={1,2,...,10},
      axis x line  = bottom,
    axis y line  = left  ,
legend cell align={left},
  legend style={nodes={scale=1, transform shape},draw=none,fill=none,at={(1.15,0.5)}}
    ]
           \addplot+ [fill=teal!30,draw=black,
            error bars/.cd,
                x dir=both,
                x explicit,
                error bar style={color=black}
        ] coordinates {
          (0.07,1) +- (0.25,0)
                    (0,2) +- (0,0)
                        (0.43,3) +- (0.5,0)
                            (0.4,4) +- (0.49,0)
                                (0.4,5) +- (0.66,0)
                                    (0.5,6) +- (0.56,0)
           (0.03,7) +- (0.18,0)
          (0.67,8) +- (0.6,0)
            (0,9) +- (0,0)
            (0,10) +- (0,0)

        };

           \addplot+ [fill=yellow!30,draw=black,
            error bars/.cd,
                x dir=both,
                x explicit,
                 error bar style={color=black}
        ] coordinates {

        (0.17,1) +- (0.64,0)
               (1.23,2) +- (0.96,0)
                (0.7,3) +- (0.64,0)
                 (0.67,4) +- (0.7,0)
                  (0.63,5) +- (0.8,0)
                   (1.1,6) +- (0.83,0)
            (0.47,7) +- (0.76,0)
              (0.87,8) +- (0.81,0)
             (0.53,9) +- (0.88,0)
             (0.77,10) +- (1.17,0)
    
        };

        \legend{
               \approach,
           \approach$_{ran}$
        }
    \end{axis}
\end{tikzpicture}

%% file: fig/rq5-3-new.tex
\begin{tikzpicture}
    \begin{axis}[
    height=7.5cm, width=6cm,
        xbar=1pt,
        bar width=5pt,
         ylabel=Participant number,
        xlabel=Total labeling time (minute),
        enlarge y limits=0.5,
        xmin=0,
        ymin=0.5,
        ymax=10.5,
        ytick style={
            /pgfplots/major tick length=0pt,
        },
        ytick={1,2,...,10},
      axis x line  = bottom,
    axis y line  = left  ,
legend cell align={left},
  legend style={nodes={scale=1, transform shape},draw=none,fill=none,at={(1.15,0.5)}}
    ]
           \addplot+ [fill=teal!30,draw=black,
            error bars/.cd,
                x dir=both,
                x explicit,
                error bar style={color=black}
        ] coordinates {
          (3.16,1)
                    (4.7,2)
                        (4.32,3)
                            (3.05,4)
                                (2.53,5)
                                    (5.73,6)
           (3.09,7)
          (6.87,8)
            (3.39,9)
            (2.26,10)

        };

           \addplot+ [fill=yellow!30,draw=black,
            error bars/.cd,
                x dir=both,
                x explicit,
                 error bar style={color=black}
        ] coordinates {

          (6.56,1)
                    (12.17,2)
                        (12.24,3)
                            (8.61,4)
                                (9.88,5)
                                    (26.62,6)
           (7.67,7)
          (24.29,8)
            (10.29,9)
            (10.06,10)
              
        };

        \legend{
                \approach,
           \approach$_{ran}$
        }
    \end{axis}
\end{tikzpicture}

%% file: sec/why.tex
\section{Discussion}
\label{sec:why}

In this section, we discuss a few notable observations drawn from the experiments and the factors to consider when applying \approach~in practical scenarios.


\subsection{The Changing Trade-off on Uncertainty, Readability, and Identifiability}
\label{sec:trade-off}

It is not hard to understand that the uncertainty, readability, and identifiability involved in effort-aware uncertainty sampling are potentially conflicting. By using an aggregated quality-effort score with equal weights among the three, we seek to arrive at the trade-off points that did not show discriminated preferences over any of those metrics. However, although the readability and identifiability of a report never change, the sampling landscape can still evolve between different timesteps due to the varying uncertainty of the reports produced by the updated neural language model. Therefore, within the landscape, the exact positions of the reports with the best quality-effort scores can shift.

\begin{figure}[!t]
\centering
\begin{subfigure}{0.2\columnwidth}
  \centering
  \includegraphics[width=\linewidth]{fig/blank_stripe.pdf} 
\end{subfigure}
\begin{subfigure}{0.5\columnwidth}
  \centering
  \includegraphics[width=\linewidth]{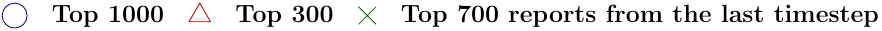} 
\end{subfigure}
\begin{subfigure}{0.2\columnwidth}
  \centering
  \includegraphics[width=\linewidth]{fig/blank_stripe.pdf} 
\end{subfigure}
\begin{subfigure}{0.3\columnwidth}
  \centering
  \includegraphics[width=\linewidth]{fig/blank_stripe.pdf} 
\end{subfigure}
\begin{subfigure}{0.3\columnwidth}
  \centering
  \includegraphics[width=\linewidth]{fig/blank_stripe.pdf} 
\end{subfigure}
\begin{subfigure}{0.3\columnwidth}
  \centering
  \includegraphics[width=\linewidth]{fig/blank_stripe.pdf} 
\end{subfigure}
\begin{subfigure}[t]{0.3\columnwidth}
\includegraphics[width=\columnwidth]{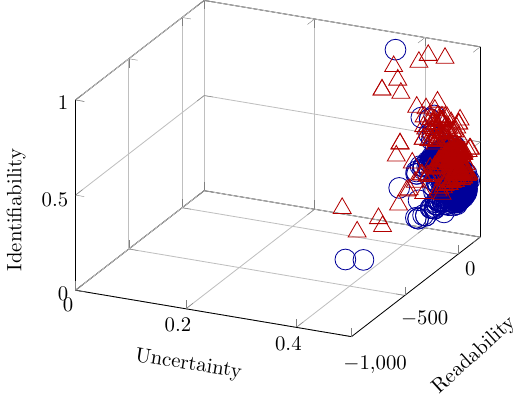}
    \subcaption{Timestep 1}
    \end{subfigure}
\begin{subfigure}[t]{0.3\columnwidth}
\includegraphics[width=\columnwidth]{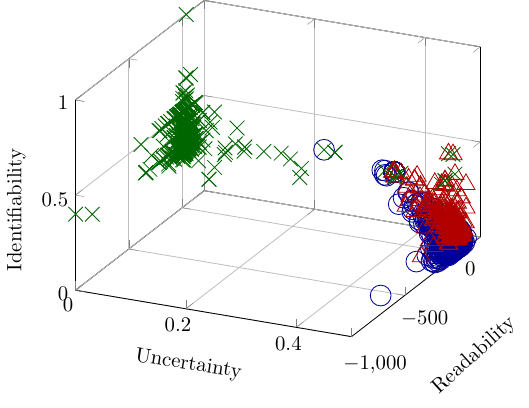}
   \subcaption{Timestep 2}
    \end{subfigure}
\begin{subfigure}[t]{0.3\columnwidth}
\includegraphics[width=\columnwidth]{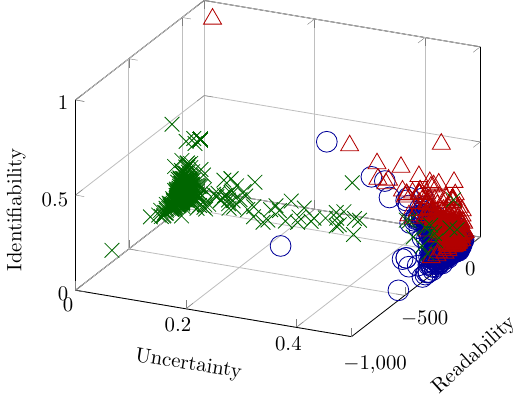}
   \subcaption{Timestep 3}
    \end{subfigure}
\begin{subfigure}[t]{0.3\columnwidth}
\includegraphics[width=\columnwidth]{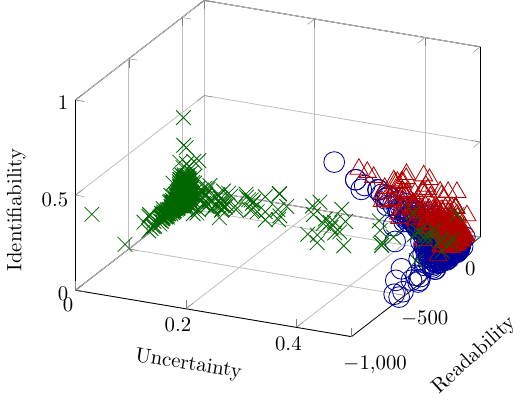}
    \subcaption{Timestep 4}
    \end{subfigure}
\begin{subfigure}[t]{0.3\columnwidth}
\includegraphics[width=\columnwidth]{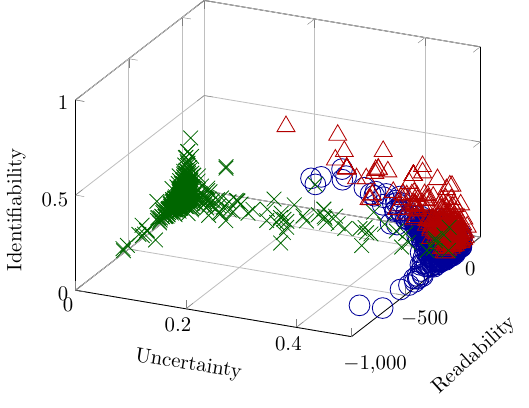}
   \subcaption{Timestep 5}
    \end{subfigure}
\begin{subfigure}[t]{0.3\columnwidth}
\includegraphics[width=\columnwidth]{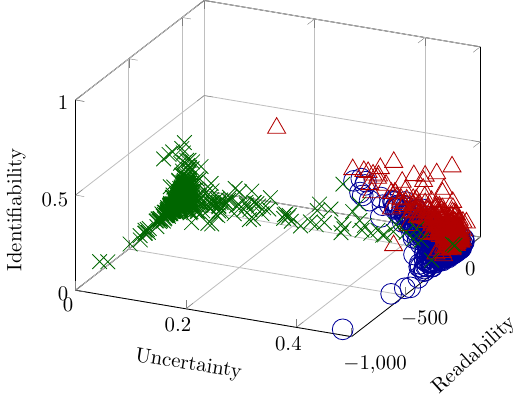}
   \subcaption{Timestep 6}
    \end{subfigure}
\begin{subfigure}[t]{0.3\columnwidth}
\includegraphics[width=\columnwidth]{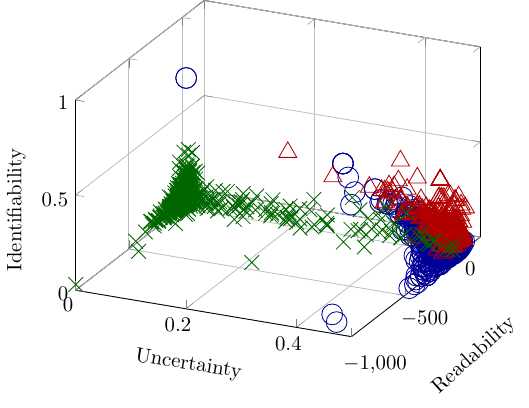}
    \subcaption{Timestep 7}
    \end{subfigure}
\begin{subfigure}[t]{0.3\columnwidth}
\includegraphics[width=\columnwidth]{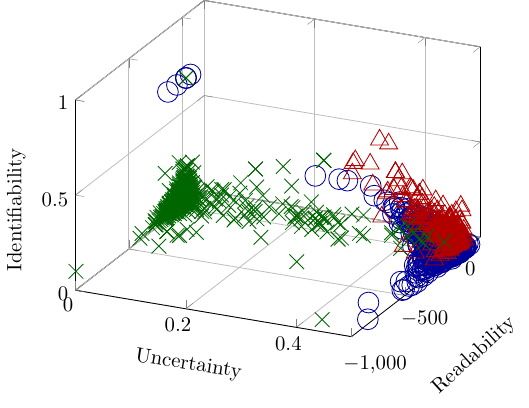}
   \subcaption{Timestep 8}
    \end{subfigure}
\begin{subfigure}[t]{0.3\columnwidth}
\includegraphics[width=\columnwidth]{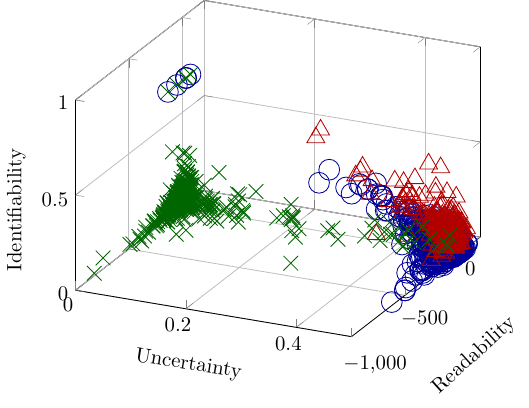}
   \subcaption{Timestep 9}
    \end{subfigure}

    \caption{The changing sampling landscapes considered by \approach~across the timesteps when $k=300$, i.e., only the top 300 reports according to the quality-effort score are queried in a timestep.}
\label{fig:discuss-1}
\end{figure}

Figure~\ref{fig:discuss-1} illustrates the changing sampling landscapes across the timesteps for one run under \approach~with a query size of 300. Since it is unrealistic to visualize more than 1 million points, for each timestep, we plot the top $1,000$ best reports based on the quality-effort score. We also highlight the 700 reports, which have not been selected from the $1,000$ best ones for querying, with their new positions in the subsequent timestep after the changed landscape. This helps us to better understand the landscape shift. The 700 points depicted from the previous timestep are selected because the top 300 points from previous timestep have already been chosen and placed into the labeled pool in that previous timestep. Thus, including them again in the current timestep would not provide meaningful insights as they are no longer part of the unlabeled set. Therefore, the top 700 points shown represent the remaining subset of the top 1000 points from the last timestep that have not yet been labeled.

Notably, a key information that we are trying to illustrate is how the uncertainty of the same points selected from timestep $t$ and using the model at timestep $t$ has been changed using the updated model at timestep $t+1$. Therefore, although we seek to illustrate the newly selected top 1000 points according to the quality-effort score at each timestep, the best 300 points from the top 1000 ones selected at timestep $t$ would have already been used to update the model for timestep $t+1$, therefore, it makes less sense to include those 300 points since they are part of the training data for the model at timestep $t+1$ (their uncertainty would be close to zero). As such, in addition to the best 1000 newly selected points, we illustrate the uncertainty of the same remaining 700 points (from 1000 selected at timestep $t+1$) using the updated model at timestep $t+1$, and compare how the uncertainty has been shifted with the same 700 points at timestep $t$.

Clearly, we see that there are obvious discrepancies in the landscape between different timesteps due to the changes of uncertainty on the unlabeled reports, hence the positions of the best $1,000$ reports in terms of the quality-effort score have also been shifted. A key information we can observe is that there is always a considerable change in terms of the uncertainty between timesteps for the remaining unselected reports (e.g., comparing the \textcolor{blue}{$\circ$} at \textit{Timestep 2} and the \textcolor{teal}{$\times$} at \textit{Timestep 3}). This makes sense, since at \textit{Timestep $n+1$} the model has been updated with the top 300 reports selected from \textit{Timestep $n$}, then the remaining 700 best reports from \textit{Timestep $n$} would also be very likely to become known by the model, hence a significant drop of the uncertainty for those reports from the updated model at \textit{Timestep $n+1$}.

Interestingly, if we look closer at the relative positions of the unselected reports in the landscape of a \textit{Timestep $n$} (the \textcolor{blue}{$\circ$}) and their corresponding positions in the landscape of the subsequent \textit{Timestep $n+1$} (the \textcolor{teal}{$\times$}), the overall change of uncertainty for those between \textit{Timestep 1} and \textit{Timestep 2} are larger than that of those between any following timesteps. This has caused a significantly bigger shift of the $1,000$ best reports in \textit{Timestep 2} compared with those in \textit{Timestep 1}. That is, the $1,000$ best reports have shifted from \textit{Timestep 1} to \textit{Timestep 2} on all three dimensions whereas those best reports between other timesteps mainly move along the dimension of uncertainty, e.g., from \textit{Timestep 2} to \textit{Timestep 3}. We found that these observations are due to the fact that the neural language model has not yet been fine-tuned with the GitHub reports when conducting the sampling at \textit{Timestep 1}, therefore it has not reached a stable state, leading to larger change on the uncertainty in \textit{Timestep 1} and hence the bigger landscape shift. The following timesteps (from \textit{Timestep 2} onwards), in contrast, would rely on a model that has already been updated with some reports, therefore the change in the landscape is smaller. The above is also the key reason that, in Section~\ref{sec:result}, there are always considerable changes in readability and identifiability under \approach~from \textit{Timestep 1} to \textit{Timestep 2}.

Another worth noting finding is that, for the same unselected reports among the best $1,000$ ones in a \textit{Timestep $n$}, the change between their uncertainty in the landscape of $n$ and that of the subsequent \textit{Timestep $n+1$} tends to shrink with more updates, e.g., the difference of uncertainty between the \textcolor{blue}{$\circ$} in \textit{Timestep 9} and the \textcolor{teal}{$\times$} in \textit{Timestep 10} is much smaller than that between the \textcolor{blue}{$\circ$} in \textit{Timestep 2} and the \textcolor{teal}{$\times$} in \textit{Timestep 3}. This is because the neural language model in \approach~has learned from increasingly more labeled reports for identifying their bug relevance, which gradually consolidates itself, hence the changes in the uncertainty of reports are reduced.

\subsection{On the Trade-off Between Human-Friendly and Model-Beneficial Data}

During model training, we realize that samples friendly to human interpretation do not necessarily contribute positively to the model’s identification performance. This section elaborates on this trade-off, showing the characteristics of data that, while potentially difficult for humans to parse, are information-rich and beneficial for training a neural language model. It is methodologically challenging to systematically isolate the most impactful training instances for two main reasons. First, it is difficult to precisely attribute improvements in model accuracy to a specific data point, as the contribution of any single report depends on the model’s state and the stage of training at which it is introduced. Second, the use of batch training complicates this analysis: when a performance increase follows a particular batch, it becomes hard to determine exactly which individual report within that batch was responsible for the improvement. Despite these challenges, we can illustrate the trade-off between readability and informational value through contrasting examples from our dataset.

\begin{figure}[!t]
\centering

\begin{subfigure}{.49\columnwidth}
  \centering
  \includegraphics[width=\linewidth]{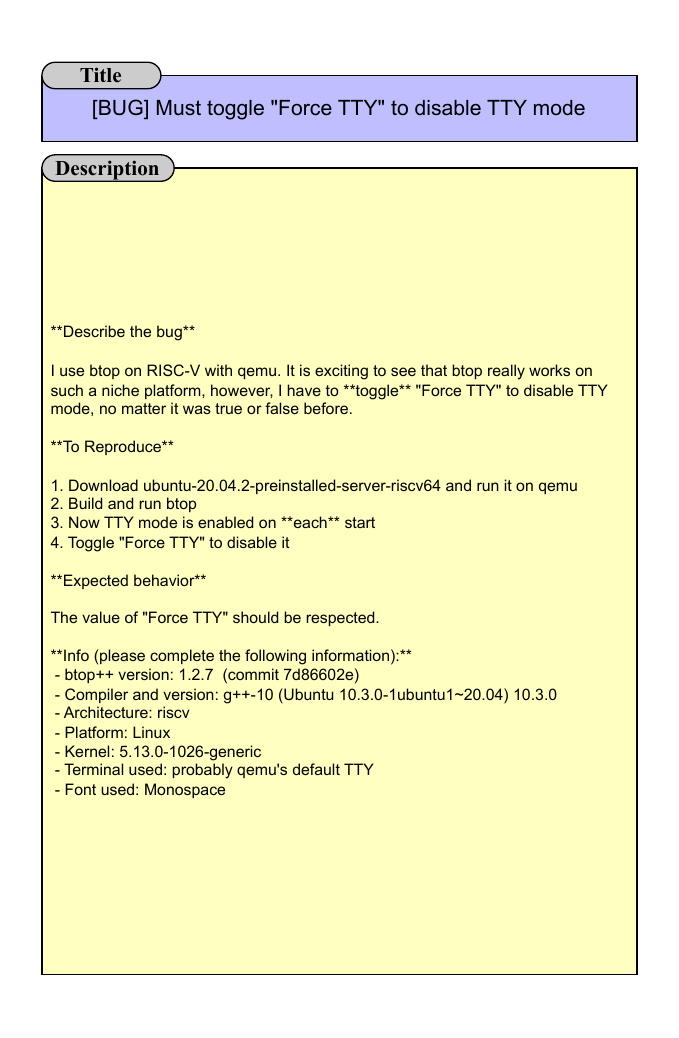} 
  \caption{Human-friendly exampled report}
  \label{fig:d2—fig1}
\end{subfigure}
~\hspace{1cm}
\begin{subfigure}{.49\columnwidth}
  \centering
  \includegraphics[width=\linewidth]{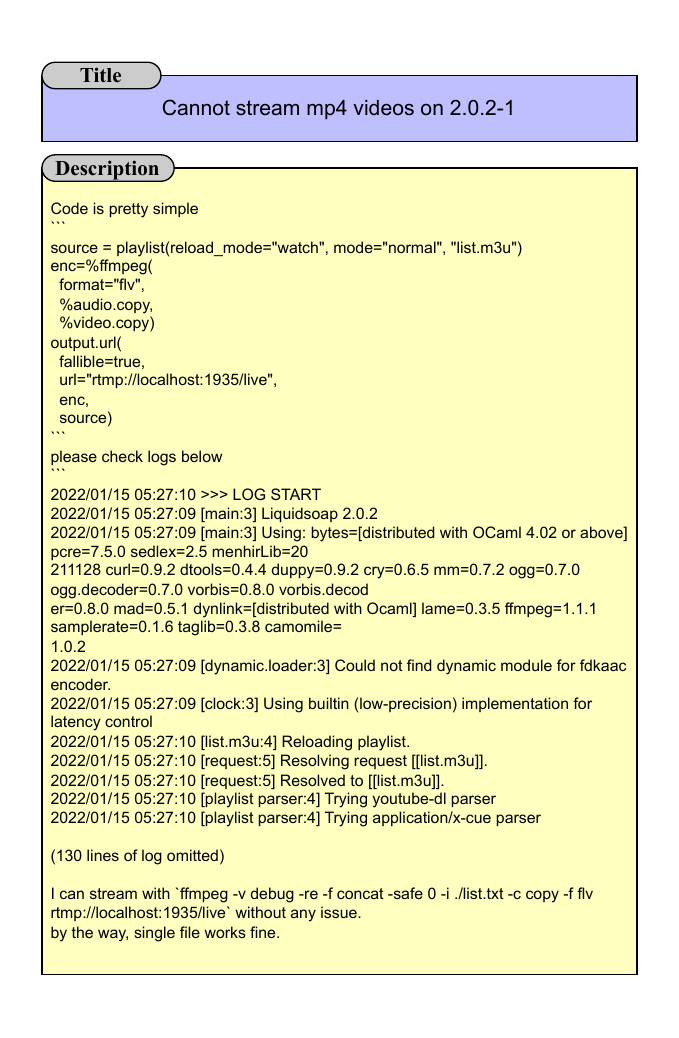} 
  \caption{Model-beneficial exampled report}
  \label{fig:d2-fig2}
\end{subfigure}

  \caption{Excerpt of the human-friendly vs. model-beneficial exampled reports}
     \label{fig:d2}
\end{figure}

Consider the well-structured bug report shown in Figure~\ref{fig:d2—fig1}. This report is clear, concise, and follows a standard template, making it very easy for a human to understand the issue. It describes the bug, provides steps to reproduce it, and lists the expected behavior and system information. From the model’s perspective, however, the information in this report may be limited. Its vocabulary and structure are conventional, offering few new patterns for a model already trained on vast quantities of similar text. While it effectively communicates the bug to a human, its informational density is relatively low and contributes minimally to enhancing the model’s ability to diagnose more complex or obscure issues.

In contrast, the second report, shown in Figure~\ref{fig:d2-fig2}, is less accessible to a human reader, consisting primarily of a short code snippet and a long, dense log file. Despite its lack of human-friendly structure, this report is valuable for the model. The log file contains a high density of specific technical information—library function calls, precise error messages (e.g., Server error: Already publishing), and stack traces (e.g., Avutil.Error(Operation not permitted)). Such data provide rich contextualized patterns, enabling the model to learn complex correlations among specific library function calls, event sequences in logs, and relevant bug types.

\subsection{On the Implications of Pseudo-labeled Reports}
\label{sec:diff-k}

While having some pseudo-labeled reports as part of the updating/training process in \approach~is certainly beneficial for the performance, it remains unclear how many such reports are sufficient. To investigate that, we examine and compare \approach~under varying numbers of pseudo-labeled reports used at each timestep, i.e., the top $s$ most similar reports would be pseudo-labeled according to each human-labeled report (by default the pseudo-labeling approach assumes $s=1$). As such, the total number of pseudo-labeled reports per timestep would be $s \times k$ where $k$ is the query size.

\begin{figure}[!t]
\centering
\begin{subfigure}{0.1\columnwidth}
  \centering
  \includegraphics[width=\linewidth]{fig/blank_stripe.pdf} 
\end{subfigure}
\begin{subfigure}{0.6\columnwidth}
  \centering
  \includegraphics[width=0.6\linewidth]{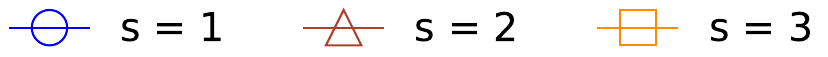} 
\end{subfigure}
\begin{subfigure}{0.1\columnwidth}
  \centering
  \includegraphics[width=\linewidth]{fig/blank_stripe.pdf} 
\end{subfigure}
\begin{subfigure}[t]{0.3\columnwidth}
\includegraphics[width=\columnwidth]{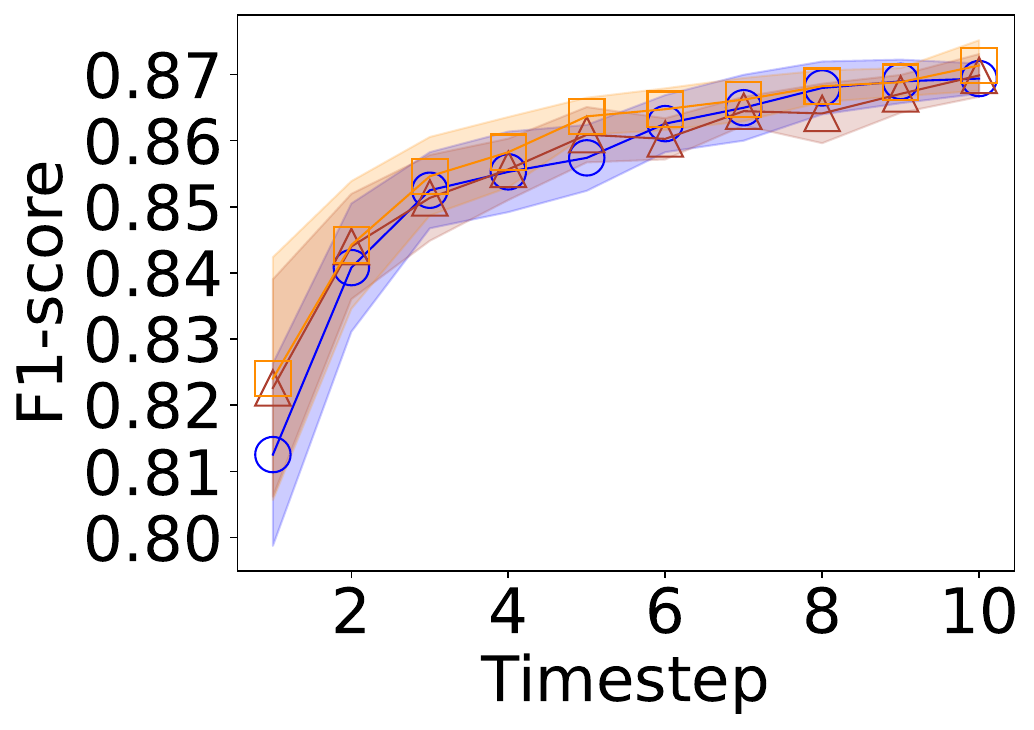}
    \subcaption{Query size 300}
    \end{subfigure}
\begin{subfigure}[t]{0.3\columnwidth}
\includegraphics[width=\columnwidth]{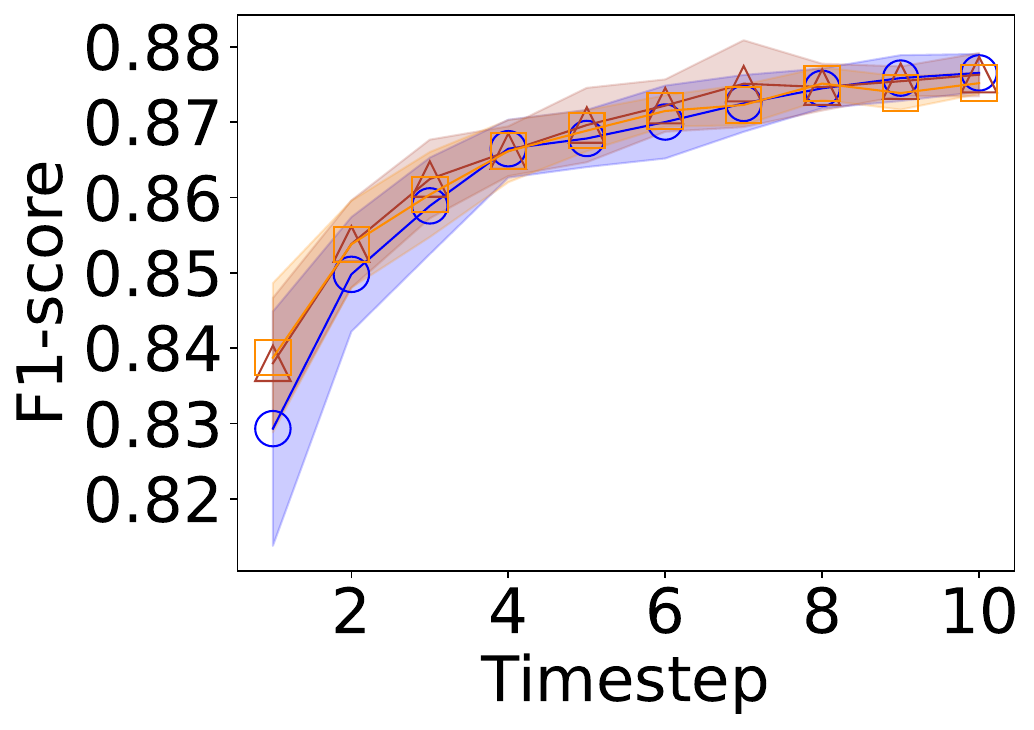}
   \subcaption{Query size 500}
    \end{subfigure}
\begin{subfigure}[t]{0.3\columnwidth}
\includegraphics[width=\columnwidth]{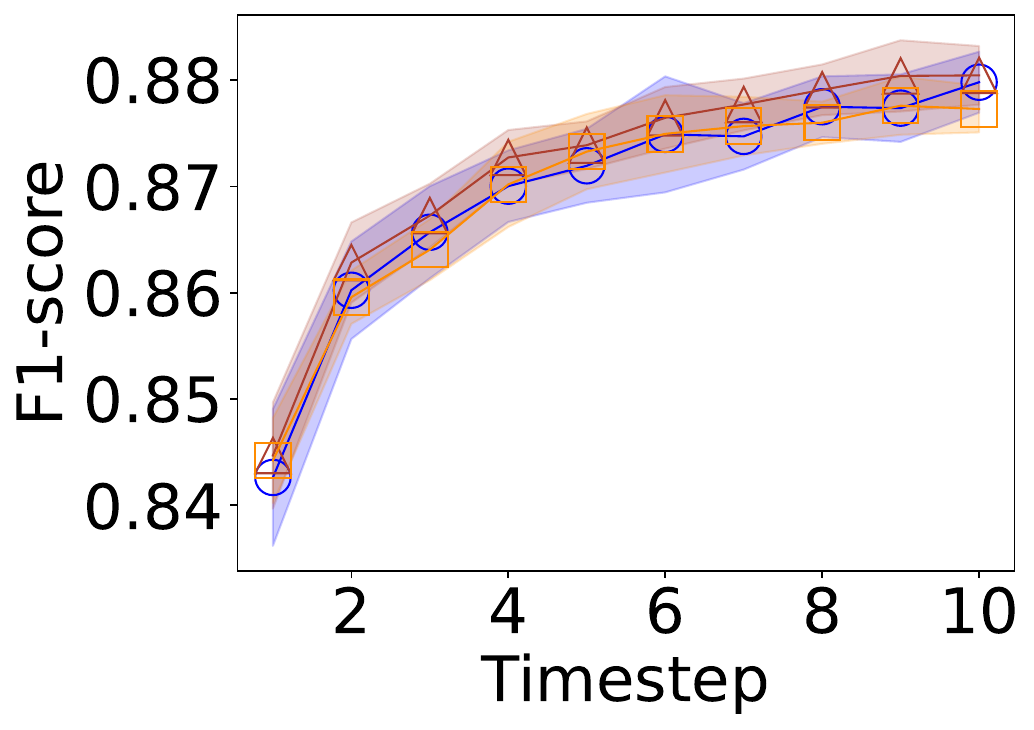}
   \subcaption{Query size 700}
    \end{subfigure}
\begin{subfigure}[t]{0.3\columnwidth}
\includegraphics[width=\columnwidth]{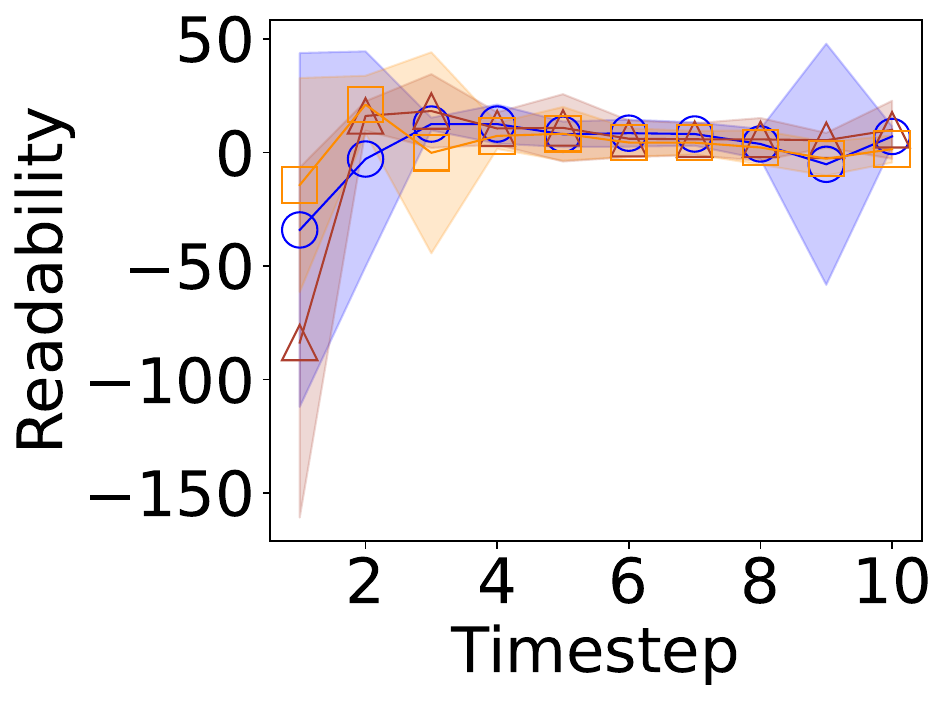}
    \subcaption{Query size 300}
    \end{subfigure}
\begin{subfigure}[t]{0.3\columnwidth}
\includegraphics[width=\columnwidth]{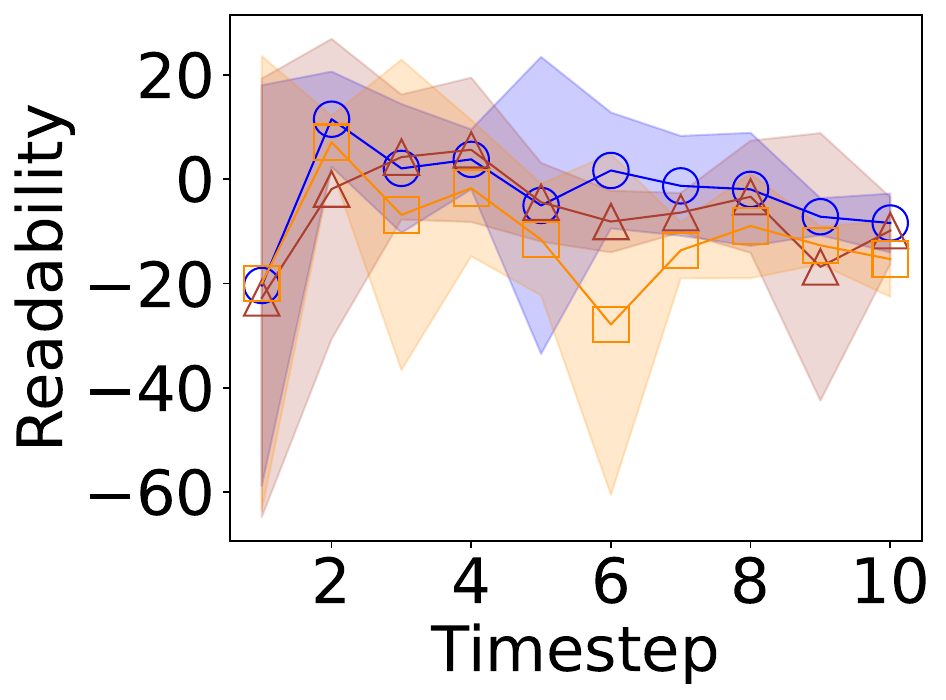}
   \subcaption{Query size 500}
    \end{subfigure}
\begin{subfigure}[t]{0.3\columnwidth}
\includegraphics[width=\columnwidth]{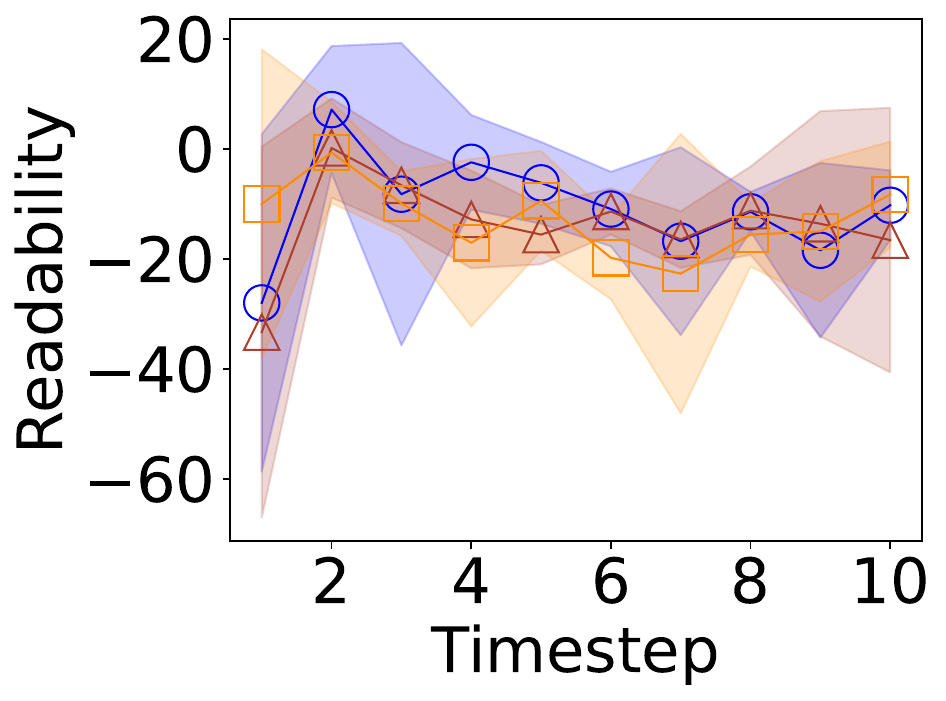}
   \subcaption{Query size 700}
    \end{subfigure}
\begin{subfigure}[t]{0.3\columnwidth}
\includegraphics[width=\columnwidth]{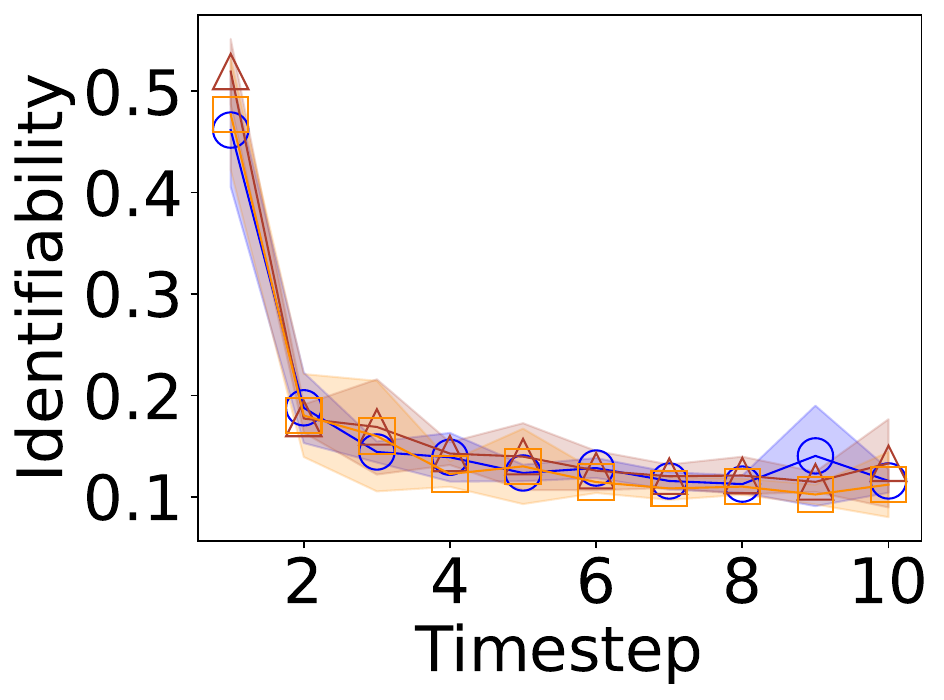}
    \subcaption{Query size 300}
    \end{subfigure}
\begin{subfigure}[t]{0.3\columnwidth}
\includegraphics[width=\columnwidth]{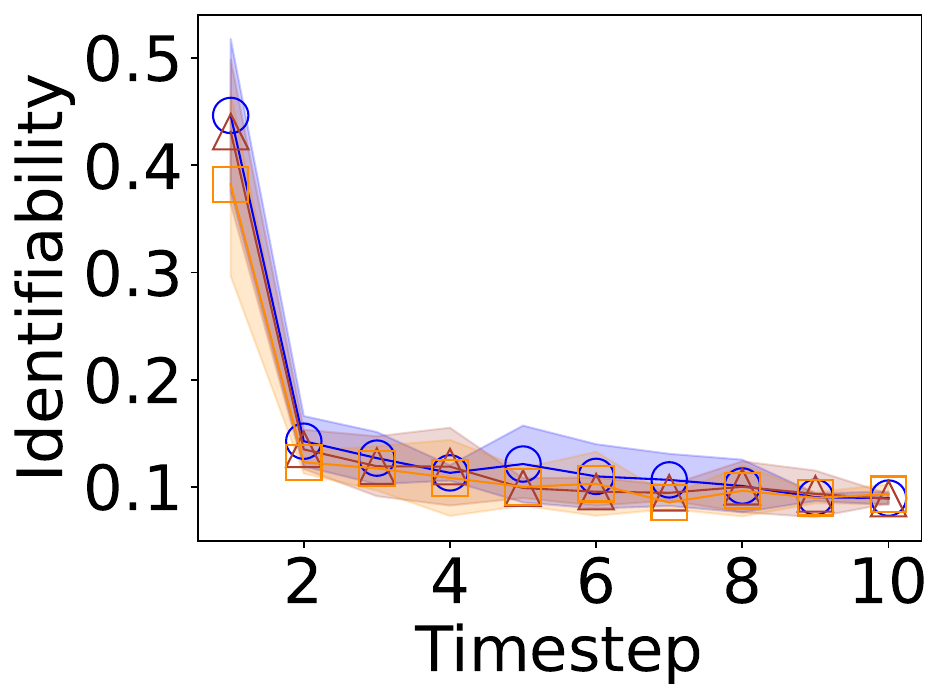}
   \subcaption{Query size 500}
    \end{subfigure}
\begin{subfigure}[t]{0.3\columnwidth}
\includegraphics[width=\columnwidth]{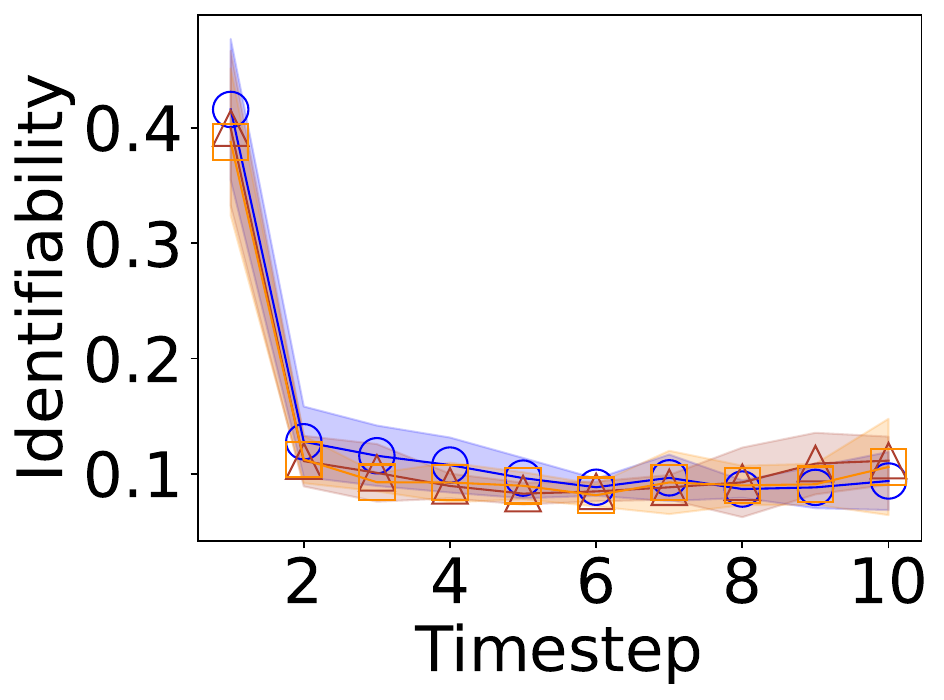}
   \subcaption{Query size 700}
    \end{subfigure}

    \caption{Comparing \approach~under different numbers of pseudo-labeled reports used at each timestep over all 10 timesteps (10 runs each). The plots show the mean and standard deviation. }
\label{fig:discuss-2}
\end{figure}

Figure~\ref{fig:discuss-2} shows the results. As can be seen, we observe no statistical significance comparing the results obtained by different $s$ values on all metrics. This means that using more pseudo-labeled reports as part of the updating/training does not help the performance but only introduces unnecessary overhead. Therefore, we suggest considering only the most similar reports to their corresponding human-labeled ones in pseudo-labeling, i.e., $s=1$.


\subsection{Additional Design Justifications}

\subsubsection{Cold Start Strategy}

While \approach~is designed to support both cold and warm start initializations as described in Section 3.3, our experimental evaluations have focused on the warm start strategy. This decision was based on several practical and methodological considerations.

The warm start approach, which involves fine-tuning a pre-trained neural language model, represents the current state-of-the-art and most common practice for NLP tasks. Models like BERT and RoBERTa are pre-trained on vast general-language corpora, which provides a powerful foundation that is more resource-efficient and often more effective than training a model from the ground up (a cold start).

Furthermore, a cold start would require training a neural language model from scratch, a process that is computationally intensive and time-consuming, especially with a dataset of over a million reports. As we will discuss in Section 6.6, even the fine-tuning process for one timestep can take nearly an hour on high-performance hardware; a full training cycle from scratch would be substantially more demanding and was considered beyond the scope of this study. Therefore, we focused on demonstrating the significant value \approach~adds to the more pragmatic and widely adopted warm start workflow. The core contributions—effort-aware uncertainty sampling and pseudo-labeling—are effectively validated by showing how they improve the standard fine-tuning process in a continuous, active learning setting.

\subsubsection{Imbalance Experiment}
\input{tab/imb}
Our primary experiments were conducted using the NLBSE'23 dataset, which is relatively balanced with a 52.6\% to 47.4\% ratio of bug reports to non-bug reports in the training set. While this provides a controlled environment for evaluation, real-world software repositories often exhibit highly skewed distributions where one class significantly outnumbers the other. To assess the robustness and practical viability of \approach~under such conditions, we conducted an additional experiment to evaluate its performance on imbalanced datasets.

We simulated various imbalance scenarios by creating five different training sets from our original data, each with a distinct ratio of bug reports to non-bug reports: 1:1 (balanced), 1:4, 1:9, 4:1, and 9:1. All other experimental parameters, including the underlying neural language model and the active learning process, were kept consistent with our main evaluation.

The results of this experiment are presented in Table~\ref{tab:ratio-comparison}. As anticipated, classification performance is influenced by the degree of class imbalance. The F1-score, a metric sensitive to the balance between precision and recall, shows a clear degradation as the data becomes more skewed. For instance, the F1-score dropped from 0.874 in the balanced 1:1 scenario to 0.843 for the 1:9 ratio and more significantly to 0.659 for the 9:1 ratio. This behavior is expected, as the model's learning process is naturally biased by the majority class. When non-bug reports are dominant (1:9), the model's precision decreases to 0.744, indicating a higher rate of false positives. Conversely, when bug reports form the vast majority (9:1), precision reaches a high of 0.947, but the low F1-score suggests a very poor recall, meaning the model struggles to correctly identify the few non-bug reports.

Interestingly, the effort metrics also reveal a notable trend. The readability score is highest (least readable) in the 9:1 scenario (15.409), suggesting that when the model is faced with a highly skewed distribution, the reports it identifies as most uncertain may also be the most complex or poorly structured. Despite these challenges, the results confirm that \approach~remains effective even in imbalanced settings. The framework's core mechanisms continue to guide the model's learning process effectively, demonstrating its resilience and suitability for deployment in practical, real-world environments where balanced data cannot be guaranteed.

\subsubsection{Performance Upper Bound}

\begin{figure}[!t]
\centering
\small 

\begin{subfigure}{0.01\columnwidth}
  \centering
  \includegraphics[width=\linewidth]{fig/blank_stripe.pdf} 
\end{subfigure}

\begin{subfigure}[t]{0.3\columnwidth}
\includegraphics[width=\columnwidth]{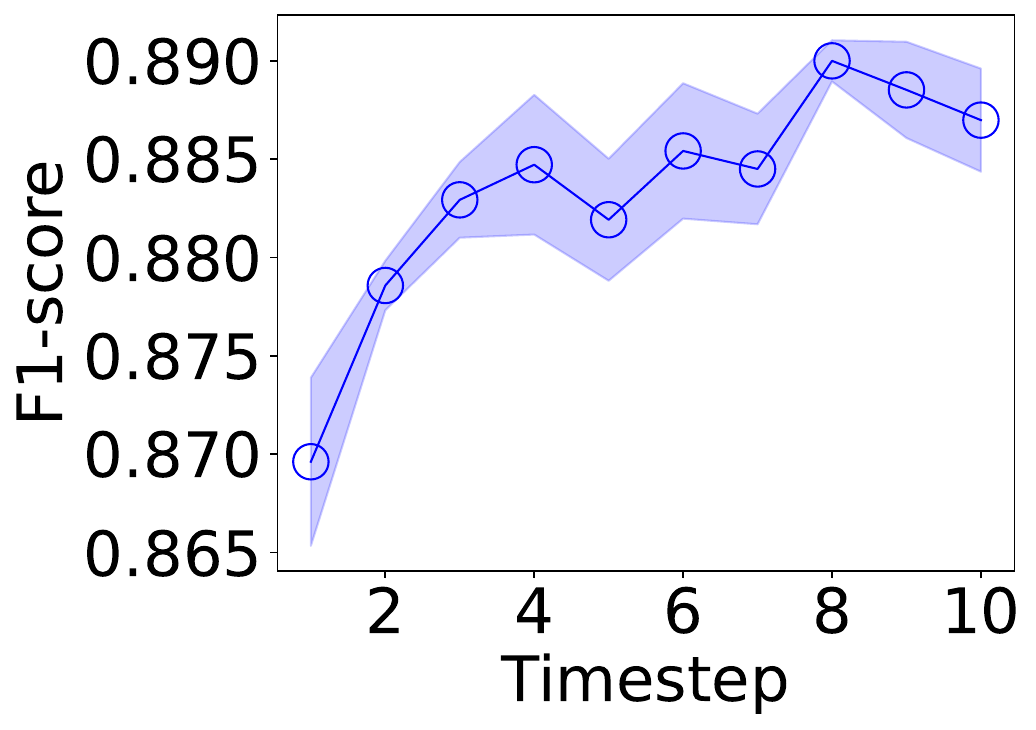}
    \subcaption{F1 Score}
    \end{subfigure}
\begin{subfigure}[t]{0.3\columnwidth}
\includegraphics[width=\columnwidth]{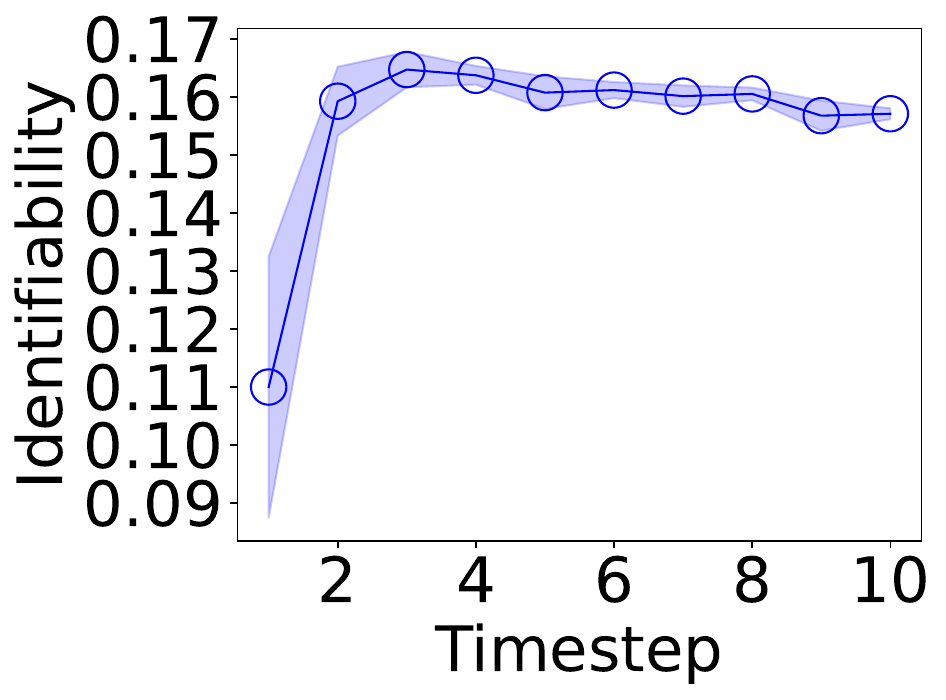}
   \subcaption{Readability}
    \end{subfigure}
\begin{subfigure}[t]{0.3\columnwidth}
\includegraphics[width=\columnwidth]{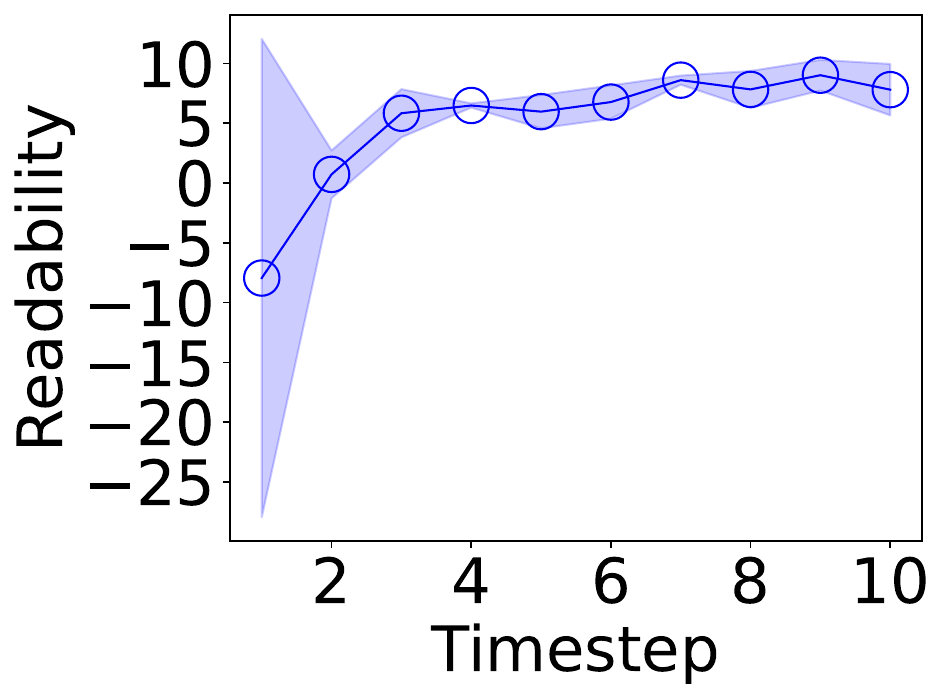}
   \subcaption{Identifiability}
    \end{subfigure}

    \caption{Performance upper bound experiment result of \approach~with metrics of F1 score, readablity and identifiability. }
 \label{fig:upper}
\end{figure}

Although \approach~consistently outperforms the baselines under practical query sizes, an exact performance upper bound remains unknown due to computational constraints. The full dataset contains over 1.27 million reports, making exhaustive fine-tuning prohibitively expensive. To approximate this bound, we conducted our largest feasible experiment. We continued this process until the performance gains became marginal with a step size of 3,000 reports, using a termination criterion of less than 0.1\% F1-score change over consecutive timesteps. This experiment yielded a final F1-score of 0.891, at which point the performance curve began to plateau over 8 timesteps (for a total of 24,000 reports), as shown in Figure \ref{fig:upper}. While we cannot claim this as the true upper bound, the results imply that \approach~has the potential to reach even higher performance with more extensive resources. We acknowledge this as a limitation and leave the precise measurement of the upper bound as future work.

\subsubsection{Diversity Sampling Method}

Our current effort-aware sampling method primarily balances model uncertainty with the readability and identifiability of reports for human labelers. A potential alternative is to incorporate sample diversity, a common strategy in active learning to ensure the queried samples are not redundant and cover a wider feature space. To investigate this, we conducted an experiment replacing our sampling method with a diversity-based approach. The results at the final timestep showed a minor 1.14\% increase in F1-score. However, this was accompanied by a significant decrease in readability (365.48\% reduction) and identifiability (52.14\% reduction). This suggests that the most diverse samples, while potentially informative for the model, are much more difficult for human developers to label. Since a core contribution of \approach~is reducing the human effort in the labeling process, we conclude that our effort-aware approach provides a better balance for effective human-machine collaboration in this context.

\subsubsection{Model Selection for Model-Agnostic Experiment and Justification}

Our model selection process for Section 5.3 was guided by the need to evaluate \approach~under varied conditions, and we chose models that represent different specializations:

\begin{itemize}
    \item A General-Purpose Text Model (RoBERTa): We included RoBERTa as it is a "robust and widely-used extension of BERT" known for its SOTA performance on a wide range of natural language understanding benchmarks. This tests \approach’s generalizability with a highly optimized, all-purpose model.

    \item A Code-Specific Model (CodeBERT): Given that bug reports on GitHub frequently contain code snippets, we chose CodeBERT, which is "a specifically designed model for code naturalness". Its ability to handle both programming and natural languages makes it highly relevant for our domain.

    \item A Task-Specific Model (RTA): To test \approach~in a highly specialized context, we used RTA, a model ``tailored for analyzing bug reports via learning a universal representation''.
\end{itemize}

While we initially performed preliminary experiments using T5, we finally excluded those results from the manuscript because the model consistently failed to generate meaningful outcomes for our specific task. Our investigation revealed several reasons for T5's poor performance:

\textbf{Architectural Mismatch for Classification: }T5 is an encoder-decoder model designed primarily for generative, text-to-text tasks (e.g., summarization, translation). Our task, bug report identification, is a binary classification problem. While T5 can be adapted for classification by training it to generate specific labels (e.g., "bug" or "not bug"), this approach is often less efficient and stable than using encoder-only models like BERT. BERT-style architectures are purpose-built for discriminative tasks, using the  token representation and a simple classification head, which is a more direct and suitable architecture for this problem.

\textbf{Training Instability and Failure to Converge: }During fine-tuning, the T5 model exhibited significant instability. Despite extensive hyperparameter tuning (adjusting learning rates, batch sizes, optimizers, and schedulers), the training loss failed to converge, and the model's accuracy on the validation set remained close to random chance. This suggests that the model architecture and its pre-training objectives are not easily adaptable to learning the specific discriminative features required to distinguish bug from non-bug reports in our dataset.

We believe this architectural mismatch is the primary reason for T5's poor performance. Encoder-only models are inherently better suited for this classification task, and our results with BERT, RoBERTa, CodeBERT, and RTA confirm this.

\subsection{Validity of Readability Metrics with Non-Natural Language Tokens}
\label{subsec:readability_discussion}

{A potential concern raised regarding our effort-aware sampling is whether the FRE score remains a valid proxy for labeling effort given that GitHub Issues frequently contain non-natural language elements, such as code snippets, stack traces, log entries, and file paths. These elements can significantly skew traditional linguistic statistics (e.g., word length and sentence count), sometimes resulting in extremely low or even negative readability scores.}

{\textbf{Processing of Non-Natural Language.} In our implementation, we do not aggressively filter out these symbols. Instead, we treat them as part of the overall textual information that a software developer must process. While tokens like version numbers or complex file paths may not follow standard grammatical structures, they contribute to the visual and cognitive complexity of the report.}

{\textbf{Validity in the Software Engineering Context.} We argue that the FRE score remains an effective indicator of labeling effort in this domain for the following reasons:}
\begin{itemize}
    \item {\textit{Correlation with cognitive load:} Non-natural language elements, particularly dense stack traces or raw logs, inherently increase the time and mental effort required for a developer to comprehend the core issue. A lower (or negative) FRE score accurately reflects this increased "reading cost." Thus, the metric's sensitivity to these tokens is a feature, not a bug, in the context of measuring labeling effort.}
    \item {\textit{Empirical consistency:} In our dataset, bug reports with extensive technical jargon and code fragments consistently received lower FRE scores compared to descriptive, text-heavy reports. This aligns with the practical experience of developers who find "noisy" issues more taxing to triage.}
\end{itemize}

{\textbf{Potential Restrictions.} However, we acknowledge certain potential restrictions. The FRE metric might be less reliable in extreme cases where a report consists \textit{exclusively} of a massive log dump without any descriptive context. In such scenarios, the metric might over-penalize the report's complexity. While our results demonstrate a significant reduction in overall labeling effort, future work could explore hybrid metrics that specifically weight code-to-text ratios to further refine the estimation of human labeling costs in highly technical repositories.}

\subsection{Scalability, Overheads, and Resources}

\subsubsection{Scalability}

{Apart from the model training, scalability can be related to the retrieval, especially the pseudo labeling process. Our current implementation utilizes an exact nearest neighbor search via optimized \texttt{pairwise\_distances} computations. While the theoretical complexity is $O(|\mathbfcal{D}_q| \cdot |\mathbfcal{D}_u|)$, the use of matrix-based operations on} {GPUs significantly accelerates this process. For scenarios with millions of reports, our framework can be seamlessly integrated with Approximate Nearest Neighbor (ANN) libraries such as FAISS or HNSW to achieve sub-linear retrieval time.} 

{Another factor that affects scalability is the updating frequency, for which \approach~relies on the number of reports needed for the update (i.e., up to 700). This lies in the trade-off between the timeliness of the model and efficiency, where we assume that the frequency can be case dependent.}

\subsubsection{Overhead of Updating/Training Time}

revision{\approach~supports warm starts by using a pre-trained neural language model, thus the key training overhead thereof would be the updating process that fine-tunes the model. On a machine of 2.25GHz CPU accelerated by a typical CUDA-enabled A100 GPU and under over one million unlabeled reports from different projects, \approach~needs $\approx$ 53 minutes to complete one cycle of the update within each timestep (with 700 query size). Further analysis of the breakdown reveals that the actual training/fine-tuning of the neural language model only takes $\approx$ 2 minutes because only 700 new reports together with any previously labeled reports are updated/trained. The main proportion of the overhead comes from the retrieval procedure: there are $\approx$ 25 minutes for the effort-aware uncertainty sampling and $\approx$ 26 minutes for pseudo labeling. We would like to stress that the above is mainly caused by the fact that there are over one million unlabeled reports to be processed therein; most of the time required is for iterating over that huge pool of reports. Certainly, the overhead is expected to be reduced with fewer accumulated unlabeled reports. The inference time for automatically identifying a new report, on the other hand, is negligible at the magnitude of milliseconds.}

{Since the interval between different timesteps is flexible, i.e., there could be only one update every a few days or even weeks, an update of less than 1 hour is often considered acceptable as \approach~can be left to run the update overnight, once the queried reports have been labeled, which is not uncommon for updating some large software systems in practice. In particular, considering the dramatic improvement in accuracy and the reduction of human labeling efforts against the state-of-the-art approaches, the benefits of \approach~outweigh its updating/training overhead.}

\subsubsection{Resources}

{With the data scale over one million (labeled and unlabeled) samples, the memory} {footprint of the process is stable, requiring approximately 3--5 GB of VRAM for the BERT model and 2--8 GB of system RAM for storing the distance matrix, depending on the specific sample size. Since the neural embedding space evolves as the neural language model in \approach~is fine-tuned, we re-calculate the embeddings in each iteration to maintain the accuracy of the pseudo-labels.}


\subsection{Practical Application of \approach}

\subsubsection{Using \approach}

\approach~works cross-project, meaning that the training, querying unlabeled reports to the software developers, pseudo-labeling reports, and prediction therein can all occur for different software projects. This provides \approach~with the ability to serve as a centralized bug report identification system for any organization. For example, companies can adopt \approach~to handle the repositories of all their projects. In particular, depending on the quality-effort score, the top $k$ unlabeled reports that would be queried might not be constrained to one project since the pool of unlabeled reports contains reports from all the projects of a company. As such, it is possible for \approach~to ask developers who are specialized in different projects for the labeling, depending on which projects the queried reports belong. Of course, when there is only one project, \approach~naturally reduces to a typical within-project setting.

The warm start property in \approach~can also largely improve its applicability, thanks to the pre-training nature of many neural language models. That is, even when applying \approach~for the very first time, it is possible to build it from a pre-trained version elsewhere, hence saving a dramatic amount of resources and time.

\subsubsection{Human Labeling Effort}

Indeed, in a case where there is plenty of labeled data, human labeling efforts can indeed be costly.  However, in our context, we target the scenarios in which all the labeled reports have been used for training, and we are hitting the point where no more labels can be used (or the labeled data is very limited). For example, in a recent large-scale study over GitHub~\cite{DBLP:journals/infsof/WangZCX22}, it has been found that 54.4\% of the issues studied have no label at all. In such a condition, there is no other way to further improve the model but to use some more labeled data, which should ideally provided by humans. To that end, active learning is a paradigm that is well-suited to a scenario that requires human-machine interaction. We agree that the human labeling process is expensive, which is why in \approach, we specifically design two strategies to relieve that:

\begin{itemize}
    \item In the \textit{Effort-aware Uncertainty Sampling},  we consider readability and identifiability, alongside uncertainty. Those two objectives aim primarily to reduce the labeling effort required by humans, as the selected reports to be labeled are often more information-intuitive.
    \item In the \textit{Pseudo-labeling}, we create pseudo labels, guided by a limited set of human-labeled data, which greatly reduce the number of labels required from humans.
\end{itemize}

The above designs have significantly reduced the human efforts needed in the active learning process while keeping promising effectiveness, as shown by the experiment results.

\subsubsection{Applicable Scenarios of \approach}

Since the neural language model is the base model in \approach, it naturally works for both single and cross-project scenarios. However, it is often more beneficial for an approach to work on cross-project cases because this allows for more diverse training data, and advanced language models can handle such complex data. Therefore, we highlight this as a key feature of \approach; however, this does not mean that it can only work on cross-project data. Our model can also be directly applied to single-project cases, especially those with a particularly large number of bug reports, which allows our approach to have sufficient data for training. \approach~can also serve as a complementary approach with tools from other software engineering tasks, e.g., the report can be used to guide testing~\cite{DBLP:conf/icse/MaCL25,10.1145/3773088}; to identify key/problematic options in configuration modeling~\cite{DBLP:conf/icse/XiangChen26,DBLP:journals/tse/GongCB25,DBLP:conf/sigsoft/Gong023,DBLP:journals/pacmse/Gong024}; to pinpoint useful information for configuration tuning~\cite{DBLP:conf/icse/ChenChen26,DBLP:conf/kbse/XiongC25,chen2024mmo,DBLP:conf/sigsoft/0001L21} and runtime self-adaptation~\cite{DBLP:conf/icse/Ye0L25,DBLP:conf/wcre/Chen22}; and to assist explainability~\cite{10.1145/3803859,DBLP:journals/tse/ChenGC25}.

\subsection{Limitations}

While the \approach~framework presents numerous advantages, it also has certain limitations and potential drawbacks that may arise:

\begin{itemize}
    \item \textbf{Model Training Dependency on GitHub Issues:} The model developed through our experiments is specifically trained on GitHub issue data. If \approach~is to be applied to other bug report repositories, it requires retraining with data from those specific repositories. This dependency on retraining can limit the immediate applicability of the model across different platforms or environments.
    
    \item \textbf{Training Data Size and Time Cost:} When retraining \approach~with large datasets, there may be significant time and computational costs associated with the process. Handling large volumes of training data can lead to extended training times, potentially impacting the efficiency and scalability of the framework. In contrast, baseline methods such as those based on TF-IDF are less computationally expensive, as they do not require the same level of deep learning infrastructure, making them faster to deploy but at the cost of reduced accuracy in identifying complex patterns. However, they may not provide as good performance as \approach due to their inability to capture nuanced semantic relationships in the text. 

    \item \textbf{Lack of Transparency in Neural Language Models:} Neural language models are often described as “black boxes,” and the predictions made by \approach~might lack transparency. This characteristic can make it challenging to understand the rationale behind the model's decisions and may affect the trustworthiness of the results. Other more interpretable models, such as decision trees, are free of this issue. Again, baseline methods such as those based on TF-IDF can provide more transparency, as they rely on straightforward statistical techniques that are easier to interpret. These methods make it clear how certain terms influence the output, which adds a level of explainability that deep learning models may lack. However, this transparency often comes at the cost of performance. This is essentially a trade-off between transparency and performance, and we chose better performance over TF-IDF-based methods, which are more transparent but offer weaker performance.

    \item \textbf{Integration Challenges:} Organizations with established systems may encounter difficulties when integrating \approach~into their existing processes. The complexity of adapting and incorporating our framework into pre-existing workflows could pose practical challenges and hinder its adoption in certain settings.
\end{itemize}

%% file: tab/imb.tex
\begin{table}[t!]
	\centering
	\setlength{\tabcolsep}{0.8mm}
	\caption{Comparing the mean and standard deviation (SD) of the metrics values for different ratios of bug reports to non-bug reports.}
	\label{tab:ratio-comparison}%
\begin{adjustbox}{width=\textwidth}
\begin{tabular}{l|llllllll|llllll}
\toprule
  \multirow{2}{*}{\textbf{Ratio}} &
  \multicolumn{2}{c}{\textbf{F1-score}} &
  \multicolumn{2}{c}{\textbf{Accuracy}} &
  \multicolumn{2}{c}{\textbf{AUC}} &
  \multicolumn{2}{c|}{\textbf{Precision}} &
  \multicolumn{2}{c}{\textbf{Readability}} &
  \multicolumn{2}{c}{\textbf{Identifiability}}\\ \cline{2-13}
   & \textbf{$r$} & \textbf{Mean (SD)} 
   & \textbf{$r$} & \textbf{Mean (SD)} 
   & \textbf{$r$} & \textbf{Mean (SD)} 
   & \textbf{$r$} & \textbf{Mean (SD)} 
   & \textbf{$r$} & \textbf{Mean (SD)} 
   & \textbf{$r$} & \textbf{Mean (SD)} \\ \midrule
  1:1  &\cellcolor{teal!20}1&\cellcolor{teal!20}0.874 ($\pm0.0050$)&\cellcolor{teal!20}1&\cellcolor{teal!20}0.869 ($\pm0.0050$)&\cellcolor{teal!20}1&\cellcolor{teal!20}0.926 ($\pm0.0040$)&3&0.881 ($\pm0.0050$)&5&6.073 ($\pm5.8000$)&\cellcolor{teal!20}1&\cellcolor{teal!20}0.189 ($\pm0.0040$)\\
  1:4  &2&0.870 ($\pm0.0023$)&2&0.851 ($\pm0.0032$)&3&0.918 ($\pm0.0043$)&4&0.804 ($\pm0.0056$)&4&9.746 ($\pm2.9677$)&3&0.156 ($\pm0.0052$)\\
  1:9  &3&0.843 ($\pm0.0053$)&3&0.809 ($\pm0.0085$)&4&0.906 ($\pm0.0114$)&5&0.744 ($\pm0.0107$)&3&9.705 ($\pm3.5003$)&5&0.147 ($\pm0.0062$)\\
  4:1  &4&0.782 ($\pm0.0115$)&4&0.802 ($\pm0.0083$)&2&0.912 ($\pm0.0051$)&2&0.928 ($\pm0.0042$)&2&13.649 ($\pm3.8468$)&2&0.179 ($\pm0.0035$)\\
  9:1  &5&0.659 ($\pm0.0240$)&5&0.725 ($\pm0.0139$)&5&0.897 ($\pm0.0121$)&\cellcolor{teal!20}1&\cellcolor{teal!20}0.947 ($\pm0.0036$)&\cellcolor{teal!20}1&\cellcolor{teal!20}15.409 ($\pm3.2403$)&4&0.164 ($\pm0.0050$)\\
\bottomrule
\end{tabular}
\end{adjustbox}
\end{table}

%% file: sec/threat.tex
\section{Threats to Validity}
\label{sec:threat}

We now discuss the potential threats in this work with respect to different aspects of validity.

\subsection{Internal Validity}
A potential threat to validity is the choice of $k$, the number of samples queried per round. We selected $k=300$, $k=500$, and $k=700$ as a pragmatic balance between effectiveness and cost, but these values may not be optimal. This choice was not rigorously optimized, and different $k$ values could yield varying results. \approach~and its underlying neural language model (e.g., BERT) contains several parameters that require pre-defining their values, which may raise threats to internal validity. To that end, we tune that to fit with our experiment infrastructure, e.g., $18$ epochs, $32$ batch size, learning rate of $3 \times 10^{-5}$ and Epsilon factor of $10^{-8}$.  Indeed, the use of BERT is only a pragmatic choice and can lead to bias. To mitigate that, in \textbf{RQ3}, we examine additional three neural language models with \approach~for confirming its model-agnostic property. Yet, hidden interactions between those settings that we have not discovered are always possible. The definition of identifiability is straight-forward. It does not mean to capture the semantics but rather, it serves as domain-specific keyword lists that help to filter those reports that are highly readable but still contain less information about bug relevance. Besides, it has the benefit of lightweight computation while fully capturing the semantics would certainly be more expensive. However, such a simple metric works really well in our large-scale dataset and we leave the capture of richer semantics as part of the future work, which might further improve the effectiveness of \approach. 

Not knowing the upper bound of this method could be another internal threat. Feeding the models with data from the dataset is the most ideal scenario. However, practically, we do not have the necessary resources to do that. This is because conducting a 10 repeated time on a standard BERT after 10 timesteps with a total of 7,000 samples has already consumed 9 hours of training. Considering the fact that the dataset has 1,275,881 reports, the time/resource required in fine-tuning increases dramatically. Not to mention the complexity of running the entire experiment, which involves varying query sizes and different modification settings. 



\subsection{Construct Validity}

The metrics in the experiment can introduce bias on construct validity. Their interpretation can vary based on the context and the specific nuances of the dataset. It is also worth noting that there's the potential risk of overemphasizing certain metrics at the expense of others, leading to a biased understanding of the true performance. To mitigate this, we use several most common performance metrics together with those that measure the efforts of labeling. We repeat the experiment 10 runs and use statistical tests to ensure the reliability of the conclusion. For our qualitative study, the participants involved might not truly represent real developers, and their judgments may be subjective, but this constraint is not uncommon for studies with humans and fits with real-world scenarios well. 
\subsection{External Validity}
Threats to external validity can be sourced from different aspects. For example, the choice of the NLBSE dataset and the model could influence the generalizability of our findings. Yet, this dataset, to the best of our knowledge, is the most comprehensive one to date while of large volume ($>1$ million reports), covers cross-projects, and contains diverse report formats. To fully evaluate \approach, we have assessed it over three neural language models and against four other state-of-the-art approaches. Yet, admittedly, adding more subjects may prove fruitful. The qualitative human study involved 10 participants, which may limit the external validity. Though this is not a significantly large number, it is a reasonable sample size to start with, provided that appropriate statistical tests are used, as suggested in the guideline of human-involved experiments in software engineering~\cite{DBLP:journals/ese/KoLB15}. 


%% file: sec/related.tex
\section{Related Work}
\label{sec:related}

We now discuss the related work in light of the contributions from \approach.

\subsection{Supervised Learning to Identify Bug Reports}

Supervising learning that assumes a model can be built with a sufficient amount of data beforehand has been a major research direction for bug report identification in the past decade. Among others~\cite{kukkar2018supervised,DBLP:journals/smr/ZhouTGG16,DBLP:conf/qrs/XiaLJW19,DBLP:conf/issre/ZhengZTCCWS21,DBLP:journals/infsof/ChoLK22,DBLP:conf/oss/Rodriguez-Perez16,kwak2022issue,DBLP:journals/tse/FanXLH20,DBLP:conf/issre/HeXF0YL20,DBLP:journals/complexity/LiQGGCG20,DBLP:journals/alr/Polpinij21,DBLP:journals/tr/DuZXZT22}, majority of the work did not distinguish the type of bug that a report is associated with, i.e., as long as they are bug related then the model should put classify them into the same category. \approach~also falls into the same type of approaches that do not distinguish the bug type. In contrast, others focus on identifying reports relevant to a particular bug type, e.g., high-impact bugs~\cite{DBLP:journals/jcst/YangLXHS17,DBLP:journals/smr/ZhouTGG16,DBLP:conf/compsac/YangLHXS16}, security bugs~\cite{DBLP:journals/infsof/JiangLSW20,DBLP:journals/ijseke/GuoCLZL19,kukkar2018supervised,das2018security}, or performance bugs~\cite{brown2021deeplabb}, \textit{etc}. Yet, regardless of whether a specific bug type is considered, the proposed approach for bug report identification can be equally applicable with no or minor amendments. In what follows, we discuss them in terms of the model proposed.

\subsubsection{Statistical Machine Learning Model}














Traditionally, report identification with machine learning has been leveraging simple but interpretable models that are paired with certain embedding methods, which most commonly include term-frequency and term frequency-inverse document frequency. Among others, \citeauthor{DBLP:journals/jcst/YangLXHS17}~\cite{DBLP:journals/jcst/YangLXHS17} and \citeauthor{kukkar2018supervised}~\cite{kukkar2018supervised} use $k$NN to aid developers with extracting information from the reports for debugging. \citeauthor{DBLP:conf/qrs/DingFYH21}~\cite{DBLP:conf/qrs/DingFYH21} use Naive Bayes (NB), and \citeauthor{DBLP:conf/compsac/YangLHXS16}~\cite{DBLP:conf/compsac/YangLHXS16} and \citeauthor{DBLP:journals/smr/ZhouTGG16}~\cite{DBLP:journals/smr/ZhouTGG16} use Naive Bayes Multinomial (NBM), respectively, to classify bug related reports. The probabilistic nature of those models offers insights into the likelihood of a report is bug related or not.


More complex models exist for bug report identification. For example, \citeauthor{DBLP:journals/alr/Polpinij21}~\cite{DBLP:journals/alr/Polpinij21} and \citeauthor{DBLP:journals/tse/FanXLH20}~\cite{DBLP:journals/tse/FanXLH20} use Support Vector Machines (SVM) while \citeauthor{DBLP:journals/ijseke/GuoCLZL19}~\cite{DBLP:journals/ijseke/GuoCLZL19}, \citeauthor{brown2021deeplabb}~\cite{brown2021deeplabb} and \citeauthor{DBLP:conf/qrs/XiaLJW19}~\cite{DBLP:conf/qrs/XiaLJW19} adopt Multilayer Perceptron (MLP). SVM and MLP demonstrate versatility in capturing intricate patterns and decision boundaries present in bug reports. Furthermore, by pairing with encoding techniques such as Doc2Vec~\cite{DBLP:conf/icml/LeM14}, the generalization abilities of SVM and MLP make them valuable for bug identification for bug reports with various formats and characteristics.


An alternative learning paradigm is ensemble learning, such as random forest. In the realm of bug report identification, those ensemble models emerge as a versatile and powerful approach, synergizing individual models to enhance overall performance and reliability as demonstrated by various research studies in this domain~\cite{DBLP:journals/complexity/LiQGGCG20,DBLP:journals/smr/ZhouTGG16,DBLP:journals/corr/abs-1905-06872}.




However, due to the restricted ability of statistical machine learning models to understand and represent the nature of texts, the above work is often ineffective in extracting and parsing the information of a submitted report.

\subsubsection{Neural Language Model}


To overcome the limitation of the statistical machine learning model, existing work has started to explore the power of neural language models for bug report identification. For example, \citeauthor{DBLP:conf/issre/ZhengZTCCWS21}~\cite{DBLP:conf/issre/ZhengZTCCWS21} incorporate attention-based Bi-directional Long Short-Term Memory (BiLSTM)---a typical language model that handles long term dependencies with improved computational efficiency---to automatically extract text features from vulnerability descriptions the submitted reports. By pairing with \texttt{FastText}~\cite{DBLP:journals/corr/JoulinGBDJM16} for word embedding, BiLSTM is designed with a tailored loss function that fits the characteristics of bug reports. Attention weights are also extracted to highlight significant text features aiding in the identification of tactical vulnerabilities. Similarly, Convolutional Neural Network (CNN), which is commonly used for image processing, has also been adopted for identifying bug reports~\cite{DBLP:journals/infsof/ChoLK22,DBLP:conf/issre/ZhengZTCCWS21,DBLP:journals/tr/DuZXZT22,DBLP:conf/issre/HeXF0YL20}, in which the word embedding is represented as a matrix of image pixels. The structure of convolutional layers helps capture local patterns and features while pooling layers behind allows reduced dimension. This approach is effective for tasks like text classification or sentiment analysis, where local relationships between words are crucial~\cite{DBLP:journals/jzusc/ChengWLHL18}. Additionally, pre-trained word embedding or convolutional layers can be used to leverage knowledge from large text corpora. In particular, \citeauthor{DBLP:journals/infsof/ChoLK22}~\cite{DBLP:journals/infsof/ChoLK22} exploit not only the information from the reports but also the key messages from the user manual for identifying whether the report is bug related. A unique characteristic of applying those CNN models, when used for bug report identification, is that they often work better with word embedding techniques such as \texttt{FastText}~\cite{DBLP:journals/corr/JoulinGBDJM16} or \texttt{GloVe}~\cite{DBLP:conf/emnlp/PenningtonSM14}. Interestingly, \citeauthor{DBLP:conf/issre/ZhengZTCCWS21}~\cite{DBLP:conf/issre/ZhengZTCCWS21}  propose a multi-modal model-based that combines image and text contained in a submitted report, which tends to improve the overall prediction performance.

Comparing to CNN and LSTM, BERT excels in capturing contextualized word representations through bidirectional context analysis. Its self-attention mechanism allows for better handling of long-range dependencies, and pre-training on large corpora enables efficient transfer of knowledge from general text classification task to bug report identification domain. The utilization of BERT as a model has been extensively explored in existing literature. For example, \citeauthor{DBLP:journals/infsof/MeherBM24}~\cite{DBLP:journals/infsof/MeherBM24} employ a heuristic annotation approach to annotate a substantial dataset of software bug reports. Subsequently, they leverage BERT for the classification of these carefully curated bug reports. The utilization of BERT proves advantageous, capitalizing on the contextual features embedded in the bug reports, facilitated by the attention mechanism. Similarly, \citeauthor{DBLP:journals/corr/abs-2202-06149}~\cite{DBLP:journals/corr/abs-2202-06149} introduce a methodology for classifying bug reports in a multi-label setting. The authors employ an off-the-shelf neural network named RoBERTa, fine-tuning it for the bug report classification task. As an optimized variant of BERT, RoBERTa utilizes an enhanced training methodology and larger datasets, resulting in superior performance compared to BERT. Their preference for RoBERTa over BERT is driven by its state-of-the-art results on benchmark datasets.

In a distinctive approach, \citeauthor{DBLP:journals/infsof/GomesTC23}~\cite{DBLP:journals/infsof/GomesTC23} present a methodology that utilizes BERT as an embedding method. They extract features using BERT and subsequently input the data into statistical machine learning models. Their findings reveal that the long-lived bug report identification achieved through BERT-based feature extraction consistently outperformed TF-IDF. Specifically, SVM and Random Forest classifiers demonstrated superior performance across nearly all datasets when utilizing BERT as the feature extraction method.

Furthermore, \citeauthor{DBLP:conf/icse/ShiMZYCCJJ022}~\cite{DBLP:conf/icse/ShiMZYCCJJ022} identify the bug report from separated dialogs by utilizing a graph-based context embedding approach to create a dialog graph. Furthermore, they employ a two-layer graph neural network to assimilate contextual information. This approach proves effective in capturing the graphical context of utterances within a single dialog, comprehending structure-level context for each vertex in a specified graph relationship, and encapsulating high-level contextual information.

Despite the adoption of various neural language models, existing work has not adopted powerful models that are compatible with pre-training capacity (e.g., BERT), which renders cross-project identification plausible. This is one of the key components in \approach. Further, the above often assumes a sufficient amount of labeled reports is readily available and one is willing to accept a significant amount of training overhead (even for fine-tuning), which might not be applicable to all real-world scenarios. \approach, on the other hand, leverage neural active learning for bug report identification via boosted human-machine teaming.


\subsection{Active Learning for Bug Report Identification}

To mitigate the prerequisite on the amount of labeled data and to relax the requirements of training on all data samples, the paradigm of active learning has been a promising direction for bug report identification~\cite{DBLP:conf/kbse/WangWCW16,DBLP:journals/infsof/GeFQGQ22,DBLP:journals/infsof/WuZCZYM21,DBLP:journals/tse/TuYM22,DBLP:conf/iwpc/ThungLL15}. \citeauthor{DBLP:conf/kbse/WangWCW16}~\cite{DBLP:conf/kbse/WangWCW16} and \citeauthor{DBLP:journals/infsof/GeFQGQ22}~\cite{DBLP:journals/infsof/GeFQGQ22} adopt active learning for identifying bug related reports where the queried samples are determined by using local neighborhoods of previously useful reports for improving model performance. Similarly, \citeauthor{DBLP:journals/infsof/WuZCZYM21}~\cite{DBLP:journals/infsof/WuZCZYM21} propose \texttt{hbrPredictor}, which also relies on active learning to identify high-impact bug related reports but they additionally exploits stronger human interaction by combining interactive machine learning with active learning.

There are also approaches that work with extended active learning. \texttt{EMBLEM}~\cite{DBLP:journals/tse/TuYM22} is an active learning approach for bug report identification, which relies on the active learning strategy that is a mix of certainty and uncertainty in the selection of reports to query. There also exists a combination of semi-supervised learning and active learning with an aim to reduce human efforts~\cite{DBLP:conf/iwpc/ThungLL15}. 

Other similar work includes HINT~\cite{DBLP:conf/icse/GaoMG000L24}, which focuses on hybrid pseudo-labeled data selection and noise-tolerant training to improve pre-trained code models by mitigating noise in pseudo-labeled data. While both MNAL and HINT leverage pseudo-labeling to enhance model performance , our direct experimental comparison reveals that MNAL achieves a better F1-score, accuracy, and recall. More fundamentally, the approaches differ in their core objectives. HINT's primary contribution is handling noisy pseudo-labels. In contrast, MNAL establishes a human-machine collaborative loop, centering on active learning and using readability and identifiability metrics to reduce the effort for human labelers, a crucial aspect not addressed by HINT.


The above work mainly targets within-project while they did not adopt a neural language model, In contrast, \approach~works for cross-project, and the selection of queried reports directly optimizes for better readability and identifiability. In addition, they did not directly select reports that are naturally more readable and identifiable which is one of the key contributions in \approach. We also enrich the samples to update the neural language model by considering pseudo-labeled reports. Such a mutualistic relation, underpinned by neural active learning, has enabled \approach~to achieve more promising results.

\subsection{Empirical Studies for Bug Report Identification}

To better design models for bug report identification, empirical studies have been conducted to understand the characteristics of the identification problem. Among others, \citeauthor{DBLP:journals/isse/PandeySHS17}~\cite{DBLP:journals/isse/PandeySHS17} examine six statistical machine learning models in identifying bug related reports for three open-source projects. Their experiments reveal that random forests and SVM with specific kernels tend to be superior to the others. Some other studies focus on techniques that deal with data unbalancing. For example, \citeauthor{DBLP:journals/tr/ZhengXWDCS21}~\cite{DBLP:journals/tr/ZhengXWDCS21} study six imbalance mitigation methods and found that the performance of the bug identification can be notably enhanced in over 78\% of cases by employing class rebalancing methods such as Rose~\cite{lunardon2014rose} and Mahakil~\cite{bennin2017mahakil} that is paired with CNN. Other work that concentrates on understanding how mislabeling can affect the model performance in identifying bug reports also exist~\cite{DBLP:journals/access/AfricVSD23,DBLP:journals/tse/WuZXL22}. They discovery that the prediction model exhibits significantly better performance on clean datasets compared to noisy datasets and employing advanced neural language models like RoBERTa is recommended for bug report identification.

We consider \approach~a general and flexible approach, which can be complementary to some of the above findings. For example, while we do not consider mislabeled reports due to human errors, some of the findings from \citeauthor{DBLP:journals/access/AfricVSD23}~\cite{DBLP:journals/access/AfricVSD23} and \citeauthor{DBLP:journals/tse/WuZXL22}~\cite{DBLP:journals/tse/WuZXL22} can serve as the foundation to take that into account, hence achieving labeling with better quality as part of the neural active learning.

%% file: sec/conclusion.tex
\section{Conclusion}
\label{sec:conclusion}

In this paper, we introduced a novel cross-project framework \approach~for automated and more effective identification of bug reports from GitHub repositories co-boosted by human-machine collaboration. Going beyond traditional active learning, \approach~introduces a mutualistic relation between humans (software developers) and machines (neural language model) for enhanced human-machine teaming when updating/training the model. This is achieved through effort-aware uncertainty sampling, which queries unlabeled reports requiring an acceptable effort level for labeling. Additionally, we proposed a pseudo-labeling approach to enrich the data for updating/training the neural language model that represents the reports in a latent space. To investigate the effectiveness of our approach, we experiment on large-scale bug report datasets containing a total of 1,275,881 reports collected from different software projects (over 127,000). The experimental results reveal that

\begin{enumerate}
    \item  The effort-aware uncertainty sampling in \approach~demonstrates competitive evolving performance while significantly reducing the effort required for manual labeling of queried reports by developers. This results in a substantial 98.1\% and 194.7\% reliability and identifiability improvement, respectively, both of which demonstrate statistical significance overall.
    \item  Pseudo-labeling in \approach~contributes to a substantial improvement in the performance of the neural language model, with minimal impact on readability and identifiability.
    \item  \approach~is prone to be model-agnostic, showing considerable performance improvement regardless of the underlying neural language model employed. This improvement comes with significantly reduced efforts required for labeling. Specifically, the superiority lies in readability and identifiability, with improvements of 78.6\% and 171.5\%, respectively, supported by significant statistical evidence.
    \item \approach~significantly outperforms other state-of-the-art approaches on various performance metrics and with up to 95.8\% and 196.0\% improvement for readability and identifiability, respectively.
    \item Through a case study with human involvement, we show that, in comparison with randomly selected reports for human labeling, \approach~achieves an effort reduction within the same time frame, leading to $3\times$ monetary savings, while also showing 74.7\% and 64.8\% improvement in qualitative readability and identifiability, respectively.
\end{enumerate}

This work is merely one step towards an enhanced human-machine collaborative approach for software data analytics, in which the convenience and benefit of human involvement are explicitly taken into account while the model is also greatly improved. Along this thread of research, the future opportunities of intervention are vast, including better handling of mislabeled reports due to human labeling error and improved efficiency of utilizing data, e.g., by means of combining few-short learning and neural active learning.
